\documentclass[final,5p,times,twocolumn,authoryear]{elsarticle}
\usepackage[utf8]{inputenc}
\usepackage{tensor}
\usepackage{hyperref}
\usepackage{amsfonts}
\usepackage{amsmath}
\DeclareMathOperator{\diag}{diag}
\usepackage{amssymb}
\usepackage{graphicx}%
\usepackage{bigints}
\usepackage{graphicx}
\usepackage{subfigure}
\usepackage{epsf}
\usepackage{bm}
\usepackage{amsmath}
\usepackage{amsfonts}
\usepackage{amssymb}
\usepackage{graphicx}
\usepackage{tabularx}
\usepackage{color}%
\usepackage{multirow}
\providecommand{\U}[1]{\protect\rule{.1in}{.1in}}
\usepackage{bigints}
\newcommand{\be}{\begin{equation}}
\newcommand{\ee}{\end{equation}}
\newcommand{\Be}{\begin{eqnarray}}
\newcommand{\Ee}{\end{eqnarray}}
\newcommand{\f}{\frac}
\newcommand{\pa}{\partial}

\newcommand{\mincir}{\raise
-3.truept\hbox{\rlap{\hbox{$\sim$}}\raise4.truept\hbox{$<$}\ }}
\newcommand{\magcir}{\raise
-3.truept\hbox{\rlap{\hbox{$\sim$}}\raise4.truept\hbox{$>$}\ }}

\newcolumntype{Y}{>{\centering\arraybackslash}X}
\providecommand{\U}[1]{\protect\rule{.1in}{.1in}}

\usepackage[utf8]{inputenc}

\journal{Physics of the Dark Universe}

\begin{document}

\begin{frontmatter}



\title{Dark matter signatures of black holes with Yukawa potential}


\author[1,2]{A. A. Ara\'{u}jo Filho}
\ead{dilto@fisica.ufc.br}
\affiliation[1]{organization={Departamento de Física Teórica and IFIC, Centro Mixto Universidad de Valencia--CSIC. Universidad
de Valencia}, addressline={Burjassot--46100, Valencia, Spain}}
\affiliation[2]{organization={Departamento de Física, Universidade Federal da Paraíba}, addressline={Caixa Postal 5008, 58051-970, João Pessoa, Paraíba, Brazil}}
\author[3]{Kimet Jusufi}
\ead{kimet.jusufi@unite.edu.mk}
\affiliation[3]{organization={Physics Department, University of Tetova},  addressline={ Ilinden Street nn, 1200, Tetova, North Macedonia}}
\author[4]{B. Cuadros--Melgar}
\ead{bertha@usp.br}
\affiliation[4]{organization={Engineering School of Lorena - University of Sao Paulo (EEL-USP)}, addressline={Estrada Municipal do Campinho Nº 100, Campinho, CEP: 12602-810, Lorena, SP - Brazil}}
\author[5,6]{Genly Leon}
\ead{genly.leon@ucn.cl}
\affiliation[5]{organization={Departamento de Matemáticas, Universidad Católica del Norte},  addressline={ Angamos 0610, Casilla 1280 Antofagasta, Chile}}
\affiliation[6]{organization={Institute of Systems Science, Durban University of Technology},  addressline={PO Box 1334, Durban 4000, South Africa}}

\begin{abstract}
This study uses a nonsingular Yukawa--modified potential to obtain a static and spherically symmetric black hole solution with a cosmological constant.  Such Yukawa--like corrections are encoded in two parameters, $\alpha$ and $\lambda$, that modify Newton's law of gravity in large distances, and a deformation parameter $\ell_0$, which plays an essential role in short distances. The most significant effect is encoded in $\alpha$, which modifies the total black hole mass with an extra mass proportional to $\alpha M$, mimicking the dark matter effects at large distances from the black hole. On the other hand, the effect due to $\lambda$ is small for astrophysical values. We scrutinize the \textit{quasinormal} frequencies and shadows associated with a spherically symmetric black hole and the thermodynamical behavior influenced by the Yukawa potential. In particular, the thermodynamics of this black hole displays a rich behavior, including possible phase transitions. We use the WKB method to probe the \textit{quasinormal} modes of massless scalar, electromagnetic, and gravitational field perturbations. In order to check the influence of the parameters on the shadow radius, we consider astrophysical data to determine their values, incorporating information on an optically thin radiating and infalling gas surrounding a black hole to model the black hole shadow image. In particular, we consider Sgr A* black hole as an example and we find that its shadow radius changes by order of $10^{-9}$, meaning that the shadow radius of a black hole with Yukawa potential practically gives rise to the same result encountered in the Schwarzschild black hole. Also, in the eikonal regime, using astrophysical data for Yukawa parameters, we show that the value of the real part of the QNMs frequencies changes by $10^{-18}$. Such  Yukawa--like corrections are, therefore, difficult to measure by observations of gravitational waves using the current technology.
\end{abstract}




\end{frontmatter}




\section{Introduction}

In modern cosmology our understanding posits that the universe displays a fundamental property of homogeneity and isotropy on its largest scales. Additionally, cosmologists propose the existence of cold dark matter as a mysterious form of matter that remains invisible and interacts solely via the gravitational force \cite{Gonzalez:2023rsd,Bond:1984fp,Trimble:1987ee}. Despite persistent efforts, direct detection of dark matter particles has proven elusive with its existence inferred fundamentally from its gravitational effects on galaxies and other expansive cosmic structures. On the other hand, another intriguing component known as dark energy is introduced to explain the observed accelerated expansion of the universe \cite{Carroll:1991mt}. This proposition garners strong support from a wealth of observational evidence \cite{SupernovaCosmologyProject:1997zqe,SupernovaSearchTeam:1998fmf,SupernovaCosmologyProject:1998vns}.

The $\Lambda$CDM paradigm has emerged as the preeminent model in modern cosmology characterized by its remarkable ability to account for a wide array of cosmological observations with excellent economy in terms of its parameter set \cite{Planck:2018vyg}. Yet, mysteries persist within fundamental physics notably concerning the elusive nature of dark matter and dark energy. In addition, the role of scalar fields in the physical description of the universe, particularly in the context of inflationary scenarios, has garnered considerable attention \cite{Guth:1980zm}. Expanding upon the $\Lambda$CDM framework, a quintessence scalar field has been incorporated in a generalized model to delve into these cosmic enigmas \cite{Ratra:1987rm,Parsons:1995kt,Rubano:2001xi,Saridakis:2008fy,Cai:2009zp,WaliHossain:2014usl,Barrow:2016qkh}. Additionally, multi--scalar field models have emerged as versatile tools capable of elucidating various cosmic epochs \cite{Elizalde:2004mq,Elizalde:2008yf,Skugoreva:2014ena,Saridakis:2016mjd,Paliathanasis:2019luv,Banerjee:2022ynv,Santos:2023eqp}. Moreover, efforts have been made to forge a unified description encompassing matter--dominated and dark energy--driven epochs leading to the development of scalar--torsion theories with a Hamiltonian foundation \cite{Leon:2022oyy}. 

Nonetheless, it is essential to acknowledge that an alternative current of thought exists in the literature advocating for a departure from this conventional path. This perspective suggests that by modifying Einstein's equations novel theories of gravity may provide compelling explanations for observed phenomena challenging the traditional framework \cite{CANTATA:2021ktz,Leon:2009rc,DeFelice:2010aj,Clifton:2011jh,Capozziello:2011et,DeFelice:2011bh,Xu:2012jf,Bamba:2012cp,
Leon:2012mt,Kofinas:2014aka,Bahamonde:2015zma,Momeni:2015uwx,Cai:2015emx,Krssak:2018ywd,Dehghani:2023yph}.
When addressing the enigma of dark matter, a vital component for explaining the curious flatness observed in galaxy rotation curves \cite{Salucci:2018eie}, one of the initial theories posited to account for this phenomenon was the Modified Newtonian Dynamics (MOND) introduced by Milgrom \cite{Milgrom:1983ca}. This theory brings about alterations to Newton's classical law of gravitation \cite{Ferreira:2009eg,Milgrom:2003,Tiret:2007kq,Kroupa:2010hf,Cardone:2010ru,Richtler:2011zk}. 
In addition, there are other intriguing proposals in the realm of dark matter. These include concepts such as superfluid dark matter \cite{Berezhiani:2015bqa} and the intriguing notion of a Bose–Einstein condensate \cite{Boehmer:2007um}, among others. 

On the other hand of the cosmic spectrum, black holes are captivating astronomical phenomena capable of testing gravitational theories under extreme conditions. Among the many fascinating aspects of black holes, one that particularly intrigues scientists is their shadowy silhouette. This outline emerges due to the immense gravitational pull exerted by the black hole distorting the paths of light rays that venture too close.

In essence, photons emitted from a brilliant source in the vicinity of a black hole face two distinct destinies: they may either succumb to the irresistible gravitational force spiralling inexorably toward the event horizon, or they may be deflected away embarking on an eternal journey into the vast cosmic expanse. Within this intricate dance critical geodesic trajectories delineate the boundary between these two outcomes known as unstable spherical orbits. By meticulously observing these pivotal paths of photons against the backdrop of the cosmos, we gain the remarkable ability to capture the elusive image of a black hole's shadow \cite{Takahashi:2004xh,Hioki:2009na,Brito:2015oca,Cunha:2015yba,Ohgami:2015nra,Moffat:2015kva,Abdujabbarov:2016hnw,Cunha:2018acu,Mizuno:2018lxz,Tsukamoto:2017fxq,Psaltis:2018xkc,Amir:2018pcu,Gralla:2019xty,Bambi:2019tjh,Cunha:2019ikd,Khodadi:2020jij,Perlick:2021aok,Vagnozzi:2022moj,Saurabh:2020zqg,Jusufi:2020cpn,Tsupko:2019pzg}.
In recent breakthroughs the Event Horizon Telescope (EHT) collaboration achieved a monumental milestone by confirming the presence of black hole shadows for two supermassive black holes, M87 and Sgr A* \cite{EventHorizonTelescope:2019dse,EventHorizonTelescope:2020qrl,EventHorizonTelescope:2021srq,EventHorizonTelescope:2022wkp}.

In this paper we investigate the phenomenology of Yukawa--like corrections on the black hole spacetime. The Yukawa effects are associated with the presence of a graviton mass leading to modifications in Newton's law of gravity, particularly at significant distances. These phenomena have revealed compelling implications in cosmology establishing connections between dark matter, dark energy, and baryonic matter \cite{Jusufi:2023xoa, Gonzalez:2023rsd}.
Theoretical and observational aspects of the Yukawa potential, dGRT gravity, and dark matter in cosmology have been studied extensively. Some examples of these studies can be found in these references, \cite{Loeb:2010gj,
anderson2017destruction, 
Kanzi:2020cyv,
Cedeno:2017sou,
Berezhiani:2009kv, 
Sakalli:2018nug}.

Our goal is to examine the effects of the Yukawa potential term in the newly discovered solution presented in this paper. The metric \eqref{exact-sol-full} is an original black hole solution due to the contribution of the deformation parameter $\ell_0$. This solution extends the black hole solution presented in \cite{Gonzalez:2023rsd} and includes short- and long-range modifications. However, we will focus more on discussing the solution's phenomenological aspects for astrophysical black holes where the impact of $\ell_0$ is minimal. The accuracy of the approximate solution will be examined using procedures belonging to the realm of perturbation methods in differential equations.
We will first check the metric's thermodynamic stability by analyzing its properties such as temperature, entropy, and specific heat. Next, we will test the geometry's dynamic stability by considering various perturbations and determining the corresponding quasinormal modes (QNM) of the system. Next, we will examine how the black hole geometry affects test particles by calculating the geodesic trajectories focusing on the lightlike case that produces the shadow of a black hole. 
Furthermore, we will investigate the connection between shadows and QNMs in the eikonal limit. As we proceed, we will use observational data to constrain the parameter $\alpha$, which characterizes the Yukawa potential contribution, and compare our findings to the Schwarzschild solution.

The paper is organized as follows. We present the black hole solution with Yukawa potential in $\S$ \ref{sectII}, pointing out its correspondence with de Rham, Gabadadze, and Tolley (dRGT) gravity and discussing the energy conditions obeyed by the apparent fluid emerging from the model. In $\S$  \ref{accuracy} we explore the accuracy of the approximate solution using procedures belonging to the realm of perturbation methods in differential equations. In $\S$  \ref{sectIII}  we study the thermodynamic properties of the system. In $\S$  \ref{sectIV} we study QNMs for scalar, electromagnetic, and gravitational field perturbations. 
Geodesic trajectories are treated in $\S$  \ref{sectVI}. Furthermore, 
$\S$  \ref{sectVII} investigates the shadow image using an infalling gas model. In addition, $\S$  \ref{sectVIII} elaborates on checking the eikonal QNMs, the correspondence with the shadow radius, and the numerical values for the eikonal QNMs. Finally, we discuss our results in $\S$  \ref{sectIX}.

\section{Black hole solution with Yukawa potential}
\label{sectII}
{We shall begin this section by pointing out the main physical motivation to include a Yukawa-modified potential in the context of black holes. Our first argument comes from cosmology. Namely, as was shown in~\cite{Jusufi:2023xoa} and~\cite{Gonzalez:2023rsd}, starting from the Yukawa potential, 
\begin{equation}
\Phi(r)=-\frac{ M m}{r}\left(1+\alpha\,e^{-\frac{r}{\lambda}}\right).
\end{equation}

One can effectively obtain the $\Lambda$CDM model in cosmology as follows. There exists an expression for the  dark matter [units $c=\hbar=G=1$] given by, 
\begin{equation}
    \Omega_{DM}= \frac{\sqrt{2 \alpha \Omega_{B,0}}}{\lambda H_0\,(1+\alpha)} 
\,{(1+z)^{3}}\,,
\end{equation}
here the subscript `0' indicates quantities evaluated at the present time, namely $z=0$.
This implies that dark matter can be understood as a consequence of the modified Newton law quantified by $\alpha$ and $\Omega_B$. If the contribution of baryonic matter disappears, i.e., $\Omega_{B,0}=0$, automatically the effect of dark matter also vanishes, i.e., $\Omega_{DM}=0$. In other words, dark matter can be viewed as an apparent effect. 
Furthermore, one can find an expression that relates baryonic matter, effective dark matter, and dark energy as,
\begin{equation}\label{DMCC}
   \Omega_{DM}(z)= \sqrt{2\, \Omega_{B,0}  \Omega_{\Lambda,0}}{(1+z)^3},
\end{equation}
where we introduced the definition,
\begin{equation}
    \Omega_{\Lambda,0}= \frac{1}{\lambda^2 H^2_0}\frac{\alpha}{(1+\alpha)^2}.
\end{equation}
Using the observational data that were reported in~\cite{Gonzalez:2023rsd}, it was shown that one can get $\Omega_{\Lambda,0}\simeq 0.69$ and $\Omega_{DM}\simeq 0.26$ from cosmological observations. This of course is a very interesting result as one can effectively obtain the $\Lambda$CDM parameters from Yukawa cosmology. A second  motivation is to naturally extend the discussion of dark matter to the black hole geometry. It might be possible, as in the case of cosmology, to analyze the effect of a  dark matter component for black holes using the same physics, namely through the Yukawa potential. }

First of all, let us review the black hole solution with the effect of dark matter recently studied in  
\cite{Gonzalez:2023rsd}. The gravitational potential we considered is modified by 
the regular Yukawa--type potential
\begin{equation}
\Phi(r)=-\frac{ M m}{\sqrt{r^2+\ell_0^2}}\left(1+\alpha\,e^{-\frac{r}{\lambda}}\right).
\end{equation}

Notice that the wavelength of
massive graviton is represented by $\lambda=\frac{\hbar}{m_g c}$ and $\ell_0$ is a deformed parameter of Planck length size. Let us see how the spacetime geometry around the 
black hole is modified in this theory. The general solution in the case of a static, spherically symmetric source reads,
\begin{equation}
ds^2
= -f(r) dt^2 +\frac{ dr^2}{f(r)} +r^2(d\theta^2+\sin^2\theta d\phi^2). \label{metric}
\end{equation}

The energy density of the modified matter can be computed from 
$
\rho(r)=\frac{1}{4\pi} \Delta \Phi(r)
$. After considering it, we obtain the following relation for the energy density,
\begin{eqnarray}
    \rho(r)=\frac{e^{-\frac{r}{\lambda}}M \alpha \mathcal{X}}{4 \pi r \lambda^2 (r^2+\ell_0^2)^{5/2}}+ \frac{3M \ell_0^2}{4 \pi (r^2+\ell_0^2)^{5/2} }
    \label{energydensity},
\end{eqnarray}
where we defined $\mathcal{X}=(2\lambda-r)\ell_0^4+(3 \lambda^2 r+2\lambda r^2-2r^3)\ell_0^2-r^5$. The energy density, therefore, consists of two terms: the first one is proportional to $\alpha$ and it is important in large distances; the second one is proportional to $\ell_0$ and plays an important role in short distances instead. In particular, if we neglect the long--range modification, i.e., as a special case of our results, only the second term remains consistent with \cite{Nicolini:2019irw}.

To find a black hole solution, let us expand only the first factor in Eq. \eqref{energydensity} in a series around $\ell_0$. After taking into account the leading order terms in $\alpha$ and $\ell_0$, we get
\begin{eqnarray}
    \rho(r)=-\frac{M \alpha}{4 \pi r \lambda^2} e^{-\frac{r}{\lambda}}+\frac{3M \ell_0^2}{4 \pi (r^2+\ell_0^2)^{5/2} }+\mathcal{O}(\ell_0^2 \alpha).
\end{eqnarray}

The first term coincides with the result obtained in \cite{Gonzalez:2023rsd}. Moreover, it is worth mentioning that the negative sign reflects that the energy conditions are violated inside the black hole. On the other hand, we assume that the Einstein field equations with a cosmological constant still hold, i.e., $G_{\mu \nu}+\Lambda g_{\mu \nu}=8\pi T_{\mu \nu}$, so that the effects of the effective dark matter are encoded in the stress--energy tensor.  Then, from the gravitational field equations we obtain,
    \begin{equation}
        \frac{r f'(r)+f(r)-1}{r^2}+\Lambda-\frac{2 M \alpha}{ r \lambda^2} e^{-\frac{r}{\lambda}}+\frac{6M \ell_0^2}{ (r^2+\ell_0^2)^{5/2} }=0, \label{eq.(5)}
    \end{equation}
    yielding the solution 
    \begin{equation}
        f(r)=1-\frac{2M r^2}{(r^2+\ell_0^2)^{3/2}}-\frac{2M \alpha (r+\lambda) e^{-\frac{r}{\lambda}}}{r \lambda}-\frac{\Lambda r^2}{3} + \frac{c_1}{r}, \label{exact-sol-full}
    \end{equation}
where $c_1$ is an integration constant. 
To our knowledge, this is a new solution and generalizes that one encountered in Ref.  \cite{Gonzalez:2023rsd}. The second term modifies the geometry due to $\ell_0$ and plays an important role in short--range distances to cure the black hole singularity and if we neglect the third, fourth, and last terms, our result matches with \cite{Nicolini:2019irw}. The third term is due to the apparent dark matter effect, while the fourth term is the contribution due to the cosmological constant. Since $\ell_0$ is of Planck length order, i.e., $\ell_0\sim 10^{-35}\,m$  \cite{Nicolini:2019irw}, in large distances and astrophysical black holes with large mass $M$, we must have $r\gg\ell_0$, then, we can neglect the effect of $\ell_0$. Setting $c_1=0$ the solution turns out to be 
 \begin{eqnarray}
       f(r)=1-\frac{2M}{r}-\frac{2M \alpha (r+\lambda) e^{-\frac{r}{\lambda}}}{r \lambda}-\frac{\Lambda r^2}{3}, \label{eq.(7)}
    \end{eqnarray}
which is in agreement with \cite{Gonzalez:2023rsd}. However, by setting $\ell_0=0$ directly in \eqref{eq.(5)} and integrating it, we obtain
\begin{align}
  f(r)= 1 +\frac{c_2}{r}-\frac{2M \alpha (r+\lambda) e^{-\frac{r}{\lambda}}}{r \lambda}-\frac{\Lambda  r^2}{3}.
   \end{align}
   
For consistency, we must recover \eqref{eq.(7)} by using the freedom to choose $c_2$ to be $c_2=-2M$. Therefore, we have to take $\ell_0=0$ and integrate the results in the same solution.
If we desire to study the properties of small black holes, i.e., black holes with small mass $M$, we can neglect the effect of $\alpha$. 
In the present work we will focus on astrophysical black holes instead. On the other hand, $\lambda$ is of Mpc order and it is important on large distances, thus, we can perform a series expansion around $\varepsilon=1/\lambda$, yielding the result \cite{Gonzalez:2023rsd},
     \begin{eqnarray}
        f(r)=1-\frac{2M (1+\alpha)}{r}+\frac{M \alpha r}{\lambda^2}-\frac{\Lambda r^2}{3}. \label{eq.(9)}
    \end{eqnarray}
It is, therefore, natural to define the true or the physical mass of the black hole to be $\mathcal{M} = M (1+\alpha)$ and write the solutions in terms of the physical mass; then, the exact solution is written as
        \begin{equation}
        f(r)=1-\frac{2\mathcal{M} r^2}{(1+\alpha)(r^2+\ell_0^2)^{3/2}}-\frac{2\mathcal{M}  \alpha (r+\lambda) e^{-\frac{r}{\lambda}}}{r (1+\alpha) \lambda}-\frac{\Lambda r^2}{3}. \label{exact-sol}
    \end{equation} and the approximate solution becomes \cite{Gonzalez:2023rsd},
 \begin{eqnarray}
        f(r)=1-\frac{2 \mathcal{M}}{r}+\frac{\mathcal{M} \alpha r}{(1+\alpha)\lambda^2}-\frac{\Lambda r^2}{3}. \label{approximated-solution}
    \end{eqnarray}

\begin{figure}
    \centering
    \includegraphics[scale=0.44]{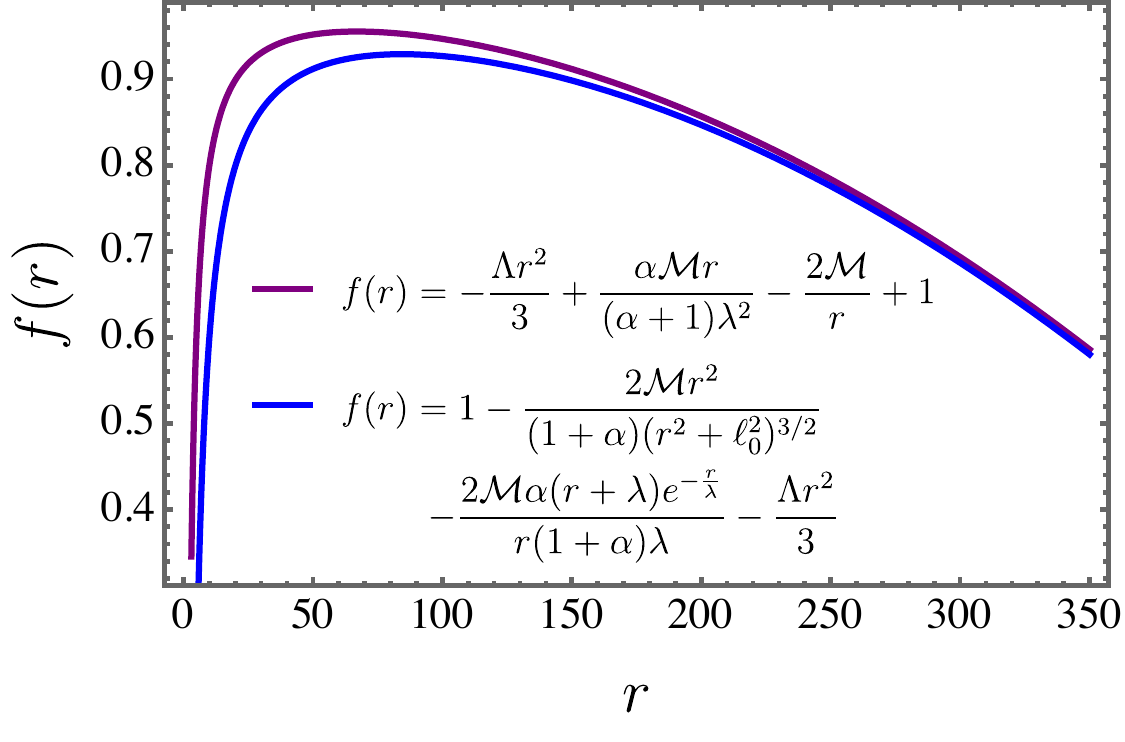}
    \caption{The metric functions in terms of $r$ for $\mathcal{M}=1$, $\lambda = 10^{5}$, $\Lambda = 10^{-5}$, $\ell_0=0$, and $\alpha = 1$.}
    \label{mmetric2}
\end{figure}

{The natural interpretation is that the extra mass term $M_{\rm dark\, matter}=M \alpha$ can be linked to the dark matter effect, which is only an apparent effect in our model. We point out that the interesting reference~\cite{DeLaurentis_2022}  argued that in order to obtain the accurate mass distribution of M87, the mass of the M87 black hole should grow due to dark matter. They of course used observational data and empirical models for dark matter,  but this result is consistent with our paper,  namely, we quite naturally obtain a shift in the mass of the system as we pointed out in the last equation and the interpretation that $\alpha$ encodes dark matter effect makes sense. The fact that the mass of the system is increased in our model yields an interesting consequence that fits a well known argument for the existence of dark matter according to which the deflection of light is increased by a point mass (which can be a black hole or a galaxy). The deflection angle yields, $${\rm Deflection\, angle}=4 \mathcal{M}/b=4M(1+\alpha)/b.$$
This relation, in principle, should also hold for galaxies and again we get an increase of the deflection angle in terms of an apparent dark matter. }

In Fig.  \ref{mmetric2} we have plotted the metric functions $f(r)$ given by Eq. \eqref{approximated-solution} (approximated solution) and Eq. \eqref{exact-sol} (complete solution). Further, if we neglect the term $\mathcal{O}(\alpha/\lambda^2)$ in \eqref{approximated-solution}, we obtain the Kottler spacetime, i.e., a  Schwarzschild black hole with a cosmological constant,
 \begin{eqnarray}
        f(r)\simeq 1-\frac{2 \mathcal{M}}{r}-\frac{\Lambda r^2}{3}.
    \end{eqnarray}

The expression above displayed in Eq. \eqref{approximated-solution} yields three distinct solutions representing an equal number of horizons. However, the most physically relevant to us is the event horizon given by,
\begin{equation}
r_{+} = \frac{1}{\lambda ^2 \Lambda }\left[\alpha  \mathcal{M}-\frac{\sqrt[3]{2} \left(\lambda ^4 \Lambda +\alpha ^2 \mathcal{M}^2\right)}{\sqrt[3]{\mathcal{A}}} -
   \frac{\sqrt[3]{\mathcal{A}}}{\sqrt[3]{2}}\right],
\end{equation}
where 
\begin{eqnarray}\notag
    \mathcal{A}&=&-2 \alpha ^3 \mathcal{M}^3 +\sqrt{\mathcal{B}}+6 (\alpha +1) \lambda ^6 \Lambda ^2 \mathcal{M}-3 \alpha  \lambda ^4 \Lambda  \mathcal{M},
\end{eqnarray}
and 
\begin{eqnarray}\notag
    \mathcal{B}&=&\left(2 \alpha ^3 \mathcal{M}^3+3 \lambda ^4 \Lambda  \mathcal{M} \left(\alpha -2 (\alpha +1) \lambda ^2 \Lambda \right)\right)^2\\
    &-&4 \left(\lambda ^4 \Lambda +\alpha ^2 \mathcal{M}^2\right)^3.
\end{eqnarray}

For the metric \eqref{metric} the Kretschmann scalar $K= R_{a b c d} R^{a b c d}$ 
is given by, 
\begin{align}
K= \frac{G^2}{c^4}  \left( f''(r)^2+\frac{4 f'(r)^2}{r^2}+\frac{4 (f(r)-1)^2}{r^4}\right).
\end{align}
Replacing the exact solution \eqref{exact-sol} and using units where $\frac{G}{c^2}=1$, we obtain, 
\begin{small}
\begin{align}
    & K|_{f(r)=\eqref{exact-sol}}  \nonumber \\
    & = \frac{4}{r^2} \Bigg[\frac{4 \ell_0^2 \mathcal{M} r}{(1+ \alpha)\left(\ell_0^2+r^2\right)^{5/2}}  \nonumber \\
    & -\frac{2\mathcal{M}}{(1+\alpha)r^2} \left(\frac{r^5}{\left(\ell_0^2+r^2\right)^{5/2}}+\frac{\alpha 
   e^{-\frac{r}{\lambda }} \left(\lambda ^2+r^2+\lambda  r\right)}{\lambda
   ^2}\right)+\frac{2 \Lambda  r}{3}\Bigg]^2 \nonumber \\
   & +\frac{4}{r^4} \left[\frac{2 \mathcal{M}}{r (1+\alpha)}
   \left(\frac{r^3}{\left(\ell_0^2+r^2\right)^{3/2}}+\frac{\alpha 
   e^{-\frac{r}{\lambda }} (\lambda +r)}{\lambda }\right)+\frac{\Lambda 
   r^2}{3}\right]^2 \nonumber \\
   & +\Bigg[\frac{2 \Lambda }{3}+\frac{2 \mathcal{M}}{(1+\alpha)} \Bigg(-\frac{11
   \ell_0^2 r^2}{\left(\ell_0^2+r^2\right)^{7/2}}+\frac{2
   r^4}{\left(\ell_0^2+r^2\right)^{7/2}} \nonumber \\
   & +\frac{2 \ell_0^4}{\left(\ell_0^2+r^2\right)^{7/2}}+\frac{\alpha 
   e^{-\frac{r}{\lambda }} (\lambda +r) \left(2 \lambda ^2+r^2\right)}{\lambda ^3
   r^3}\Bigg)\Bigg]^2. 
\end{align}
\end{small}
However, we use the approximated solution 
\eqref{approximated-solution} such that, 
\begin{small}
\begin{align}
    K|_{f(r)=\eqref{approximated-solution}}= \frac{8 \Lambda ^2}{3}+\frac{8 \mathcal{M}^2 \left(\frac{\alpha
   ^2 r^4}{(\alpha +1)^2 \lambda ^4}+6\right)}{r^6}+\frac{8
   \alpha  \Lambda  \mathcal{M}}{(\alpha +1) \lambda ^2 r}.
\end{align}
\end{small}

Evaluating in $r_{+}$ we acquire a nonzero value for $K$ as follows,
\begin{small}
\begin{align}
& K_+= \frac{8}{3} \Lambda ^2 \Bigg[\frac{3 \alpha  \mathcal{M}}{(\alpha +1) \left(-\sqrt[3]{\frac{2}{\mathcal{A}}} \left(\lambda ^4 \Lambda +\alpha ^2
   \mathcal{M}^2\right)-\sqrt[3]{\frac{\mathcal{A}}{2}}+\alpha  \mathcal{M}\right)} \nonumber \\
   & +\frac{3 \lambda ^{12} \Lambda ^4 \mathcal{M}^2 \left(\frac{\alpha ^2 \left(\sqrt[3]{\frac{2}{\mathcal{A}}}
   \left(\lambda ^4 \Lambda +\alpha ^2 \mathcal{M}^2\right)+\sqrt[3]{\frac{A}{2}}-\alpha \mathcal{M}\right)^4}{(\alpha +1)^2 \lambda ^{12} \Lambda ^4}+6\right)}
   {\left(\sqrt[3]{\frac{2}{\mathcal{A}}} \left(\lambda ^4 \Lambda +\alpha ^2 \mathcal{M}^2\right)+\sqrt[3]{\frac{\mathcal{A}}{2}} - \alpha \mathcal{M}\right)^6}+1\Bigg].
\end{align}
\end{small} 

As $r$ approaches to zero, we have
\begin{equation}
    K \simeq    \frac{48 \mathcal{M}^2}{r^6}+\mathcal{O}\left(\frac{1}{r}\right),
\end{equation}
reproducing the Schwarzschild's value at $r=0$.

\subsection{Correspondence with the black hole solution in dRGT gravity }

Up to now we have seen how the black hole solution could be obtained by assuming the form of Einstein field equations with a cosmological constant, where the stress--energy was further modified due to Yukawa potential. From a field equation point of view -- or better to say from the level of action -- one can obtain a similar result. As we shall see, we shall point out an interesting link between our black hole solution and the spherical black hole solutions within the framework of dRGT massive gravity, elucidated in the work of de Rham et al. \cite{deRham:2010kj}. In this theoretical framework it was shown that the Einstein field equations must satisfy \cite{deRham:2010kj},
\begin{eqnarray}
    G^{\mu}_{\nu} + m^2_g X^{\mu}_{\nu} = 8 \pi G T^{\mu (m)}_{\nu}.
\end{eqnarray}

Here the stress--energy tensor $T^{\mu (m)}_{\nu}$ is computed by using the matter Lagrangian. This equation is derived from the action of dRGT massive gravity, which has the form,
\begin{equation}
S = \frac{1}{16 \pi G}\int d^4 x \sqrt{-g} \left[R + m_g^2 \,\mathcal{U}(g,f) \right] + S_m \,.
\end{equation}
In this context $R$ represents the Ricci scalar, $m_g$ denotes the graviton mass, and $\mathcal{U}$ stands for the self--interacting potential of graviton. It is essential to structure $U(g,f)$ in a specific form to prevent the emergence of the Boulware--Deser ghost \cite{deRham:2010kj}
\begin{align*}
&\mathcal{U} \equiv \mathcal{U}_2 + \alpha_3 \mathcal{U}_3 + \alpha_4 \mathcal{U}_4 \,, \\
&\mathcal{U}_2 \equiv  [\mathcal{K}]^2 - [\mathcal{K}^2] \,, \\
&\mathcal{U}_3 \equiv [\mathcal{K}]^3 - 3 [\mathcal{K}][\mathcal{K}^2] + 2 [\mathcal{K}^3] \,, \\
&\mathcal{U}_4 \equiv [\mathcal{K}]^4 - 6 [\mathcal{K}]^2 [\mathcal{K}^2] + 3[\mathcal{K}^2]^2 + 8 [\mathcal{K}][\mathcal{K}^3] - 6 [\mathcal{K}^4] \,,
\end{align*}
where the tensor $\mathcal{K}^{\mu}_{\nu}$ is defined as \cite{deRham:2010kj,Ghosh:2015cva},
\begin{equation}
\mathcal{K}^{\mu}_{\nu} \equiv \delta^{\mu}_{\nu} - \sqrt{g^{\mu\lambda} \partial_{\lambda} \varphi^a \partial_{\nu} \varphi^b f_{ab}},
\end{equation}
in which
\begin{eqnarray}
    [\mathcal{K}] = \mathcal{K}^{\mu}_{\mu}\,\,\,\, \text{and}\,\,\, (\mathcal{K}^i)^{\mu}_{\nu} = \mathcal{K}^{\mu}_{\rho_1} \mathcal{K}^{\rho_1}_{\rho_2} \ldots \mathcal{K}^{\rho_i}_{\nu},
\end{eqnarray}
the physical metric is denoted by $g_{\mu\nu}$ while $f_{\mu\nu}$ serves as a reference (fiducial) metric and $\varphi^a$ represents the Stückelberg fields. In this study we adopt the unitary gauge, where $\varphi^a = x^{\mu} \delta^a_{\mu}$, thus \cite{deRham:2010kj},
\begin{equation}
    \sqrt{g^{\mu\lambda} \partial_{\lambda} \varphi^a \partial_{\nu} \varphi^b f_{ab}} = \sqrt{g^{\mu\lambda} f_{\lambda\nu}}.
\end{equation}
The massive graviton tensor $X^{\mu}_{\nu}$ is 
given by \cite{deRham:2010kj},
\begin{align}
X^{\mu}_{\nu} &= \mathcal{K}^{\mu}_{\nu} - [\mathcal{K}] \delta^{\mu}_{\nu} - \bar{\alpha}\left[(\mathcal{K}^2)^{\mu}_{\nu} - [\mathcal{K}]\mathcal{K}^{\mu}_{\nu} +\frac{1}{2}\delta^{\mu}_{\nu} ([\mathcal{K}]^2 - [\mathcal{K}^2])\right] \nonumber \\
&~~ + 3 \beta \left[(\mathcal{K}^3)^{\mu}_{\nu} - [\mathcal{K}](\mathcal{K}^2)^{\mu}_{\nu} +\frac{1}{2}\mathcal{K}^{\mu}_{\nu} ([\mathcal{K}]^2 - [\mathcal{K}^2])\right. \nonumber \\
&~~ \left. - \frac{1}{6} \delta^{\mu}_{\nu} ([\mathcal{K}]^3 - 3 [\mathcal{K}][\mathcal{K}^2] + 2[\mathcal{K}^3]) \right] \,,
\end{align}
where $\alpha_3 = \frac{\bar{\alpha} - 1}{3}$ and $\alpha_4 = \frac{\beta}{4} + \frac{1 - \bar{\alpha}}{12}$.  The terms of order $\mathcal{O}(\mathcal{K}^4)$ disappear under the fiducial metric ansatz $f_{\mu\nu} =\diag(0, 0, C^2,  C^2 \sin^2 {\theta})$,
where $C$ is a positive constant. In this sense the static and spherically symmetric solution is given by \cite{deRham:2010kj, Ghosh:2015cva},
 \begin{eqnarray}
        f(r)=1-\frac{2 \mathcal{M}}{r}-\frac{\Lambda r^2}{3}+\gamma \,r+\zeta,
    \end{eqnarray}
for $\zeta=0$ and $\gamma=\mathcal{M}\alpha/\lambda^2$ along with $\Lambda \sim m_g^2$. 

\subsection{Energy conditions}

Let us consider now the modifications introduced by the Yukawa potential as an apparent type of matter described by a fluid. If we calculate the stress--energy tensor corresponding to this fluid using Eq.  \eqref{approximated-solution}, we obtain,
\begin{eqnarray} 
T^t _t = T^r _r &=& \frac{{\cal M}\alpha}{4\pi(1+\alpha)\lambda^2 r} \,,  \label{tema} \\
T^\theta _\theta = T^\phi _\phi
&=& \frac{{\cal M}\alpha}{8\pi(1+\alpha)\lambda^2 r} \,.\label{temb}
\end{eqnarray}
As stated in~\cite{Simpson:2018tsi}, a necessary aspect to accomplish the null energy condition (NEC) is that both $\rho+p_{||}\geq 0$ and $\rho+p_\perp \geq 0$ are valid. For the apparent fluid we are modelling we identify $\rho=-T^t _t$, $p_{||}=T^r _r$, and $p_\perp = T^\theta _\theta = T^\phi _\phi$ and using the expressions \eqref{tema} and \eqref{temb}, we obtain
\begin{eqnarray}
\rho+p_{||} &=& 0 \,, \\
\rho+p_\perp &=& -\frac{{\cal M}\alpha}{8\pi(1+\alpha)\lambda^2 r}\,,
\end{eqnarray}
in which the latter relation clearly violates NEC as long as $\alpha>0$. As mentioned in~\cite{alma991027056009703276}, this result is enough to assume that weak, strong, and dominant energy conditions are equally violated. 

We can also check if the exact solution given in Eq. \eqref{exact-sol} can alleviate this violation. Calculating both conditions we have,
\begin{align}
\rho+p_{||} &= 0 \,, \label{neca} \\
\rho+p_\perp &= -\frac{\mathcal{M}\alpha (r+\lambda) e^{-r/\lambda}}{8\pi\lambda^3 (1+\alpha) r} + \frac{15 \mathcal{M}\ell_0 ^2 r^2}{8\pi(1+\alpha) (\ell_0 ^2+r^2)^{7/2}}.\label{necb}
\end{align}
We will mathematically show that it is possible to achieve a fine--tuning such that Eq. \eqref{necb} becomes zero or positive. However, physically it is practically impossible due to the order of magnitude of $\ell_0$, as we shall see. 

Indeed, for small values of the radial coordinate, we can approximate Eq. \eqref{necb} as
\begin{equation}
\rho+p_\perp\simeq -\frac{\alpha  \mathcal{M}}{8 \pi  (1+\alpha)\lambda ^2 r}+\mathcal{O}\left(r\right).
\end{equation}
Such that $\rho+p_\perp\geq 0$ only for $\alpha \leq 0$ at the interior of the black hole. 
Nevertheless, classical energy conditions can be fulfilled by this solution for large $r$, since we have 
\begin{equation}
\rho+p_\perp\simeq \frac{15 \ell_0^2 \mathcal{M}}{8 \pi (1+\alpha) r^5}-\frac{\alpha  \mathcal{M} e^{-\frac{r}{\lambda }} (\lambda +r)}{8 \pi  \lambda ^3 (1+\alpha)r}.
   \end{equation}
Such that the negative term can be exponentially suppressed and
$\rho+p_\perp\geq 0$.   
 The required condition is 
 \begin{equation}
     15 \ell_0^2\geq \frac{\alpha  r^4 e^{-\frac{r}{\lambda }} (\lambda +r)}{\lambda ^3}.
 \end{equation}
 Therefore, the strategy is to show that the maximum of 
 \begin{equation}
     h(r)= \frac{\alpha  r^4 e^{-\frac{r}{\lambda }} (\lambda +r)}{\lambda ^3}
 \end{equation} is less or equal than $15 \ell_0^2$.
 The extreme values of $h(r)$ satisfy the following relation,
 \begin{equation}-\frac{\alpha  r^3 e^{-\frac{r}{\lambda }} \left(-4 \lambda ^2+r^2-4 \lambda  r\right)}{\lambda^4}=0.
 \end{equation}
 The possible physical solutions are $r=0$, $r=2 \left(1+\sqrt{2}\right)\lambda$, or $\alpha=0$. Because $h''(2 \left(1+\sqrt{2}\right)\lambda)=-32 \left(10+7 \sqrt{2}\right) e^{-2 \left(1+\sqrt{2}\right)} \alpha<0$, the maximum is obtained at $ r=2 \left(1+\sqrt{2}\right)\lambda$ and the required condition turns to be
 \begin{equation}
\lambda \neq 0, \alpha \leq \frac{15  e^{2+2 \sqrt{2}}\ell_0^2}{16 \left(99+70 \sqrt{2}\right) \lambda ^2} \lesssim \frac{0.591938\ell_0^2}{\lambda ^2}= a_{\star}. 
 \end{equation}

As expected, $a_\star$ is practically zero.  In this manner, classical energy conditions are not fulfilled by the apparent fluid emerging from the Yukawa black hole solution -- unless that $\alpha<0$. At this point we should emphasize that the stress--energy tensor that we obtained corresponds to an apparent effect (it mimics the dark matter effect due to the modification of the gravitational potential,  \textit{i.e.}, it is a screening effect in large distances) and not to real matter. Alternatively, we can consider this system as made of a composition of several fluids as Refs.~\cite{Visser:2019brz,Boonserm:2019phw} do and remark that the resultant does not necessarily preserves NEC.  Nevertheless, the violation of NEC by this composite -- but not physical fluid representing the Yukawa modified black hole -- does not rule out the study of its interesting properties, especially in cosmological contexts where $\alpha$ has been estimated around $0.4$~\cite{Gonzalez:2023rsd}. Moreover, as we will see in $\S$ \ref{sectVII},  we can constrain $\alpha$ using EHT observations and it can also assume negative values preventing NEC violation.

In the next section we will discuss the accuracy of the approximate solution with which we will work throughout this paper.

\section{Accuracy of the approximate solution}
\label{accuracy}

The approach described here involves identifying an order parameter $\varepsilon= {\hbar}^{-1}{m_g c}$ to tackle a problem that depends on two different scales: (i) a ``slow" variable that has a solution similar to the outer solution in a boundary layer problem, and (ii) a ``fast" system that describes rapid evolution occurring over shorter radii, which can be solved by changing the scale of the system. This approach is similar to the inner solution of a boundary layer problem and involves seeking the solution of each subsystem by using regular perturbation expansion. 

For singularly perturbed problems the subsystems will have simpler structures than the whole problem. Combining the results of these two limiting problems, we can obtain information about the dynamics for small values of $\varepsilon$. These procedures belong to the realm of perturbation methods in differential equations \cite{fenichel, Fusco1989SlowmotionMD, Berglund, verhulst2006method, holmes2012introduction, kevorkian2013perturbation, AwrejcewiczJan2014ANDS} and there can be used to construct uniformly valid approximations of the solutions of perturbation problems using those ones that satisfy the original expressions in the limit of $\varepsilon\rightarrow 0$.

For this analysis we use the black hole physical mass $\mathcal{M}$ and we properly write the corresponding solution. For simplicity we set $\ell_0=0$. Hence, Eq. \eqref{eq.(5)} is recast as, 
\begin{small}
 \begin{align}
P(\varepsilon)[f(r)] & :=        \frac{r f'(r)+f(r)-1}{r^2}+\Lambda-\frac{2 \mathcal{M} \alpha}{(1+ \alpha) r } \varepsilon^2 e^{-\varepsilon r}=0,
    \end{align} 
\end{small}
where we introduce $\varepsilon=1/\lambda$. 
Associated with this problem we have the zeroth--order problem, 
\begin{small}
 \begin{align}
P(0)[f(r)] & :=   \frac{r f'(r)+f(r)-1}{r^2}+\Lambda=0.
    \end{align} 
\end{small}
For convenience we redefine $f(r)= g(r)/r$ and pose the problems $\tilde{P}(\varepsilon)[g(t)]= r^2 P(\varepsilon)[g(r)/r]$ and $\tilde{P}(0)[g(t)]= r^2 P(0)[g(r)/r]$, namely,
\begin{small}
 \begin{align}
\tilde{P}(\varepsilon)[g(r)] & :=    g'(r)-1+\Lambda {r^2} -\frac{2 \mathcal{M} \alpha}{(1+ \alpha)} \varepsilon \left(\varepsilon r\right) e^{-\varepsilon r}=0, \label{eq.(22)}
    \end{align} 
\end{small}
and 
\begin{small}
 \begin{align}
\tilde{P}(0)[g(r)] & :=    g'(r)-1+\Lambda {r^2}=0.
    \end{align} 
\end{small}
For both problems $P(\varepsilon)[g(r)]=0$ and $P(0)[g(r)]=0$ we choose the 
initial layer condition as, 
\begin{equation}
  \lim_{r\rightarrow 0^+} g(r)= -2 \mathcal{M}.
\end{equation}

\subsection{Inner solution}

In our case Eq. \eqref{eq.(22)} is written as, 
\begin{align}
    & g'(r)= 1 - \Lambda {r^2} +\frac{2 \mathcal{M} \alpha}{(1+ \alpha)} r \varepsilon^2 e^{-\varepsilon r}, \; g(0)= -2 \mathcal{M}.
\end{align}
By replacing 
\begin{align}
    g_{\text{inner}}(r)= g_0 (r) + \varepsilon g_1(r) + \varepsilon^2 g_2(r) + \varepsilon^3 g_3(r)  + \varepsilon^4 g_4(r) + \ldots 
\end{align}
 we have 
 \begin{align}
    &  g_0' (r) + \varepsilon g_1'(r) + \varepsilon^2 g_2'(r) + \varepsilon^3 g_3'(r) + \varepsilon^4 g_4'(r) + \ldots \nonumber \\
    & = 1 - \Lambda {r^2} +\frac{2 \mathcal{M} \alpha}{(1+ \alpha)} r \left[\varepsilon ^2-r \varepsilon ^3+\frac{r^2 \varepsilon
   ^4}{2}+ \ldots\right], \\
    & g_0(0)= -2 \mathcal{M}, \quad g_j(0)= 0, \,\,\,\, j\geq 1. 
\end{align}
Now comparing equal powers of $\varepsilon$, we obtain the following systems,
\begin{small}
 \begin{align}
    &  g_0' (r)  = 1 - \Lambda {r^2}, \; g_0(0)= -2 \mathcal{M}  \nonumber \\
    & \implies g_0(r)= -2 \mathcal{M} + r - \frac{\Lambda}{3} r^3, \\
    &  g_1' (r)  = 0,  \; g_1(0)=0 \implies g_1(r)=0, \\
    & g_2' (r)  =  \frac{2 \mathcal{M} \alpha}{(1+ \alpha)} r, \;  g_2(0)=0  \implies g_2(r)= \frac{ \mathcal{M} \alpha}{(1+ \alpha)} r^2,\\
    & g_3' (r)  = -\frac{2 \mathcal{M} \alpha}{(1+ \alpha)} r^2,  \; g_3(0)=0 \implies  g_3' (r)  = -\frac{2 \mathcal{M} \alpha}{3(1+ \alpha)} r^2, \\
  & g_4' (r)  = \frac{\mathcal{M} \alpha}{(1+ \alpha)} r^3, \;  g_4(0)=0 \implies  g_4 (r)  = \frac{\mathcal{M} \alpha}{4(1+ \alpha)} r^4.
\end{align}
\end{small}
Plugging all together we recover the inner solution 
\begin{align}
g_{\text{inner}}(r)  & = r -2 \mathcal{M} - \frac{1}{3}\Lambda  r^3-\left(\frac{\alpha  \mathcal{M}
   r }{\alpha +1}\right) \varepsilon^2\nonumber\\
   & +\frac{2 \alpha  \mathcal{M} r^3}{3 (\alpha +1)} \varepsilon^3 -\frac{\left(\alpha  \mathcal{M}
   r^4\right)}{4 (\alpha +1)}  \varepsilon^4. \label{eq.(50)}
\end{align}
Then, using the relation $\varepsilon=1/\lambda$ and $f=g/r$ we get 
\begin{align}
f_{\text{inner}}(r)  & = 1 -2 \frac{\mathcal{M}}{r} - \frac{1}{3}\Lambda  r^2-\left(\frac{\alpha  \mathcal{M}
   r }{\alpha +1}\right) \left(\frac{1}{\lambda}\right)^2\nonumber\\
   & +\frac{2 \alpha  \mathcal{M} r^2}{3 (\alpha +1)} \left(\frac{1}{\lambda}\right)^3 -\frac{\left(\alpha  \mathcal{M}
   r^3\right)}{4 (\alpha +1)}  \left(\frac{1}{\lambda}\right)^4. \label{eq.(51)}
\end{align}

\subsection{Outer solution}

Defining 
\begin{equation}
  G(\tau)= g(\tau/\varepsilon),
\end{equation}
from \eqref{eq.(22)} we obtain the ``slow system'',
\begin{equation}
\label{eq:1.2}
 \frac{d G(\tau)}{d \tau}=- \Lambda \epsilon^{-3} \tau^2 + \epsilon^{-1}  +\frac{2 \mathcal{M} \alpha}{(1+ \alpha)} \tau  e^{-\tau}, 
\end{equation}
attained after the rescaling $\tau=\varepsilon r$, where we introduced the small parameter $\varepsilon=1/\lambda$. 
Using the substitution 
\begin{align}
    G(\tau)=  \varepsilon^{-3}  G_{-3} (\tau) +  \varepsilon^{-2}  G_{-2} (\tau) +  \varepsilon^{-1}  G_{-1} (\tau) +  G_{0} (\tau),
\end{align}
we yield
 \begin{align}
  &\varepsilon^{-3}  G_{-3}' (\tau) +  \varepsilon^{-2}  G_{-2}' (\tau) +  \varepsilon^{-1}  G_{-1}' (\tau) +  G_{0}' (\tau) \nonumber \\
  & = -\Lambda \epsilon^{-3} \tau^2 + \epsilon^{-1}  +\frac{2 \mathcal{M} \alpha}{(1+ \alpha)} \tau  e^{-\tau}.
\end{align}
Now, comparing equal powers of $\varepsilon$ we have these systems,
\begin{align}
&    G_{-3}' (\tau) = -  \Lambda  {\tau^2} \implies   G_{-3} (\tau)=  - \frac{\Lambda \tau^3}{3} + c_{-3}, \\
&   G_{-2}' (\tau) =0 \implies G_{-1}= c_{-2}, \\
  &   G_{-1}' (\tau) =1 \implies G_{-1}(\tau)= \tau + c_{-1},
\\
 & G_{0}' (\tau) =\frac{2 \mathcal{M} \alpha}{(1+ \alpha)} \tau  e^{-\tau}  \nonumber \\
 & \implies  G_{0} (\tau)= -\frac{2 \alpha  \mathcal{M} e^{-\tau } (\tau +1)}{\alpha +1}+c_0.
\end{align}
Replacing all the terms we arrive at, 
\begin{align}
    & G(\tau)=  \varepsilon^{-3}  \left( - \frac{\Lambda \tau^3}{3} + c_{-3}\right) + \varepsilon^{-2}  c_{-2}+ \varepsilon^{-1} \left(\tau + c_{-1}\right) \nonumber \\
    & +  \left( -\frac{2 \alpha  \mathcal{M} e^{-\tau } (\tau +1)}{\alpha +1}+c_0\right).
\end{align}
We have the matching condition given by,
\begin{align}
& \lim_{r\rightarrow \infty} g_{\text{inner}}(r) = \lim_{\tau\rightarrow 0} G(\tau)\nonumber \\
& =  \varepsilon^{-3} c_{-3} + \varepsilon^{-2}  c_{-2}+ \varepsilon^{-1} c_{-1}  +  \left(c_0-\frac{2 \alpha  \mathcal{M}}{\alpha+1}\right). 
\end{align}
Comparing terms of the same power in $\varepsilon$ with \eqref{eq.(50)}, we require that
\begin{equation}
   c_{-3}= c_{-2}= c_{-1}=0, \quad c_0= -2 \mathcal{M}+\frac{2 \alpha  \mathcal{M}}{\alpha +1}.
\end{equation}
Thus,
\begin{align}
    & G(\tau)=  -\varepsilon^{-3}  \frac{\Lambda \tau^3}{3} + \varepsilon^{-1} \tau -2 \mathcal{M}  -\frac{2 \alpha  \mathcal{M} e^{-\tau } \tau}{\alpha +1}.
\end{align}
Replacing \begin{align}
    g(r):= G(\varepsilon r)= r -2 \mathcal{M}  -\frac{\Lambda r^3}{3}  -\frac{2 \alpha  \mathcal{M} e^{-\varepsilon r} \varepsilon r}{\alpha +1}
\end{align} 
in $f(r)=g(r)/r$ and using $\varepsilon=1/\lambda$ we obtain, 
\begin{align}
    & f_{\text{outer}}(r)= 1 - \frac{2 \mathcal{M}}{r}  -\frac{\Lambda r^2}{3}  -\frac{2 \alpha  \mathcal{M} e^{-\frac{r}{\lambda}}}{\lambda(\alpha +1)}.
\end{align}

\subsection{Multiple scales - Derivative expansion method}

The method of multiple scales determines solutions to perturbed oscillators by suppressing resonant forcing terms that would yield spurious secular terms in the asymptotic expansions. It assumes that the solution can be expressed as a function of multiple (just two for our purposes) radial variables, which are introduced to keep a well--ordered expansion, $x(r) = X(r, \tau)$, where $r$ is the regular (or ``fast") radial variable and $\tau$ is a new variable describing the ``slow--radial" dependence of the solution. The idea is to use any freedom that is in the $\tau$-dependence to minimize the approximation's error and, whenever possible, to remove secular terms \cite{fenichel, Fusco1989SlowmotionMD, Berglund, verhulst2006method, holmes2012introduction, kevorkian2013perturbation, AwrejcewiczJan2014ANDS}.

Eq. \eqref{eq.(22)} depends explicitly on two scales, $r$ and $\varepsilon r$. In this way, for $\varepsilon \ll 1$, we set the fast $\tau_A= r$ and slow $\tau_B= \varepsilon r$ scales. 
Thereby,
	\begin{equation}
	    \frac{\mathrm{d}}{\mathrm{d} r}=\frac{\partial}{\partial \tau_{A}}+\varepsilon \frac{\partial}{\partial \tau_{B}}. 
	\end{equation}
We denote the partial derivatives with respect to $\tau_A$ and $\tau_B$ with the indexes $A$ and $B$, namely, $\frac{\partial f}{\partial \tau_A}= f_{, A}$. Substituting the expression
\begin{equation}
	g=g_{0}(r, \varepsilon r)+\varepsilon g_{1}(r, \varepsilon r)+ \varepsilon^2 g_{2}(r, \varepsilon r) + \ldots
 \end{equation}
in the Eq. \eqref{eq.(22)} and identifying the dependencies on $\tau_A$ and $\tau_B$, the differential operator is written as,
\begin{align}
 &   \tilde{P}(\varepsilon)[g_{0}+\varepsilon g_{1}+ \varepsilon^2 g_{2} + \ldots]  :=  \nonumber\\
 &   \left[\frac{\partial}{\partial \tau_{A}}+\varepsilon \frac{\partial}{\partial \tau_{B}} \right] \left(g_{0}+\varepsilon g_{1}+ \varepsilon^2 g_{2} +  \ldots\right) \nonumber\\
 & -1+\Lambda {\tau_A^2} -\frac{2 \mathcal{M} \alpha}{(1+ \alpha)} \varepsilon \tau_B e^{-\tau_B} \nonumber\\
 & = g_{0, A}  -1+\Lambda {\tau_A^2} \nonumber \\
 & + \varepsilon  \left(g_{1,A} + g_{0,B}-\frac{2 \mathcal{M} \alpha}{(1+ \alpha)} \tau_B e^{-\tau_B}\right) \nonumber \\
 & +\varepsilon ^2 \left(g_{2,A}+ g_{1,B}\right) + \varepsilon^3   g_{2,B} + \ldots \,\,\,.
    \end{align}
The initial condition is given by,
\begin{align}
  &  g(0) =g_{0}(0,0)+\varepsilon g_{1}(0,0)+ \varepsilon^2 g_{2}(0,0) + \ldots \\
\implies 
 & -2 \mathcal{M} + \varepsilon \cdot 0 + \varepsilon^2 \cdot 0 + \varepsilon^3 \cdot 0 + \ldots \nonumber\\
 & = g_{0}(0,0)+\varepsilon g_{1}(0,0)+ \varepsilon^2 g_{2}(0,0) + \ldots \,\,\,.
 \end{align}
 With it we have the following problems.

 \textbf{Order $\varepsilon^0$}:
\begin{align}
   &  g_{0, A}  -1+\Lambda {\tau_A^2}=0, \;  g_{0}(0,0)= -2 \mathcal{M}.
\end{align}
 with solution 
 \begin{equation}
 g_0(\tau_A, \tau_B)=\tau_{A} +\frac{\Lambda  \tau_{A}^3}{3} + c_1(\tau_{B}), \quad   c_1(0)=-2 \mathcal{M}.
 \end{equation}
 
 \textbf{Order $\varepsilon^1$}:
\begin{align}
   &  g_{1,A} + g_{0,B}-\frac{2 \mathcal{M} \alpha}{(1+ \alpha)} \tau_B e^{-\tau_B}=0, \nonumber \\
   & g_{0}(0,0)= -2 \mathcal{M}, \quad  g_{1}(0,0)= 0.
\end{align}
In such a way that, 
\begin{align}
   &  g_{1,A} + c_{1}'(\tau_B)-\frac{2 \mathcal{M} \alpha}{(1+ \alpha)} \tau_B e^{-\tau_B}=0, \nonumber \\
   & c_1(0)= -2 \mathcal{M}, \quad  g_{1}(0,0)= 0.
\end{align}

 \textbf{Order $\varepsilon^2$}:
\begin{align}
   &  g_{2,A}+ g_{1,B}=0, \quad g_{1}(0,0)= 0, \quad g_{2} (0,0)=0.
\end{align}

We can eliminate higher--order terms (and possibly resonant terms) by choosing $ g_{1}= g_{2}= \ldots = 0$, and solving 
\begin{align}
    &  c_{1}'(\tau_B)-\frac{2 \mathcal{M} \alpha}{(1+ \alpha)} \tau_B e^{-\tau_B}=0, \quad  c_1(0)= -2 \mathcal{M}.
\end{align}
With
\begin{align}
    c_1(\tau_{B})= \frac{2 \alpha  \mathcal{M} e^{-\tau_{B}}
    \tau_{B}}{\alpha +1}+\frac{2 \alpha  \mathcal{M} e^{-\tau_{B}}}{\alpha +1}-\frac{2 (2 \alpha +1) \mathcal{M}}{\alpha +1}.
\end{align}
we have, 
\begin{align}
& g(\tau_A, \tau_B)= -\frac{2 (2 \alpha
   +1) \mathcal{M}}{\alpha +1}+\tau_A + \frac{\Lambda  \tau_A^3}{3} \nonumber \\
   & +\frac{2 \alpha  \mathcal{M}
   e^{-\tau_B} \tau_B}{\alpha +1}+\frac{2
   \alpha  \mathcal{M} e^{-\tau_B}}{\alpha +1}.
\end{align}
Finally, 
\begin{align}
r f(r) & =-\frac{2 (2 \alpha +1) \mathcal{M}}{\alpha +1}+r +\frac{\Lambda  r^3}{3} \nonumber \\
&  +\frac{2 \alpha  \mathcal{M} r \varepsilon
    e^{-r \varepsilon }}{\alpha +1}+\frac{2 \alpha  \mathcal{M} e^{-r \varepsilon
   }}{\alpha +1}. \label{eq.(47)}
   \end{align}
In this case with multiple scales we provide the exact solution of the model. 
The regular expansion method, on the other hand, provides the solution,
\begin{small}
\begin{align}
 r f(r)  & = \frac{1}{3} \left(-6 \mathcal{M}+\Lambda  r^3+3 r\right)-\frac{\alpha  \mathcal{M}
   r^2 \varepsilon ^2}{\alpha +1}  +\frac{2 \alpha  \mathcal{M} r^3 \varepsilon
   ^3}{3 (\alpha +1)}, \label{eq.(48)}
\end{align}
\end{small}
with the error term, 
\begin{align}
& r(f(r)|_{\eqref{eq.(47)}}- f(r)|_{\eqref{eq.(48)}})  = -\frac{2 \alpha  \mathcal{M}-2 \alpha  \mathcal{M} e^{-r \varepsilon
   }}{\alpha +1} \nonumber \\
& + \frac{2 \alpha  \mathcal{M} r \varepsilon  e^{-r \varepsilon
   }}{\alpha +1}  +\frac{\alpha  \mathcal{M} r^2 \varepsilon
   ^2}{\alpha +1}-\frac{2 \alpha  \mathcal{M}
   r^3 \varepsilon ^3}{3 (\alpha +1)} .
   \end{align}
From Eq.\eqref{eq.(48)} the expansion \eqref{eq.(50)} is recovered. Finally, the main topic of discussion was centred around how the approximated solution for $f(r)$, given by Eq. \eqref{approximated-solution}, accurately represents the exact solution \eqref{exact-sol} as the parameters $\varepsilon=1/\lambda$ and $\ell_0$ approach zero. This is particularly significant when both horizon solutions are close. The error is bounded by 
\begin{align}
    \Big|f(r)|_{\eqref{exact-sol}}- f(r)|_{\eqref{approximated-solution}}\Big| \leq  C_3    \left|\frac{2 \alpha  \mathcal{M} r^2 \varepsilon ^3}{3 (\alpha +1)}\right|, 
\end{align}
where $C_3$ is a constant. 

Once we have established that the approximation is reliable, we can further investigate the solution characteristics with $f(r)$ as defined by Eq. \eqref{approximated-solution}. {We will begin by checking the
  thermodynamic stability of the solution through the study of its main
  properties in the next section.}


\section{Thermodynamic properties}
\label{sectIII}

\subsection{Hawking temperature}

The metric defined in Eq. \eqref{approximated-solution} possesses a timelike Killing vector denoted as $\xi = \partial/\partial t$. This metric gives rise to a conserved quantity inherently associated with $\xi$. We can construct this conserved quantity using the Killing vector as follows,
\begin{equation}
\nabla^\nu(\xi^\mu\xi_\mu) = -2\kappa\xi^\nu.
\end{equation}
Here $\nabla_\nu$ represents the covariant derivative and $\kappa$ remains constant along the orbits defined by $\xi$. This implies that the Lie derivative of $\kappa$ with respect to $\xi$ is identically zero,
\begin{equation}
\mathcal{L}_\xi\kappa = 0.
\end{equation}
Remarkably, $\kappa$ retains its constancy across the horizon and is called surface gravity. In the coordinate basis the timelike Killing vector components are expressed as $\xi^{\mu} = (1, 0, 0, 0)$. The surface gravity for the metric ansatz in Eq. \eqref{approximated-solution} is given by,
\begin{equation}
\kappa = {\left.\frac{f^{\prime}(r)}{2} \right|_{r = {r_{+}}}}.
\end{equation}

Hawking's groundbreaking work, as presented in Refs. \cite{hawking1975particle,Hawking:1975vcx}, revealed that black holes emit radiation. This one is associated with a temperature known as the \textit{Hawking} temperature, expressed as $T = \kappa/2\pi$ \cite{Filho:2022zdh,Sedaghatnia:2023fod,Filho:2023ydb,Filho:2023yly}. In our case, it can be expressed as follows
\begin{equation}
\label{tem1}
T = \frac{1}{4\pi} \left( \frac{\alpha  \mathcal{M}}{(\alpha +1) \lambda ^2}+\frac{2 \mathcal{M}}{r_{+}^2}-\frac{2 \Lambda }{3}  r_{+}\right),
\end{equation}
or in terms of the event horizon,
\begin{equation}
\label{Trh}
T=\frac{1}{4\pi {\cal C}}\left(\frac{\cal J}{r_+}-\frac{\Lambda r_+ {\cal K}}{3}\right)\,,
\end{equation}
with 
\begin{eqnarray}
{\cal J} &=& 2(\alpha+1)\lambda^2 +\alpha r_+ ^2, \\
{\cal K} &=& 6 (\alpha+1) \lambda^2 -\alpha r_+, \\
{\cal C} &=& 2 (\alpha+1) \lambda^2 -\alpha r_+ ^2\,.
\end{eqnarray}

\begin{figure}
    \centering
    \includegraphics[scale=0.45]{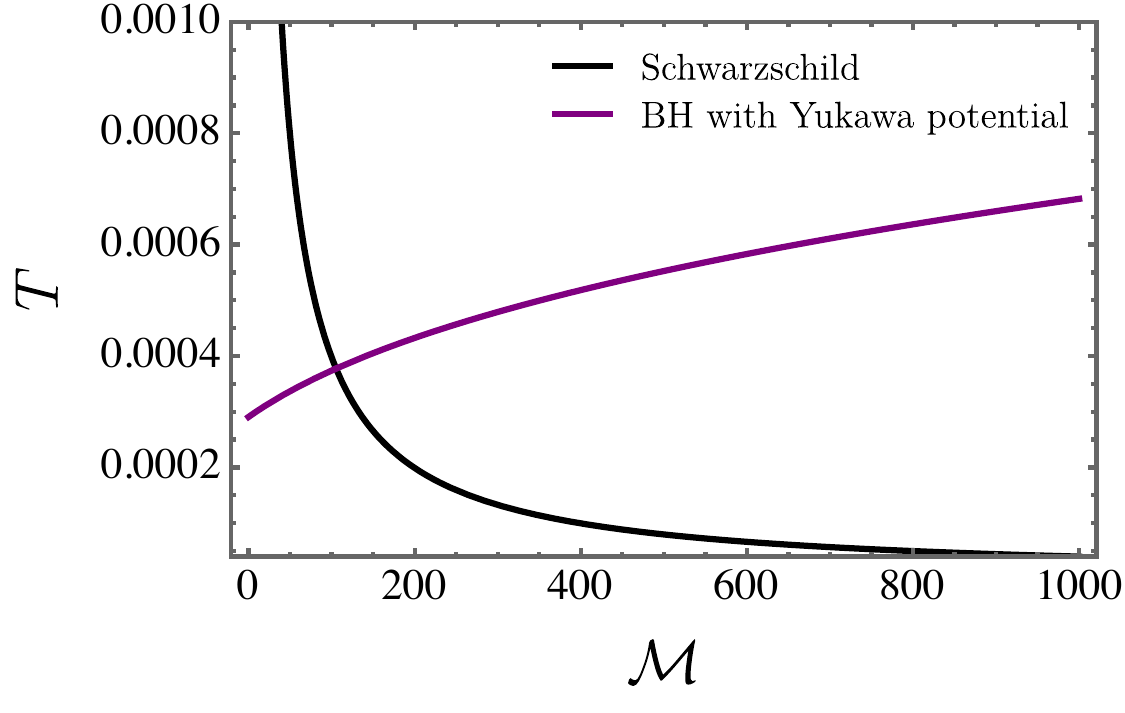}
    \caption{\textit{Hawking} temperature of Yukawa--corrected black hole as a function of $\mathcal{M}$ for $\lambda = 10^{5}$, $\Lambda = 10^{-5}$, and $\alpha = 1$ in comparison with the Schwarzschild case.}
    \label{hhtemperature}
\end{figure}

\begin{figure}
    \centering
    \includegraphics[scale=0.45]{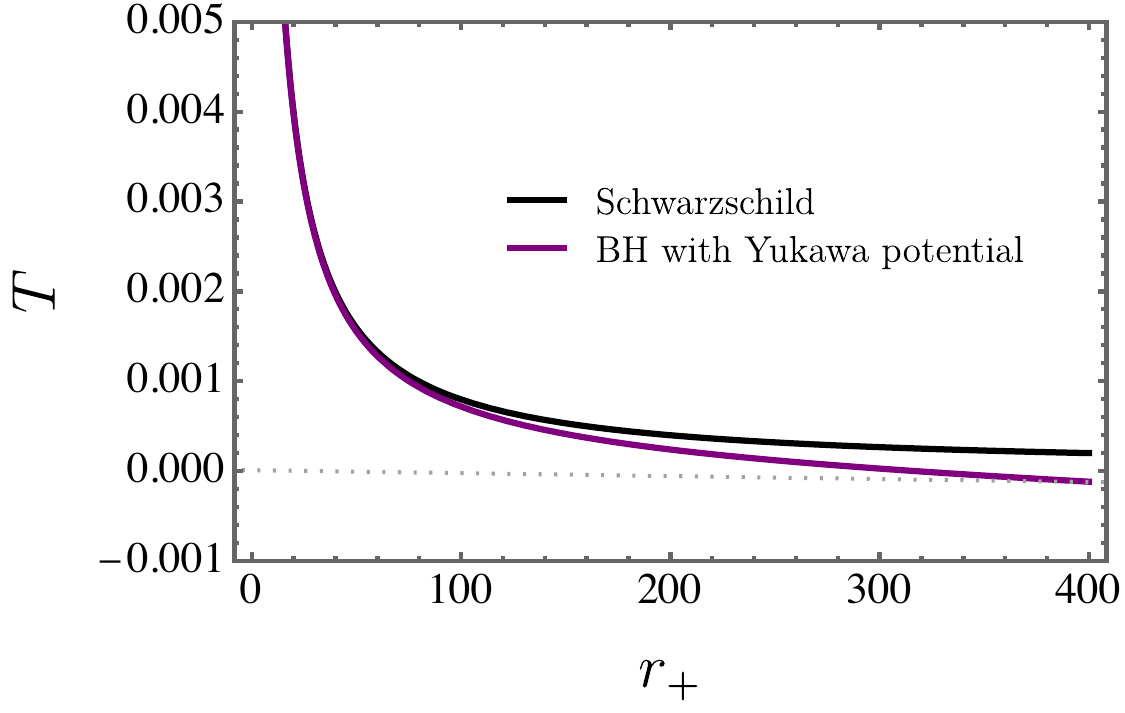}
    \caption{\textit{Hawking} temperature of Yukawa--corrected black hole as a function of $r_{+}$ for $\lambda = 10^{5}$, $\Lambda = 10^{-5}$, and $\alpha = 1$ in comparison with the Schwarzschild case.}
    \label{hhtemperatureasr}
\end{figure}

In Figs. \ref{hhtemperature} and  \ref{hhtemperatureasr}, we plot the \textit{Hawking} temperature as a function of mass $\mathcal{M}$ and event horizon $r_+$, respectively, considering  $\lambda = 10^{5}$, $\Lambda = 10^{-5}$, and $\alpha = 1$ for both Schwarzschild and Yukawa--corrected black holes. From Fig.  \ref{hhtemperatureasr} we can see that for small event horizons Yukawa black hole temperature mimics Schwarzschild case. However, whereas Schwarzschild temperature goes asymptotically to zero, Yukawa black hole temperature reaches zero and turns to be negative at certain value of $r_+$. According to the third law of thermodynamics, we can interpret this point as the value where the event horizon meets the cosmological horizon, {\it i.e.}, the extreme black hole solution, after which the temperature loses its physical meaning.

\subsection{Entropy}

Certainly, when we consider $f(r_{+})=0$, a precise expression for $\mathcal{M}$ can be obtained and it is given by,
\begin{equation}
\mathcal{M}= \frac{(\alpha +1) \lambda ^2 r_{+} \left(\Lambda  r_{+}^2-3\right)}{3 \alpha  r_{+}^2-6 (\alpha +1) \lambda ^2}.
\end{equation}

Furthermore, following the first law of thermodynamics, the \textit{Hawking} temperature is defined as,
\begin{align}
&\Tilde{T}_{f}  = \frac{\mathrm{d}\mathcal{M}}{\mathrm{d}S}= \frac{1}{2\pi r_{+}} \frac{\mathrm{d}\mathcal{M}}{\mathrm{d}r_{+}} \nonumber \\
& = \frac{1}{2\pi r_{+}}  \left( \frac{(\alpha +1) \lambda ^2 \left(6 (\alpha +1) \lambda ^2+\alpha  \Lambda  r_{+}^4+\mathcal{Y}\right)}{3 \left(\alpha  r_{+}^2-2 (\alpha +1) \lambda ^2\right)^2} \right).
\end{align}
where $\mathcal{Y}=3 r_{+}^2 \left(\alpha -2 (\alpha +1) \lambda ^2 \Lambda \right)$.
This temperature contradicts the value presented in Eq.  \eqref{tem1}. In line with the approach detailed in Ref. \cite{Ma:2014qma}, under the assumption of the validity of the area law, we employ the corrected first law of thermodynamics to determine the temperature for regular black holes. The following first law determines the corrected temperature,
\begin{equation}
\Upsilon(r_{+},\mathcal{M},\alpha,\lambda,\Lambda) \mathrm{d}\mathcal{M} = \overset{\nsim}{T}_{f}\, \mathrm{d}S,
\end{equation}
where $\overset{\nsim}{T}_{f}$ represents the corrected version of the \textit{Hawking} temperature obtained through the first law of thermodynamics and $S$ denotes the entropy.

It is important to emphasize that the function $\Upsilon(r_{+},\mathcal{M},\alpha,\lambda,\Lambda)$, dependent on the terms of the mass function, not only defines the first law for regular black holes in the specific scenario discussed here, but also plays a crucial role in establishing the first law for various other classes of regular black holes (see \cite{Maluf:2018lyu}).

As indicated in Ref. \cite{Ma:2014qma}, the general formula for $\Upsilon(r_{+},\mathcal{M},\alpha,\lambda,\Lambda)$ is as follows,
\begin{equation}
\label{correction}
\Upsilon(r_{+},\mathcal{M},\alpha,\lambda,\Lambda) = 1 + 4\pi \int^{\infty}_{r_{+}} r^{2} \frac{\partial T^{0}_{0}}{\partial  \mathcal{M}} \mathrm{d}r.
\end{equation}
Here the notation $T^{0}_{0}$ corresponds to the stress--energy component associated with the energy density. Therefore, Eq.  \eqref{correction} can explicitly be evaluated as given below,
\begin{align}
& \Upsilon(r_{+},\mathcal{M},\alpha,\lambda,\Lambda) \nonumber\\
& = \frac{3 \left(\alpha  r_{+}^2-2 (\alpha +1) \lambda ^2\right)^2 \left(\frac{\alpha  \mathcal{M}}{(\alpha +1) \lambda ^2}+\frac{2 \mathcal{M}}{r_{+}^2}-\frac{1}{3} 2 \Lambda  r_{+}\right)}{4 \pi  (\alpha +1) \lambda ^2 \mathcal{Z}},
\end{align}
where $\mathcal{Z}=6 (\alpha +1) \lambda ^2+\alpha  \Lambda  r_{+}^4+3 r_{+}^2 \left(\alpha -2 (\alpha +1) \lambda ^2 \Lambda \right)$.
After these remarks all \textit{Hawking} temperatures are in agreement with each other
\begin{equation}
\overset{\nsim}{T}_{f}=T = \Upsilon(r_{+},\mathcal{M},\alpha,\lambda,\Lambda) \Tilde{T}_{f}.
\end{equation}

In this manner, we may write the entropy $S$ as given below,
\begin{equation}
S = \int \frac{\Upsilon(r_{+},\mathcal{M},\alpha,\lambda,\Lambda)}{\overset{\nsim}{T}_{f}} \mathrm{d} \mathcal{M} = \pi r_{+}^{2} = \frac{A}{4}\,,
\end{equation}
and we recover the usual area law.

\subsection{Heat capacity}

Indeed, when it comes to exploring additional thermodynamic quantities, the heat capacity is of particular interest. It can be expressed as,
\begin{equation}
C_V = T \frac{dS}{dT} = 2\pi r_+ ^2 \frac{\cal D}{\cal E}\,,
\end{equation}
where
\begin{eqnarray}
{\cal D} &=& 12(1+\alpha)^2(\Lambda r_+ ^2 -1) \lambda^4 - 8 \Lambda \alpha r_+ ^4 (1+\alpha) \lambda^2 \nonumber \\ 
&&+\alpha^2 r_+ ^4 (\Lambda r_+ ^2 +3)\,, \\
{\cal E} &=& 12(1+\alpha)^2(\Lambda r_+ ^2 +1) \lambda^4 - 24 \alpha r_+ ^2 (1+\alpha) \lambda^2 \nonumber \\ 
&&+\alpha^2 r_+ ^4 (\Lambda r_+ ^2 -3)\,.
\end{eqnarray}

\begin{figure}
    \centering
    \includegraphics[scale=0.45]{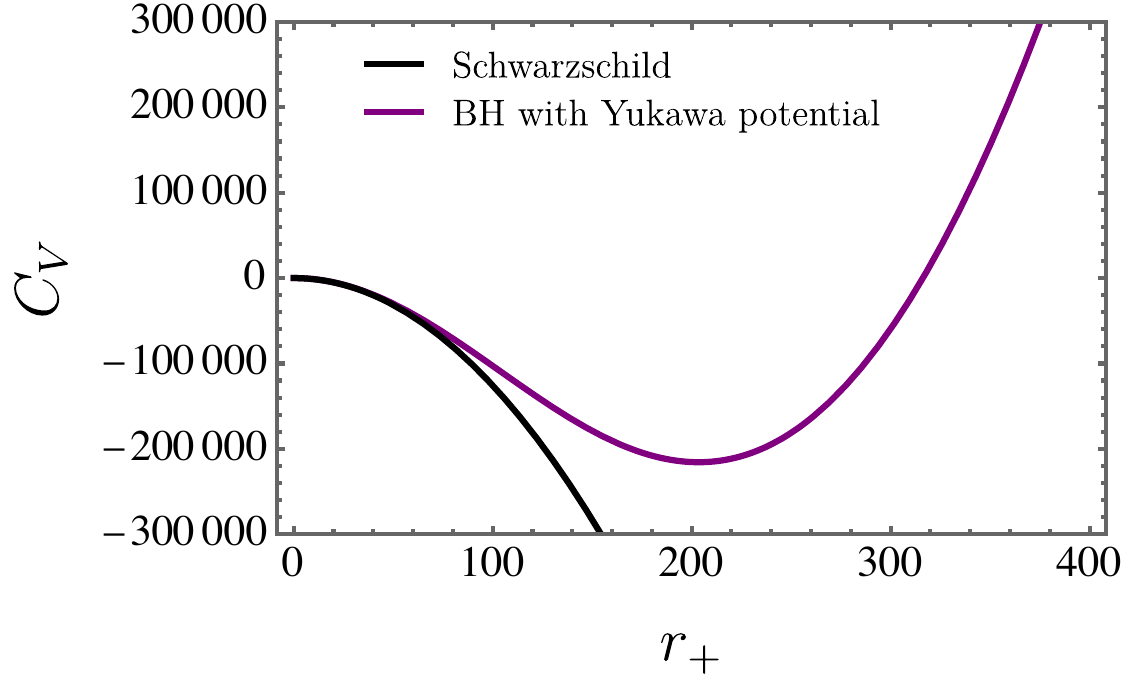}
    \caption{Heat capacity as a function of $r_{+}$ for $\lambda = 10^{5}$, $\Lambda = 10^{-5}$, and $\alpha = 1$ for small values of $r_+$.}
    \label{heat1}
\end{figure}

\begin{figure}
    \centering
    \includegraphics[scale=0.45]{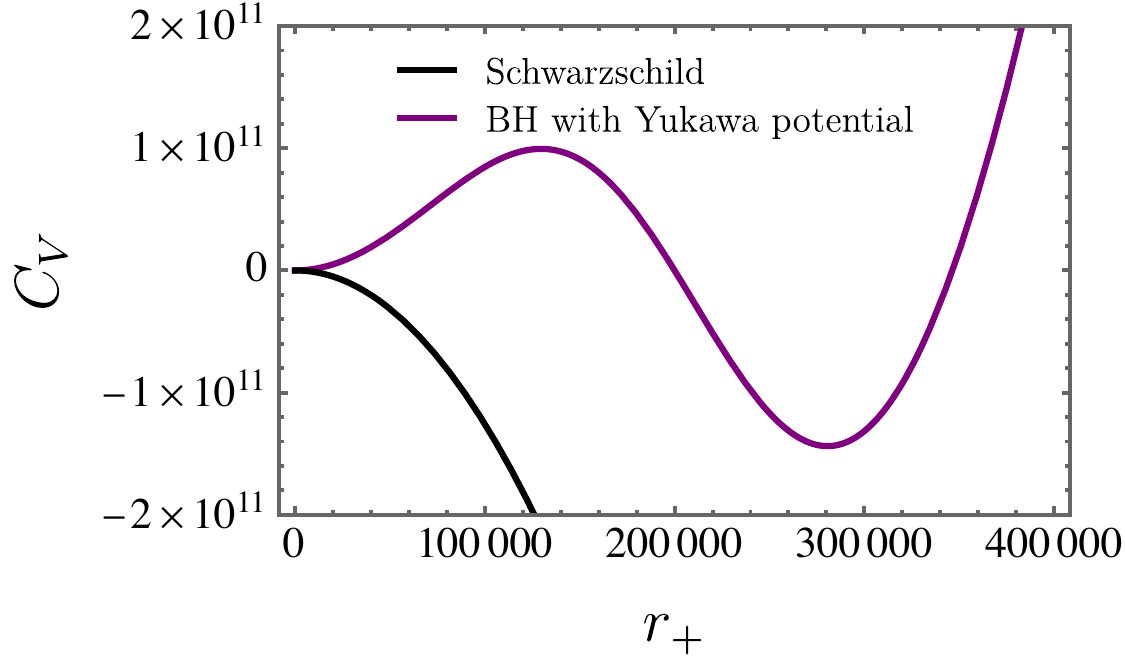}
    \caption{The heat capacity as a function of $r_{+}$ for $\lambda = 10^{5}$, $\Lambda = 10^{-5}$, and $\alpha = 1$ for large values of $r_+$.}
    \label{heat2}
\end{figure}

In Figs. \ref{heat1} and \ref{heat2} we show the heat capacity as a function of the event horizon radius for small and large values of the event horizon, respectively. From Fig. \ref{heat1} we notice that, for small black holes, the Yukawa corrected heat capacity follows Schwarzschild solution, however, as $r_+$ grows, it departs and displays a very rich behavior (see Fig. \ref{heat2}) which includes first order phase transitions at the points where $C_{V}=0$. Moreover, in the regions where $C_{V}<0$ the solution is thermodynamically unstable.  

{Although our results show some instabilities and phase transitions in the Yukawa corrected black hole, they say nothing about the dynamic stability of the object. Thus, in the next section, in order to test this type of stability, we should perturb the background solution and obtain the corresponding quasinormal frequencies.}

\section{Quasinormal frequencies}
\label{sectIV}

During the ringdown phase the remarkable phenomenon of \textit{quasinormal} modes arises showcasing unique oscillation patterns unaffected by initial perturbations. These modes reflect the system's inherent characteristics stemming from the natural oscillations of spacetime, irrespective of specific initial conditions. Contrary to \textit{normal} modes, which are related to closed systems, \textit{quasinormal} modes correspond to open systems. Consequently, they gradually lose energy by emitting gravitational waves. From a mathematical perspective they are described as poles of the complex Green function.

Solutions to the wave equation within a system governed by the background metric \(g_{\mu\nu}\) must be found to ascertain their frequencies. Nonetheless, obtaining analytical solutions for these modes often poses significant challenges. Various methodologies for getting solutions for these modes have been explored in scientific literature. The WKB (Wentzel--Kramers--Brillouin) method is among the most prevalent. This approach traces its roots to the pioneering contributions of Will and Iyer \cite{Iyer:1986np,Iyer:1986nq}. Later enhancements extending up to sixth order were achieved by Konoplya \cite{Konoplya:2003ii} and up to thirteenth order by Matyjasek and Opala~\cite{Matyjasek:2017psv}.

\subsection{Scalar perturbations}

\begin{figure}
    \centering
    \includegraphics[scale=0.5]{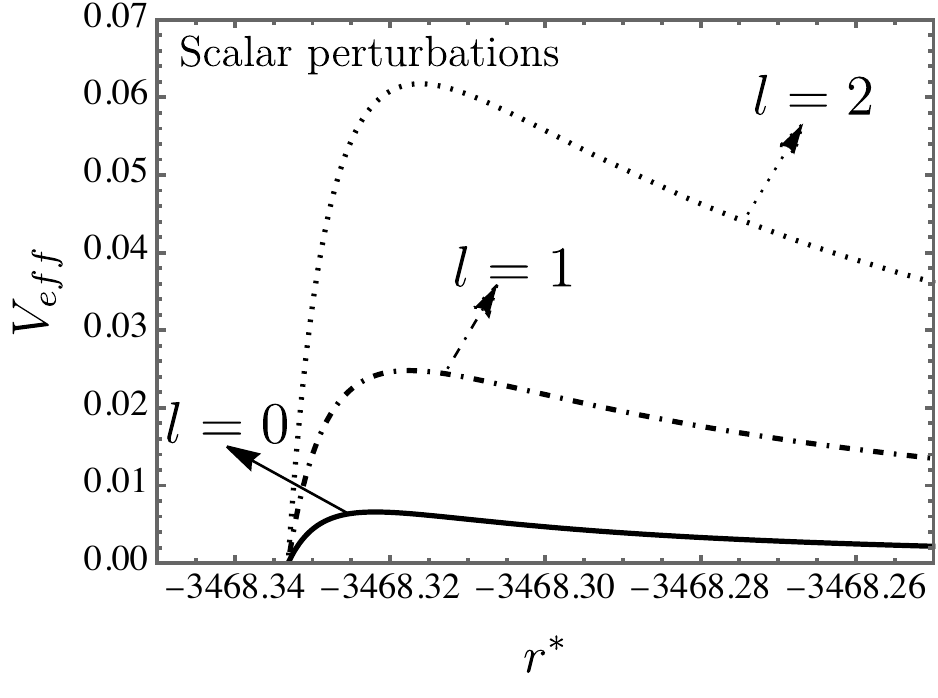}
    \caption{The effective potential $V_{eff}$ is depicted as a function of the tortoise coordinate $r^{*}$ considering different values of $l$ for $\mathcal{M}=\alpha=1$, $\lambda = 10^{5}$, and $\Lambda = 10^{-5}$.}
    \label{effectivepotentialyukawa}
\end{figure}

For our calculations we specifically examine perturbations via the scalar field incorporating the Klein--Gordon equation in the setting of curved spacetime,
\begin{equation}
\frac{1}{\sqrt{-g}}\partial_{\mu}(g^{\mu\nu}\sqrt{-g}\partial_{\nu}\Phi) = 0.\label{KL}
\end{equation}
Though the investigation of \textit{backreaction} effects in this context is compelling, this manuscript steers its attention elsewhere. We concentrate primarily on examining the scalar field as a minor perturbation. The existing spherical symmetry in the scenario enables us to decompose the scalar field in a particular way detailed further below,
\begin{equation}
\Phi(t,r,\theta,\varphi) = \sum^{\infty}_{l=0}\sum^{l}_{m=-l}r^{-1}\Psi_{lm}(t,r)Y_{lm}(\theta,\varphi),\label{decomposition}
\end{equation}
where the spherical harmonics are denoted by $Y_{lm}(\theta,\varphi)$. By inserting the decomposition of the scalar field, as illustrated in Eq. \eqref{decomposition} into Eq. \eqref{KL}, the equation takes a Schrödinger--like form. This transformation imbues the equation with wave-like attributes rendering it especially apt for our analysis,
\begin{equation}
-\frac{\partial^{2} \Psi}{\partial t^{2}}+\frac{\partial^{2} \Psi}{\partial r^{*2}} + V_{eff}(r^{*})\Psi = 0.\label{schordin11ger}
\end{equation}
The potential \(V_{eff}\) is commonly referred to as the \textit{Regge--Wheeler} potential or the effective potential. It encapsulates essential details about the black hole's geometry. We also introduce the tortoise coordinate $r^{*}$, which covers the full expanse of spacetime stretching as $r^{*}\rightarrow \pm \infty$. This is represented by $\mathrm{d} r^{*} = \sqrt{[1/f(r)^{2}]}\mathrm{d}r$. With some algebraic rearrangements we can express the effective potential as
\begin{equation}
V_{eff} = f(r)\left( \frac{l (l+1)}{r^2}+\frac{\frac{\alpha  \mathcal{M}}{\lambda^2 (1+\alpha)}+\frac{2 \mathcal{M}}{r^2}-\frac{2 \Lambda  r}{3}}{r} \right). \label{effectivepotenrtial}
\end{equation}
To better comprehend Eq. \eqref{effectivepotenrtial}, we provide Fig.  \ref{effectivepotentialyukawa} for $\mathcal{M}=\alpha=1$, $\lambda = 10^{5}$, and $\Lambda = 10^{-5}$. 
We observe that a barrier-like shape forms when considering positive values for the cosmological constant $\Lambda$, $\lambda$, and $\alpha$. Notice that, as $l$ increases, the height of $V_{eff}$ also rises.

In order to calculate the \textit{quasinormal} modes, we focus on the WKB method. Next, our main goal is to obtain stationary solutions for the system. To achieve this we propose that the function \(\Psi(t,r)\) can be represented as \(\Psi(t,r) = e^{-i\omega t} \psi(r)\), where \(\omega\) signifies the frequency. By adopting this representation, we can seamlessly isolate the time-independent aspect of Eq. \eqref{schordin11ger} as outlined below,
\begin{equation}
\frac{\partial^{2} \psi}{\partial r^{*2}} - \left[  \omega^{2} - V_{eff}(r^{*})\right]\psi = 0.\label{timeindasdependent}
\end{equation}
To effectively solve Eq. \eqref{timeindasdependent} it is essential to consider the relevant boundary conditions meticulously. For our particular scenario solutions meeting these conditions are distinguished by their purely ingoing behaviour near the horizon,
\[
    \psi^{\text{in}}(r^{*}) \sim 
\begin{cases}
    C_{l}(\omega) e^{-i\omega r^{*}} & ( r^{*}\rightarrow - \infty)\\
    A^{(-)}_{l}(\omega) e^{-i\omega r^{*}} + A^{(+)}_{l}(\omega) e^{+i\omega r^{*}} & (r^{*}\rightarrow + \infty).\label{boundaryconditions11}
\end{cases}
\]
The complex constants $A^{(+)}_{l}(\omega)$, $C_{l}(\omega)$, and $A^{(-)}_{l}(\omega)$ are crucial for our following analysis. They are foundational in examining the \textit{quasinormal} modes of a black hole, characterized by frequencies \(\omega_{nl}\) that meet the condition $A^{(-)}_{l}(\omega_{nl})=0$. These modes showcase a distinctive behaviour, manifesting as purely outgoing waves at spatial infinity and exclusively ingoing waves near the event horizon. The integers $n$ and $l$ delineate the overtone and multipole numbers, respectively. Furthermore, it is worth mentioning that the \textit{quasinormal} modes spectrum is anchored on the eigenvalues of Eq.  \eqref{timeindasdependent}. To delve into these frequencies we utilize the WKB method, a semi-analytical approach reminiscent of quantum mechanics.

Furthermore, the WKB approximation, initially introduced by Schutz and Will \cite{1985ApJ...291L..33S}, has emerged as an indispensable method for determining \textit{quasinormal} modes, especially when studying particle scattering around black holes. This technique has been refined over the years with significant contributions by Konoplya \cite{Konoplya:2003ii, Konoplya:2004ip}. It is crucial, however, to recognize that the applicability of this method is contingent on the potential assuming a barrier--like form and levelling off to constant values as $r^{*} \to \pm \infty$. By aligning the solution power series with the peak potential turning points, the \textit{quasinormal} modes can be accurately derived \cite{Santos:2015gja}. Given these premises the sixth order WKB formula is expressed as,
\begin{equation}
\frac{i(\omega^{2}_{n}-V_{0})}{\sqrt{-2 V''_{0}}} - \sum^{6}_{j=2} \Lambda_{j} = n + \frac{1}{2}.
\end{equation}
In other words Konoplya's formulation for the \textit{quasinormal} modes encompasses various components. Specifically, the term $(V''_{0}$ denotes the second derivative of the potential, calculated at its zenith $(r_{0}$). Additionally, the constants \(\Lambda_{j}\) are influenced by the effective potential and its derivatives at this peak. Remarkably, recent progress in this domain has unveiled a 13th-order WKB approximation pioneered by Matyjasek and Opala \cite{Matyjasek:2017psv}, which markedly elevates the precision of \textit{quasinormal} frequency computations.

Notice that the \textit{quasinormal} frequencies associated with the scalar field have a negative imaginary component. This defining characteristic suggests that these modes experience exponential decay over time, representing the energy dissipation through scalar waves. This feature agrees with previous studies investigating scalar, electromagnetic, and gravitational perturbations in spherically symmetric configurations \cite{Konoplya:2011qq,Berti:2009kk,Heidari:2023bww,2023InJPh.tmp..228C}.

Table  \ref{TABI} shows quasinormal frequencies calculated via $6$-th order WKB approximation for spin-zero particles considering different values of $\lambda$ and fixing the other parameters as $\mathcal{M}=\alpha=1$ and $\Lambda=10^{-10}$. We can notice that as $\lambda$ grows, the real and imaginary parts of the frequency converge to a stable value, which corresponds to the Schwarzschild quasinormal frequency.

\begin{table*}[tbp]
\begin{tabular}{|l|l|l|l|l|}
\hline
 \multicolumn{1}{|c|}{ spin 0 } &  \multicolumn{1}{c|}{  $l=1, n=0$ } & \multicolumn{1}{c|}{  $l=2, n=0$ } & \multicolumn{1}{c|}{ $l=2, n=1$ }\\\hline
   $\lambda$ & \hspace{1,5cm}$\omega_{n}$ & \hspace{1,5cm}$\omega_{n}$ & \hspace{1,5cm}$\omega_{n}$   \\ \hline
$10^{1}$ & 0.300356 - 0.1005030$i$ & 0.495007 - 0.0995914$i$ & 0.474429 - 0.304355$i$  \\ 
$10^{2}$ & 0.292984 - 0.0977892$i$ & 0.483756 - 0.0967944$i$ & 0.463953 - 0.295714$i$    \\
$10^{3}$ & 0.292910 - 0.0977619$i$ & 0.483643 - 0.0967664$i$ & 0.463848 - 0.295628$i$  \\ 
$10^{4}$ & 0.292910 - 0.0977616$i$ & 0.483642 - 0.0967661$i$ & 0.463847 - 0.295627$i$  \\ 
$10^{5}$ & 0.292910 - 0.0977616$i$ & 0.483642 - 0.0967661$i$ & 0.463847 - 0.295627$i$  \\\hline
\end{tabular}
\caption{Quasinormal frequencies calculated via $6$-th order WKB approximation for spin-zero particles varying $\lambda$  with fixed parameters,  $\mathcal{M}=\alpha=1$ and $\Lambda=10^{-10}$.}
\label{TABI}
\end{table*}

\subsection{Electromagnetic perturbations}
In this section we move forward with the examination of the electromagnetic field's propagation. To accomplish this we recall the wave equations obeyed by a test electromagnetic field,
\begin{equation}
\frac{1}{\sqrt{-g}}\partial_{\nu}\left[ \sqrt{-g} g^{\alpha \mu}g^{\sigma \nu} \left(A_{\sigma,\alpha} -A_{\alpha,\sigma}\right) \right]=0.
\end{equation}
The four-potential, denoted as $A_{\mu}$, can be expanded in 4-dimensional vector spherical harmonics in the following manner \cite{Toshmatov:2017bpx},
\begin{small}
\begin{align}\notag
& A_{\mu }\left( t,r,\theta ,\phi \right)  \nonumber \\
&=\sum_{l ,m} 
\begin{bmatrix} 
f(t,r)Y_{l m}\left( \theta ,\phi \right) \\
h(t,r)Y_{l m}\left( \theta ,\phi \right) \\
\frac{a(t,r)}{\sin \left( \theta \right) }\partial _{\phi }Y_{l
m}\left( \theta ,\phi \right) + k(t,r)\partial _{\theta }Y_{l m}\left( \theta ,\phi \right)\\
-a\left( t,r\right) \sin \left( \theta \right) \partial _{\theta }Y_{l
m}\left( \theta ,\phi \right)+ k(t,r)\partial _{\varphi }Y_{l m}\left( \theta ,\phi \right)
\end{bmatrix}.%
\end{align}%
\end{small}
In this expansion $Y_{l m}(\theta,\phi)$ represents the spherical harmonics. It is worth noting that the first term on the right-hand side exhibits a parity of $\left(-1\right)^{l +1}$ (referred to as the axial sector), while the second term has a parity of $\left(-1\right)^l$ (known as the polar sector). When we straightforwardly substitute this expansion into the Maxwell equations, we can derive a second-order differential equation for the radial component (see in particular \cite{Toshmatov:2017bpx}),
\begin{equation}
\frac{\mathrm{d}^{2}\Psi \left( r_{\ast }\right) }{\mathrm{d}r_{\ast }^{2}}+\left[ \omega
^{2}-V_{E}\left( r_{\ast }\right) \right] \Psi \left( r_{\ast }\right) =0.
\end{equation}%
For both the axial and polar sectors we obtain a second-order differential equation for the radial part with the relation $r_{\ast} = \int f^{-1}(r)dr$ denoting the tortoise coordinate. The mode $\Psi(r_{\ast})$ is a linear combination of the functions $a(t,r)$, $f(t,r)$, $h(t,r)$, and $k(t,r),$ but the specific functional dependence varies according to the parity. For the axial sector the mode is expressed as follows,
\begin{equation}
a(t,r)=\Psi \left( r_{\ast }\right),
\end{equation}
whereas for the polar sector it is given by,
\begin{equation}
\Psi \left( r_{\ast }\right) =\frac{r^{2}}{%
l (l +1)}\left[ \partial _{t}h(t,r)-\partial _{r}f(t,r)\right].
\end{equation}
The corresponding  effective potential in our case is found to be,
\begin{eqnarray}
    V_{eff}(r)=f(r) \left(\frac{l(l+1)}{r^2}\right),
\end{eqnarray}
with 
\begin{eqnarray}
\nonumber
        f(r)=1-\frac{2 \mathcal{M}}{r}+\frac{\mathcal{M} \alpha r}{(1+\alpha)\lambda^2}-\frac{\Lambda r^2}{3}.
    \end{eqnarray}

\begin{figure}
    \centering
    \includegraphics[scale=0.5]{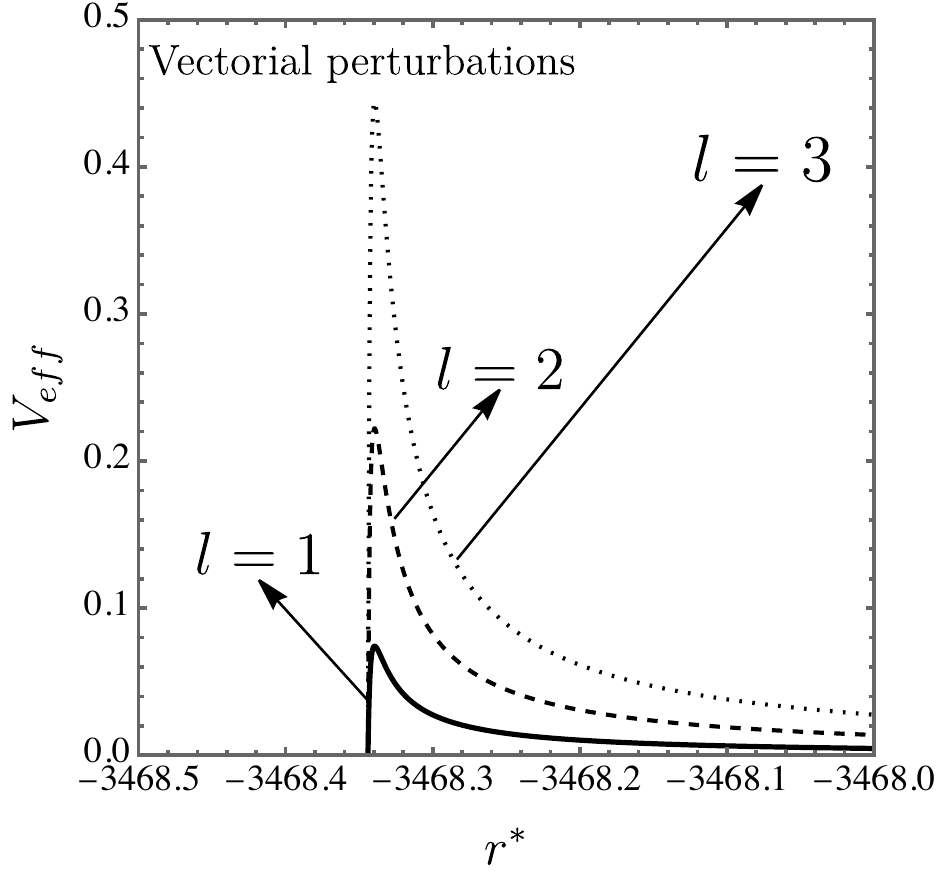}
    \caption{The effective potential $V_{eff}$ concerning vectorial perturbations is depicted as a function of the tortoise coordinate $r^{*}$ considering different values of $l$ as well as  $\mathcal{M}=\alpha=1$, $\lambda = 10^{5}$, and $\Lambda = 10^{-5}$.}
    \label{vectorialperturbations}
\end{figure}

In Fig.  \ref{vectorialperturbations} we display the effective potential, considering the vectorial perturbations for different values of $l$. As we did before, we have considered $\mathcal{M}=\alpha=1$, $\lambda = 10^{5}$, and $\Lambda = 10^{-5}$. We can see from this figure that the barrier grows as $l$ increases.

Table  \ref{TABII} shows quasinormal frequencies performed via $6$-th order WKB approximation for spin-one particles for several values of $\lambda$ keeping the fixed parameters $\mathcal{M}=\alpha=1$ and $\Lambda=10^{-10}$. As mentioned in the scalar case, we again notice that as $\lambda$ increases, both parts of the frequencies tend to the Schwarzschild asymptotical value.

\begin{table*}[tbp]
\begin{tabular}{|l|l|l|l|l|}
\hline
 \multicolumn{1}{|c|}{ spin 1 } &  \multicolumn{1}{c|}{  $l=1, n=0$ } & \multicolumn{1}{c|}{  $l=2, n=0$ } & \multicolumn{1}{c|}{ $l=2, n=1$ }\\\hline
   $\lambda$ & \hspace{1,5cm}$\omega_{n}$ & \hspace{1,5cm}$\omega_{n}$ & \hspace{1,5cm}$\omega_{n}$   \\ \hline
$10^{1}$ & 0.253368 - 0.0953556$i$ & 0.467604 - 0.0978352$i$ & 0.445768 - 0.299454$i$  \\ 
$10^{2}$ & 0.248244 - 0.0926643$i$ & 0.457694 - 0.0950395$i$ & 0.436627 - 0.290816$i$    \\
$10^{3}$ & 0.248192 - 0.0926373$i$ & 0.457594 - 0.0950114$i$ & 0.436535 - 0.290729$i$  \\ 
$10^{4}$ & 0.248191 - 0.0926370$i$ & 0.457593 - 0.0950112$i$ & 0.436534 - 0.290728$i$  \\ 
$10^{5}$ & 0.248191 - 0.0926371$i$ & 0.457593 - 0.0950112$i$ & 0.436534 - 0.290728$i$  \\\hline
\end{tabular}
\caption{Quasinormal frequencies performed via $6$-th order WKB approximation for spin-one particles varying $\lambda$   with fixed parameters, $\mathcal{M}=\alpha=1$ and $\Lambda=10^{-10}$.}
\label{TABII}
\end{table*}

\subsection{Tensorial perturbations}
Let us now investigate the QNMs for tensorial (gravitational) perturbations. In particular, we shall follow closely Ref.\cite{Kim:2004ju}, where the gravitational perturbations were investigated.  In general, one can write the axially symmetric spacetime as,
\begin{eqnarray}\notag
\mathrm{d}s^2 &=& -e^{2\nu}\mathrm{d}t^2 + e^{2\psi}(\mathrm{d}\phi - q_1\mathrm{d}t - q_2\mathrm{d}r - q_3\mathrm{d}\theta )^2\\
&+& e^{2\mu_2}\mathrm{d}r^2 + e^{2\mu_3}\mathrm{d}\theta^2. 
\end{eqnarray}
In the case of unperturbed black hole spacetime one can write
\be e^{2\nu} = f(r), \quad
e^{-2\mu_2}=\left( 1 - \f{2m(r)}{r} \right) = \f{\Delta}{r^2}, 
\ee
along with
\be
\Delta = r^2 - 2 m(r) r, \quad e^{\mu_3} = r, \quad e^\psi = r
\sin\theta, \ee and \be q_1=q_2=q_3=0, \ee
where the metric \eqref{approximated-solution} has been rewritten as,
\begin{eqnarray}
    f(r)=1-\frac{2 m(r)}{r},
\end{eqnarray}
such that the mass function is given by,
\begin{eqnarray}
    m(r)=\mathcal{M}-\frac{\mathcal{M} \alpha r^2}{2 \lambda^2 (1+\alpha)}+\frac{\Lambda r^3}{6}.
\end{eqnarray}

It is well known that axial perturbations are characterized by
$q_1$, $q_2$, and $q_3$. Note that for the linear perturbations
$\delta\nu, \delta\psi, \delta\mu_2, \delta\mu_3$ one has
polar perturbations with even parity. This type of perturbation will not be considered in the present paper. Now from Einstein's equations we have, \be ( e^{3\psi+\nu-\mu_2-\mu_3}
Q_{23} )_{,3} = - e^{3\psi-\nu-\mu_2+\mu_3} Q_{02,0},\ee where
$x^2 = r, x^3 = \theta$ and $Q_{AB}=q_{A,B}-q_{B,A}, Q_{A0} =
q_{A,0}-q_{1,A}$ \cite{Kim:2004ju}. In particular, this can further be written as, \be
\f{\sqrt{f(r)}}{\sqrt{\Delta}}\f{1}{r^3\sin^3\theta}\f{\pa
Q}{\pa\theta} = - (q_{1,2} - q_{2,0})_{,0},\ee with $Q$ given by \be
Q(t,r,\theta) = \Delta Q_{23}\sin^3\theta = \Delta
(q_{2,3}-q_{3,2})\sin^3\theta. \ee

Having in mind another important equation, \be ( e^{3\psi+\nu-\mu_2-\mu_3} Q_{23} )_{,2} =
e^{3\psi-\nu+\mu_2-\mu_3} Q_{03,0}, \ee it can be shown that \be
\f{\sqrt{f(r)}\sqrt{\Delta}}{r^3\sin^3\theta}\f{\pa Q}{\pa\theta} =
(q_{1,3} - q_{3,0})_{,0}.\ee

We can further show that by using $Q(r,\theta) = Q(r)C^{-3/2}_{l+2}(\theta)$, in which we have the Gegenbauer
function $C^\nu_n (\theta)$ that satisfies the following differential equation \cite{Kim:2004ju},
\be
\left[ \f{\mathrm{d}}{\mathrm{d}\theta}\sin^{2\nu}\theta \f{\mathrm{d}}{\mathrm{d}\theta} +
n(n+2\nu)\sin^{2\nu}\theta \right] C^\nu_n (\theta) = 0, \ee then,
\be r \sqrt{f(r) \Delta} \f{\mathrm{d}}{\mathrm{d}r} \left( \f{\sqrt{f(r) \Delta}
}{r^3} \f{\mathrm{d}Q}{\mathrm{d}r} \right) - \mu^2
\f{f(r)}{r^2}Q + \omega^2Q = 0,  \ee 
with
$\mu^2=(l-1)(l+2)$. Finally, if we introduce $Q=rZ$ along with $\f{\mathrm{d}}{\mathrm{d}r_*}=
\sqrt{f(r) \Delta}\f{1}{r}\f{\mathrm{d}}{\mathrm{d}r} $, we obtain
\be
\left( \f{\mathrm{d}^2}{\mathrm{d}r^{*2}} +
\omega^2 - V(r) \right) Z = 0, 
\ee 
with the following potential,
\begin{equation}
    V_{eff}(r)=f(r) \left(\frac{l(l+1)}{r^2}-\frac{6 m(r)}{r^3}+\frac{2 m'(r)}{r^2} \right),
\end{equation}
or 
\begin{equation}
    V_{eff}(r)=f(r) \left(\frac{l(l+1)}{r^2}-\frac{6 \mathcal{M}}{r^3}+\frac{\mathcal{M}\alpha}{\lambda^2 (1+\alpha)r} \right).
\end{equation}
Notice the exciting result that the cosmological constant does not affect the potential for gravitational perturbations. In addition, in the case $\alpha \to 0$ we get the well-known potential for the Schwarzschild black hole.

\begin{figure}
    \centering
    \includegraphics[scale=0.4]{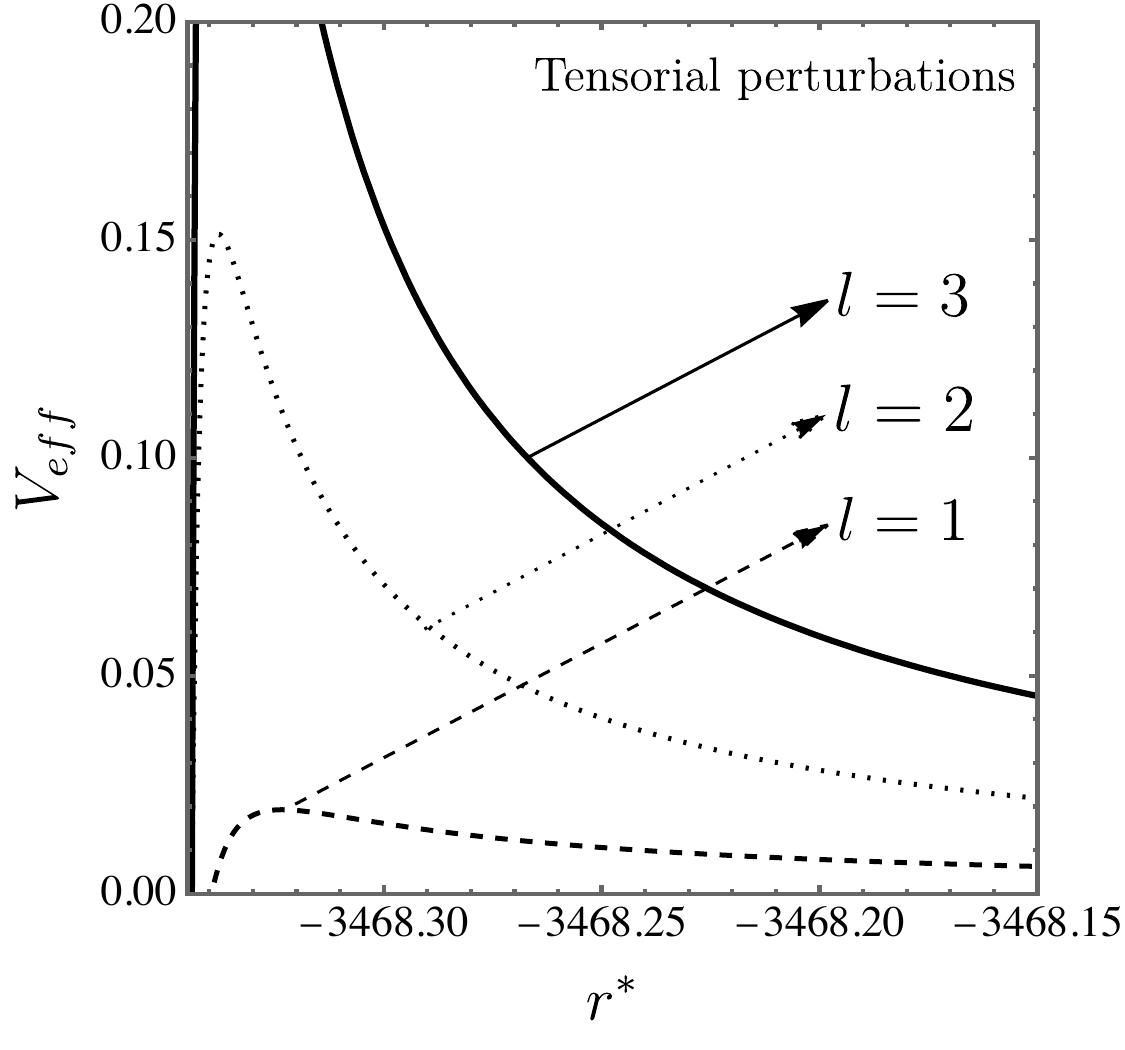}
    \caption{The effective potential $V_{eff}$ for tensorial perturbations is plotted as a function of the tortoise coordinate $r^{*}$ considering different values of $l$ and fixing $\mathcal{M}=\alpha=1$, $\lambda = 10^{5}$, and $\Lambda = 10^{-5}$.}
    \label{tensorialperturbations}
\end{figure}

In Fig.  \ref{tensorialperturbations} we display the effective potential concerning tensorial perturbations for diverse values of $l$, considering $\mathcal{M}=\alpha=1$, $\lambda = 10^{5}$, and $\Lambda = 10^{-5}$. We notice the general behavior of an increasing peak as the multipole number $l$ grows. 

\begin{table*}[tbp]
\begin{tabular}{|l|l|l|l|l|}
\hline
 \multicolumn{1}{|c|}{ spin 2 } &  \multicolumn{1}{c|}{  $l=2, n=0$ } & \multicolumn{1}{c|}{  $l=2, n=1$ } & \multicolumn{1}{c|}{ $l=3, n=0$ }\\\hline
   $\lambda$ & \hspace{1,5cm}$\omega_{n}$ & \hspace{1,5cm}$\omega_{n}$ & \hspace{1,5cm}$\omega_{n}$   \\ \hline
$10^{1}$ & 0.381934 - 0.0913639$i$  & 0.353395 - 0.281224$i$ & 0.612734 - 0.0953797$i$  \\ 
$10^{2}$ & 0.373703 - 0.0889158$i$ & 0.346368 - 0.273557$i$ & 0.599577 - 0.0927294$i$    \\
$10^{3}$ & 0.373620 - 0.0888912$i$ & 0.346297 - 0.273481$i$ & 0.599445 - 0.0927028$i$  \\ 
$10^{4}$ & 0.373619 - 0.0888910$i$ & 0.346297 - 0.273480$i$ & 0.599443 - 0.0927025$i$  \\ 
$10^{5}$ & 0.373619 - 0.0888910$i$ & 0.346297 - 0.273480$i$ & 0.599443 - 0.0927025$i$  \\\hline
\end{tabular}
\caption{Quasinormal frequencies computed via $6$-th order WKB approximation for spin--two particles with varying $\lambda$ and fixed parameters $\mathcal{M}=\alpha=1$ and $\Lambda=10^{-10}$.}
\label{TABIII}
\end{table*}

Table  \ref{TABIII} presents the numerical values for the real part and the imaginary part for some gravitational modes obtained through the 6-th order WKB method. For smaller values of $\lambda$ the value of the real part is shown to be more significant compared to the Schwarzschild case; however, the more precise astrophysical values we use for $\lambda$, the closer the Schwarzschild case is obtained. This means that the Yukawa black hole corrections and the Schwarzschild case are indistinguishable. 

{Our results in this section show that Yukawa black hole metric is stable against scalar, electromagnetic, and gravitational perturbations since the imaginary part of all the frequencies we found is always negative. Following our aim to study the outcome of the gravitational field of this black hole, in the next section we will discuss the effect on the test particle motion in the neighborhood of this geometry focusing in the lightlike trajectories.}

\section{Geodesic trajectories}
\label{sectVI}

\subsection{The complete case}

\begin{figure}
    \centering
    \includegraphics[scale=0.52]{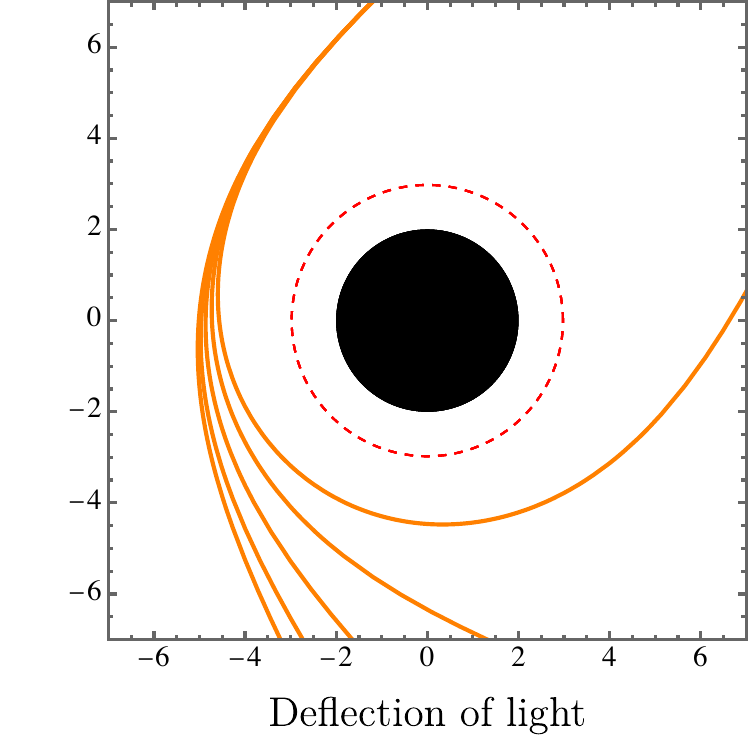}
    \caption{The light path for the complete metric case when $\Lambda = 10^{-10}$,  $\mathcal{M} = \alpha = 1$, and $l_{0}=0$ for different values of $\lambda$. The black circle represents the event horizon and the dashed points correspond to the photon sphere of the black hole. The greater the value of $\lambda$, the more pronounced the deviation of the light becomes.}
    \label{lightpath}
\end{figure}

In this section our primary focus lies in understanding the behavior resulting from the geodesic equation. This involves expressing it as follows,
\begin{equation}
\frac{\mathrm{d}^{2}x^{\mu}}{\mathrm{d}s^{2}} + \Gamma\indices{^\mu_\alpha_\beta}\frac{\mathrm{d}x^{\alpha}}{\mathrm{d}s}\frac{\mathrm{d}x^{\beta}}{\mathrm{d}s} = 0. \label{geodesicfull}
\end{equation}
Taking into account Eq. \eqref{exact-sol} the above equation leads to four partial differential equations
\begin{small}
\begin{align}
 & \frac{\mathrm{d}\theta^{\prime}}{\mathrm{d}s} = \sin (\theta ) \cos (\theta ) \left(\varphi '\right)^2-\frac{2 \theta ' r'}{r},
\\
& \frac{\mathrm{d}\varphi^{\prime}}{\mathrm{d}s}= -\frac{2 \varphi ' \left(r'+r \theta ' \cot (\theta )\right)}{r}, \\
\begin{split}
&\frac{\mathrm{d}t^{\prime}}{\mathrm{d}s} = -\frac{1}{\frac{2 M r^2}{\left(l_{0}^2+r^2\right)^{3/2}}+\frac{2 \alpha  M e^{-\frac{r}{\lambda }} (\lambda +r)}{\lambda  r}+\frac{\Lambda  r^2}{3}-1}\\
& \times r' t' \left[M \left(\frac{4 r}{\left(l_{0}^2+r^2\right)^{3/2}}-\frac{6 r^3}{\left(l_{0}^2+r^2\right)^{5/2}}-\frac{2 \alpha  e^{-\frac{r}{\lambda }}}{r^2} \right.\right. \\
 &\left.\left. 
-\frac{2 \alpha  e^{-\frac{r}{\lambda }}}{\lambda ^2}-\frac{2 \alpha  e^{-\frac{r}{\lambda }}}{\lambda  r}\right)+\frac{2 \Lambda  r}{3}\right],
\end{split}
\end{align}
\begin{align}
\begin{split}
&\frac{\mathrm{d}r^{\prime}}{\mathrm{d}s} = \frac{e^{-\frac{r}{\lambda }}}{6 \lambda  \left(l_{0}^2+r^2\right)^{3/2}} \\
& \times \left\{   \frac{1}{r \left(\frac{2 M r^2}{\left(l_{0}^2+r^2\right)^{3/2}}+\frac{2 \alpha  M e^{-\frac{r}{\lambda }} (\lambda +r)}{\lambda  r}+\frac{\Lambda  r^2}{3}-1\right)^2} \right.\\
& \left. \times \left[\left( 2 M \left(-\frac{2 r}{\left(l_{0}^2+r^2\right)^{3/2}}+\frac{3 r^3}{\left(l_{0}^2+r^2\right)^{5/2}}+\frac{\alpha  e^{-\frac{r}{\lambda }}}{r^2} \right.\right.\right.\right. \\
& \left.\left.\left.\left. +\frac{\alpha  e^{-\frac{r}{\lambda }}}{\lambda ^2}+\frac{\alpha  e^{-\frac{r}{\lambda }}}{\lambda  r}\right)-\frac{2 \Lambda  r}{3} \right) \right. \right. \\
& \left.\left. \times \left[ \left(  -l_{0}^2 \sqrt{a^2+r^2} \left(6 \alpha  M (\lambda +r)+\lambda  r e^{r/\lambda } \left(\Lambda  r^2-3\right)\right)  \right.\right.\right.\right. \\
& \left.\left.\left.\left. -r^{2} \left( 6 M \left( \alpha  \lambda  \sqrt{l_{0}^2+r^2}+\alpha  r \sqrt{l_{0}^2+r^2}+\lambda  r e^{r/\lambda }    \right)  \right.\right.\right.\right.\right.\\
&\left.\left.\left.\left.\left. +  \lambda  r \sqrt{a^2+r^2} e^{r/\lambda } \left(\Lambda  r^2-3\right) \right)  \right) r^{\prime 2} \right]\right.\right. \\
& \left.\left. - \frac{1}{r}  \times \left[  2 M \left(-\frac{2 r}{\left(l_{0}^2+r^2\right)^{3/2}}+\frac{3 r^3}{\left(l_{0}^2+r^2\right)^{5/2}}+\frac{\alpha  e^{-\frac{r}{\lambda }}}{r^2} \right.\right.\right.\right.\\
& \left.\left.\left.\left. +\frac{\alpha  e^{-\frac{r}{\lambda }}}{\lambda ^2}  
 +\frac{\alpha  e^{-\frac{r}{\lambda }}}{\lambda  r}\right)-\frac{2 \Lambda  r}{3}     \right] \right.\right.\\
 &\left.\left. \times \left[   -l_{0}^2 \sqrt{l_{0}^2+r^2} \left(6 \alpha  M (\lambda +r)+\lambda  r e^{r/\lambda } \left(\Lambda  r^2-3\right)\right) \right.\right.\right. \\
 & \left.\left.\left.  + \lambda  r \sqrt{l_{0}^2+r^2} e^{r/\lambda }  (\Lambda  r^2-3)\right]  \right] t^{\prime 2} \right. \\
 & \left.  -2\times \left[  l_{0}^2 \sqrt{l_{0}^2+r^2} \left(6 \alpha  M (\lambda +r)+\lambda  r e^{r/\lambda } \left(\Lambda  r^2-3\right)\right)  \right.\right. \\
 & \left. \left.    +r^2 \left(6 M \left(\alpha  \lambda  \sqrt{l_{0}^2+r^2}+\alpha  r \sqrt{l_{0}^2+r^2} \right. \right.\right.\right.\\
 &\left.\left.\left.\left. +\lambda  r e^{r/\lambda }\right)+\lambda  r \sqrt{l_{0}^2+r^2} e^{r/\lambda } \left(\Lambda  r^2-3\right)\right)    \right] \theta^{\prime 2} \right. \\
 & \left. -2 \left[ l_{0}^2 \sqrt{l_{0}^2+r^2} \left(6 \alpha  M (\lambda +r)+\lambda  r e^{r/\lambda } \left(\Lambda  r^2-3\right)\right) \right.\right. \\
 & \left. \left. + r^{2} \left( 6 M \left(\alpha  \lambda  \sqrt{l_{0}^2+r^2}+\alpha  r \sqrt{l_{0}^2+r^2}+\lambda  r e^{r/\lambda }\right) \right.\right.\right.\\ & \left.\left.\left. \lambda  r \sqrt{l_{0}^2+r^2} +e^{r/\lambda } \left(\Lambda  r^2-3\right)  \right) \right] \sin ^2(\theta ) \varphi^{\prime 2} \right\}.
 \end{split}
\end{align}
\end{small}
In Fig.  \ref{lightpath}, we display the light path for $l_{0}=0$. Here the more $\lambda$ increases, the larger the deviation becomes, represented by the orange lines (light path). The black circle represents the event horizon and the dashed points are the photon sphere of the black hole.

\subsection{The approximated case}
Let us discuss the case when the geodesic trajectories are obtained from the approximated solution of $f(r)$ in Eq. \eqref{approximated-solution} when the parameters $\varepsilon=1/\lambda$ and $\ell_0$ approach zero. With these approximations the equations are simplified significantly, as we can see in what follows,
\begin{small}
\begin{align}
&\frac{\mathrm{d}\theta^{\prime}}{\mathrm{d}s}= \sin (\theta ) \cos (\theta ) \left(\varphi '\right)^2-\frac{2 \theta ' r'}{r},
\\
&\frac{\mathrm{d}\varphi^{\prime}}{\mathrm{d}s} = -\frac{2 \varphi ' \left(r'+r \theta ' \cot (\theta )\right)}{r},\\
& \frac{\mathrm{dt^{\prime}}}{\mathrm{d}s} = \frac{r' t' \left(\frac{\alpha  \mathcal{M}}{(\alpha +1) \lambda ^2}+\frac{2 \mathcal{M}}{r^2}-\frac{2 \Lambda }{3} r\right)}{\frac{\Lambda  r^2}{3}-\frac{\alpha  \mathcal{M} r}{(\alpha +1) \lambda ^2}+\frac{2 \mathcal{M}}{r}-1},
\\
\begin{split}
& \frac{\mathrm{d}r^{\prime}}{\mathrm{d}s} =\\
&  \frac{\left(r'\right)^2 \left(-6 (\alpha +1) \lambda ^2 \mathcal{M}+2 (\alpha +1) \lambda ^2 \Lambda  r^3-3 \alpha  \mathcal{M} r^2\right)}{2 r \left(6 (\alpha +1) \lambda ^2 \mathcal{M}+(\alpha +1) \lambda ^2 \Lambda  r^3-3 \alpha  \mathcal{M} r^2-3 (\alpha +1) \lambda ^2 r\right)} \\
& - \frac{1}{18 (\alpha +1)^2 \lambda ^4 r^3}\\
& \times \left\{  \left[  6 (\alpha +1) \lambda ^2 \mathcal{M}+(\alpha +1) \lambda ^2 \Lambda  r^3-3 \alpha  \mathcal{M} r^2-3 (\alpha +1) \lambda ^2 r  \right]  \right. \\
& \left. \times \left[ \left.-6 (\alpha +1) \lambda ^2 \mathcal{M}+2 (\alpha +1) \lambda ^2 \Lambda  r^3-3 \alpha  \mathcal{M} r^2\right)t^{\prime 2} \right] \right. \\
& \left. \times \left[  +6r^3\alpha +1\lambda ^2\left(\theta^{\prime 2} +\sin ^2(\theta ) \varphi^{\prime 2} \right.  \right]
\right\}.
\end{split}
\end{align}
\end{small}

\begin{figure}
    \centering
    \includegraphics[scale=0.33]{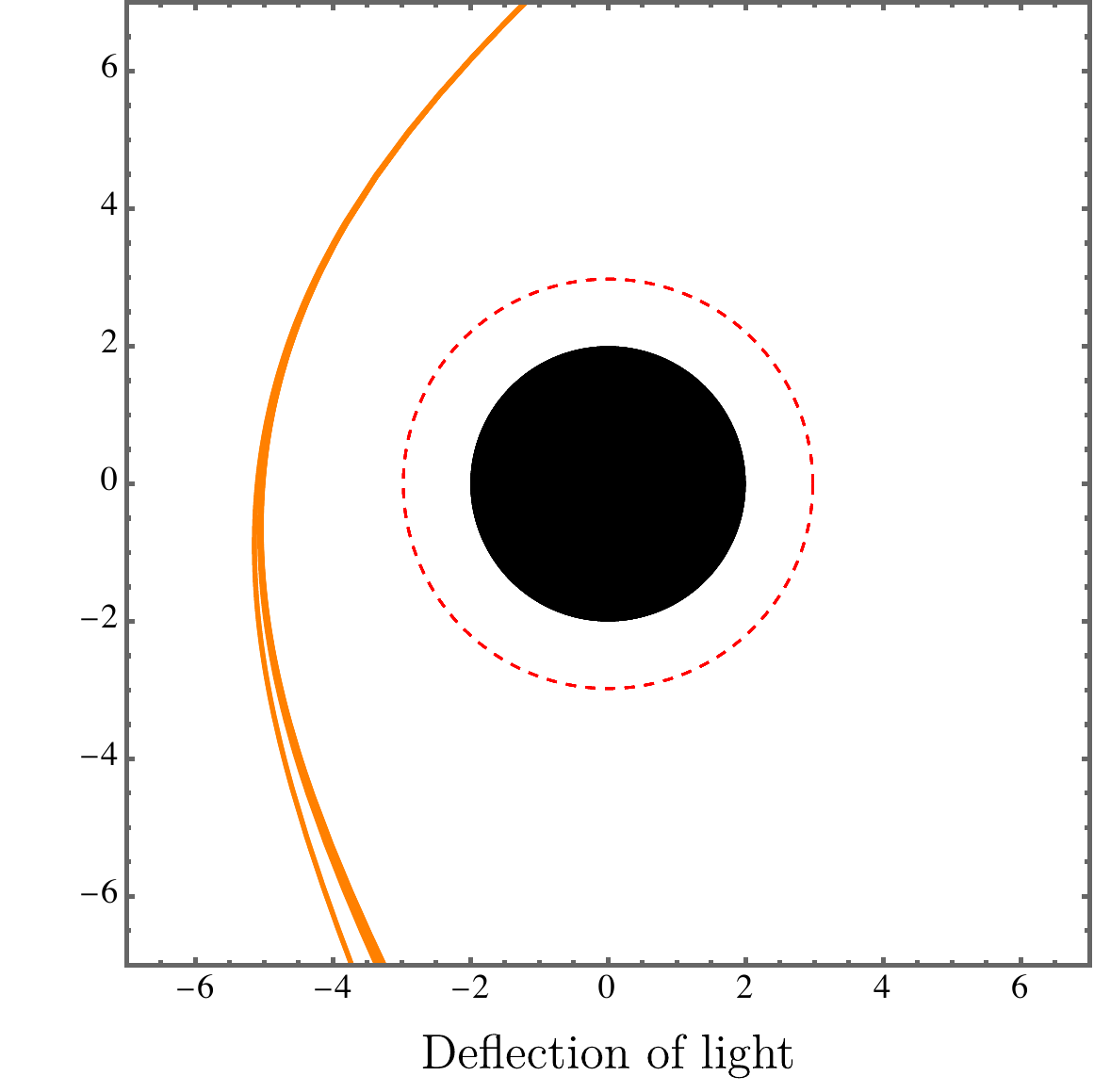}
    \caption{In the particular scenario where $\Lambda = 10^{-10}$, $\mathcal{M} = \alpha = 1$, and $l_{0} = 0$ we can observe the trajectory of light for various values of $\lambda$. The event horizon is denoted by a black circle, while the dashed points mark the photon sphere surrounding the black hole. }
    \label{lightpath2}
\end{figure}

In Fig.  \ref{lightpath2} we display the behavior of the light path close to the black hole. Intriguingly, after around $\lambda > 30$ the deflection of light is practically not altered for such a parameter.

{Light paths corresponding to the set of last unstable orbits around the compact object, known as photon sphere, give rise to an interesting phenomenon that will be studied in the next section, namely the shadow of a black hole. }

\section{Photon spheres and Shadows: the case of optically thin radiating and infalling gas surrounding a black hole}
\label{sectVII}

To fully grasp particle dynamics and light ray patterns near black holes, it is essential to understand photon spheres. These spheres are crucial in discerning the shadows, shedding light on the effects of the Yukawa potential in the spacetime under consideration.

Understanding the significance of the photon sphere, commonly labelled as the critical orbit, in the context of black hole dynamics requires, for instance, the application of the Lagrangian method for calculating null geodesics. We aim to ascertain the effects of parameters, specifically \(M,\Lambda,\alpha,\lambda\), on the photon sphere. To elucidate further, we state
\begin{equation}
\mathcal{L} = \frac{1}{2} g_{\mu\nu}\Dot{x}^{\mu}\Dot{x}^{\nu}.
\end{equation}
When fixing the angle at $\theta=\pi/2$, the previously mentioned expression simplifies to
\begin{equation}
g_{00}^{-1} E^{2} + g_{11}^{-1} \Dot{r}^{2} + g_{33}^{-1} L^{2} = 0, \label{separation}
\end{equation}
where $L$ represents the angular momentum and $E$ denotes the energy. Subsequently, Eq.  \eqref{separation} is given by, 
\begin{equation}
\Dot{r}^{2} = E^{2} - \left(  \frac{\alpha  \mathcal{M} r}{(\alpha +1) \lambda ^2}-\frac{2 \mathcal{M}}{r}-\frac{\Lambda  r^2}{3}+1 \right)\left(  \frac{L^{2}}{r^{2}} \right),
\end{equation}
with 
\begin{equation}\label{potgeo}
\mathcal{V} \equiv \left(  \frac{\alpha \mathcal{M} r}{(\alpha +1) \lambda ^2}-\frac{2 \mathcal{M}}{r}-\frac{\Lambda  r^2}{3}+1 \right)\left(  \frac{L^{2}}{r^{2}} \right)\,,
\end{equation}
being the effective potential. To determine the critical radius, the equation $\partial \mathcal{V}/\partial r = 0$ must be solved, which can be expressed as,
\begin{align}\label{rps}
    r_{\rm c}=\frac{\sqrt{(1+\alpha)\left(6\mathcal{M}^2\lambda^2\alpha+\lambda^4(1+\alpha)\right)}-\lambda^2(1+\alpha)}{\mathcal{M}\alpha}.
\end{align}
Here a remarkable feature can be singled out, the cosmological constant does not affect the critical orbits.

Moreover, a unique physical solution exists for this equation, denoted as $r_{c}$ -- the photon sphere radius. Recent literature, such as Refs. \cite{Filho:2023qxu}, has tackled similar studies within dark matter contexts. Also, it is noteworthy to highlight that the emergence of two-photon spheres has been recently documented in the Simpson--Visser solution context \cite{Tsukamoto:2021caq,Tsukamoto:2022vkt} among others \cite{Guerrero:2022msp}.
The left side of Table  \ref{photonspheres} shows various results for the shadow radius corresponding to different values of $\lambda$ while keeping $\alpha=1$. In this scenario as $\lambda$ increases, the shadow radius decreases. Given that this parameter is directly tied to the Yukawa potential, our findings suggest that for astrophysical values of the parameters the overall influence on the shadow radius is very small. In terms of the graviton mass $\lambda=\hbar/(m_g c)$, this means the smaller the graviton mass is, the stronger the effect becomes. As the graviton mass decreases, the shadow radius also decreases. 

Examining shadows in the context of Yukawa potential and black hole structures is of paramount importance. Defined by the distinct outline of a black hole set against a bright background, these shadows offer perspectives on spacetime geometry and gravitational phenomena near the black hole. By analyzing these shadows, we can glean valuable information to test and enhance theoretical models further authenticating the nature of gravity. To streamline our analysis we introduce two new parameters as follows,
\begin{equation}
\label{newparameters}
\xi  = \frac{L}{E}
\text{ and }
\eta  = \frac{\mathcal{K}}{{E^2}},
\end{equation}
where $\mathcal{K}$ is the so--called Carter constant. After performing algebraic manipulations we obtain, 
\begin{equation}
{\xi ^2} + \eta  =  \frac{r_{c}^2}{f(r_{c})}.
\end{equation}

In the endeavor to deduce the radius of the shadow we shall make use of the celestial coordinates $X$ and $Y$ \cite{Singh:2017vfr,Heidari:2023bww,Filho:2023ycx,Heidari:2023egu,Filho:2023etf} as follows, $X=-\xi$ and $Y=\pm\sqrt{\eta}$. It is natural to define the true or the physical mass of the black hole to be $\mathcal{M} = M (1+\alpha)$. Furthermore, in the case of large distances with a cosmological constant it has been shown that the size of the BH shadow can explicitly depend on the radial coordinate of the distant observer noted as $r_O$. Following \cite{Gonzalez:2023rsd} we have
\begin{eqnarray}
\sin^2\alpha_{\rm sh} = \frac{r^2_{c}}{f(r_{c})}\frac{f(r_O)}{r_O}\,.
\label{eq:alphashnotasymptoticallyflat}
\end{eqnarray}
In the physically relevant small-angle approximation it is easy to see that the shadow size is given by,
\begin{eqnarray}
R_{\rm SH} = r_{c} \sqrt{\frac{f(r_O)}{f(r_{c})}}\,.
\label{eq:rshnotasymptoticallyflat}
\end{eqnarray}
It is easy to show that the shadow radius can be written as \cite{Gonzalez:2023rsd},
\begin{equation}
 R_{\rm SH}=\frac{r_{c}}{ \sqrt{f(r_{c})}}\sqrt{1-\frac{2 GM_{\rm BH}}{c^2 r_O}+\frac{G M_{\rm BH} \alpha r_O}{c^2 (1+\alpha) \lambda^2}-\frac{1}{3}\Lambda r_O^2},
\end{equation}
where $r_O$ is the distance to the black hole. 

Now let us consider a straightforward model, specifically an optically sparse accretion flow that emits radiation around the object. We will employ a numerical method called Backward Raytracing to uncover the shadow created by this radiating flow. To determine the intensity distribution within the emitting region, we must make certain assumptions about the radiative processes and emission mechanisms in play. In particular, the observed specific intensity $I_{\nu 0}$ at the observed photon frequency $\nu_\text{obs}$ at the point $(X,Y)$ of the observer's image (usually measured in $\text{erg} \text{s}^{-1} \text{cm}^{-2} \text{str}^{-1} \text{Hz}^{-1}$) is given by, \cite{Bambi:2013nla,Saurabh:2020zqg}
\begin{eqnarray}
    I_{obs}(\nu_{obs},X,Y) = \int_{\gamma}\mathrm{g}^3 j(\nu_{e})dl_\text{prop},  
\end{eqnarray}
 
In this analysis we are examining a simplified scenario involving accreting gas. Our assumption is that the gas undergoes radial free fall and its four-velocity, under static and spherically symmetric conditions, simplifies to the following \cite{Bambi:2013nla, Saurabh:2020zqg},
\begin{eqnarray}
u^t_{e} & = & \frac{1}{f(r)}, \;
u^r_{e} = -\sqrt{\frac{g(r)}{f(r)}\left(1-f(r)\right)}, \nonumber \\
u^{\theta}_{e} & = & 0, \;
u^{\phi}_{e} = 0, 
\end{eqnarray}
where 
\begin{eqnarray}
f(r)=g(r)=1-\frac{2 \mathcal{M}}{r}+\frac{\mathcal{M} \alpha r}{(1+\alpha)\lambda^2}-\frac{\Lambda r^2}{3}.
\end{eqnarray}
Using the four-velocity for the photons we can establish a connection between the radial and temporal components of the four-velocity,
\begin{eqnarray}
    \frac{k^r}{k^t} = \pm f(r) \sqrt{g(r)\bigg(\frac{1}{f(r)}-\frac{b^2}{r^2}\bigg)}.
\end{eqnarray}
The sign +(-) represents the direction in which the photon is either approaching or moving away from the massive object, respectively. As a result, the redshift function $\mathrm{g}$ is expressed as follows,
\begin{eqnarray}
  \mathrm{g} = \frac{1}{\frac{1}{f(r)} \pm \frac{k_r}{k_t}\sqrt{\frac{g(r)}{f(r)}\bigg(1-f(r)\bigg)}}.
\end{eqnarray}
Regarding the specific emissivity, we adopt a straightforward model in which the emission is monochromatic, having an emitter's rest-frame frequency denoted as $\nu_{\star}$. Additionally, the emission exhibits a radial profile that follows an inverse square law, i.e., $1/r^2$,
\begin{eqnarray}
    j(\nu_{e}) \propto \frac{\delta(\nu_{e}-\nu_{\star})}{r^2},
\end{eqnarray}
where $\delta$ is the Dirac delta function along with the proper length which can be written as, 
\begin{eqnarray}
    dl_{\text{prop}} = k_{\alpha}u^{\alpha}_{e}d\lambda = -\frac{k_t}{\mathrm{g}|k^r|}dr.
\end{eqnarray}
By integrating the intensity across all observable frequencies, we arrive at the observed flux,
\begin{eqnarray} \label{inten}
    F_{obs}(X,Y) \propto -\int_{\gamma} \frac{\mathrm{g}^3 k_t}{r^2k^r}dr.
\end{eqnarray}

\begin{table}[!h]
\begin{center}
\begin{tabular}{c c c c c c  } 
 \hline\hline
 $\mathcal{M}$ & $\alpha$ & $\Lambda$ [m$^{-2}$] & $\lambda/\mathcal{M}$  & $R_{SH}/\mathcal{M}$  \\ [0.2ex] 
 \hline 
 1.0 & 1.0 & $10^{-52}$ & $10^{5}$ &  6.363961027   \\ 
 
 1.0 & 1.0 & $10^{-52}$ & $10^{6}$ &  5.209126608   \\ 

 1.0 & 1.0 & $10^{-52}$ & $10^{8}$  & 5.196153722   \\
 
 1.0 & 1.0 & $10^{-52}$ &  $10^{13}$ & 5.196152422   \\
 
 1.0 & 1.0 & $10^{-52}$ &  $10^{15}$ & 5.196152422   \\
 [0.2ex] 
 \hline \hline
\end{tabular}
\caption{\label{photonspheres} The shadow radius is presented for varying values of mass and the parameter $\lambda$. We have set $r_0/\mathcal{M}=10^{10}$. }
\label{TABIV}
\end{center}
\end{table}

\begin{figure*}
    \includegraphics[scale=0.5]{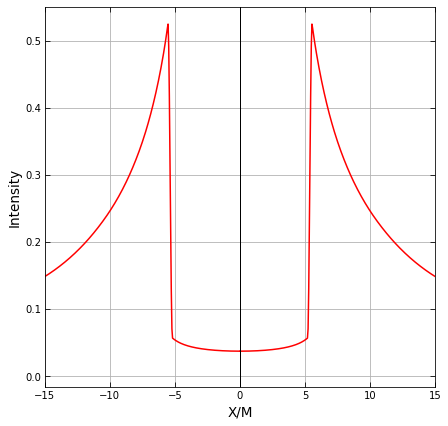}
     \includegraphics[scale=0.5]{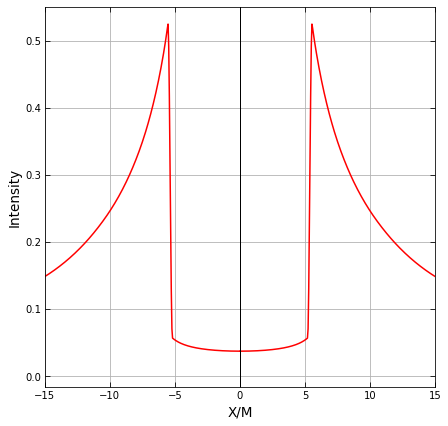}
    \caption{Plots of the intensity from the infalling gas for the Yukawa black hole (left) and the Schwarzschild black hole (right). If we use real astrophysical data, the result is practically the same. }
    \label{FIG10}
\end{figure*}

\begin{figure*}
      \includegraphics[scale=0.77]{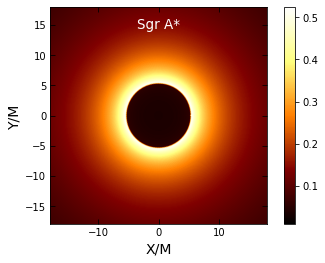}
       \includegraphics[scale=0.77]{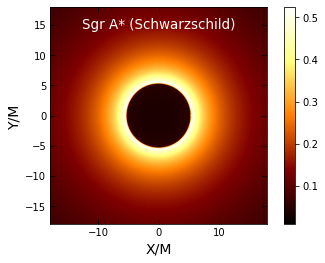}
    \caption{Plots of shadow images of the Yukawa black hole (left) and the Schwarzschild black hole (right) using data for Sgr A*. We have used real astrophysical data for $\lambda$ and $\alpha$ and the results show that the images are practically indistinguishable.}
    \label{FIG11}
\end{figure*}

In order to estimate the shadow radius for the Sgr A* black hole, we used real astrophysical values, $\Lambda \simeq 10^{-52} \rm 
m^{-2}$. Also, for this black hole we can take (here we have written the black hole mass in geometric units using the fact that $M_{\rm Sun}=1.5$ km),
\begin{eqnarray}
    M_{\rm BH}^{\rm Sgr A}=4 \times 10^6 M_{\rm Sun}=6 \times 10^9 \rm m.
\end{eqnarray}
For the distance we have $r_O=8.3 $ kpc and  $\lambda \sim 2693 $ Mpc or 
\begin{eqnarray}
    \frac{\lambda } { M_{\rm BH}^{\rm Sgr A}} =1.382406667 \times 10^{16},
\end{eqnarray}
along with $\alpha=0.416$ \cite{Gonzalez:2023rsd}. We work in geometric units; hence, we use $M_{\rm Sun}=1.5\,km$. For the shadow radius we obtain 
\begin{eqnarray}
    R_{\rm SH}/M_{\rm BH}^{\rm Sgr A}=5.196152423.
\end{eqnarray}
For the vacuum solution or the Schwarzschild black hole we know that 
\begin{eqnarray}
    R^{\rm Schwarzschild}_{\rm SH}/M_{\rm BH}^{\rm Sgr A}=3 \sqrt{3}\simeq 5.196152424.
\end{eqnarray}
This means that the change in the shadow radius compared to the Schwarzschild black hole is of the order 
\begin{eqnarray}
    \delta R_{\rm SH}=\frac{R^{\rm Schwarzschild}_{\rm SH}- R_{\rm SH}}{M_{\rm BH}^{\rm Sgr A}}\sim 10^{-9}.
\end{eqnarray}
In Fig.  \ref{FIG10} and Fig.  \ref{FIG11} we have shown the intensity of the infalling gas and the shadow images of the Yukawa black hole. Finally, let us briefly mention that recent studies  \cite{Jusufi:2024ifp,Jusufi:2024rba} pointed out the possibility of varying graviton mass. This means that $\lambda=\hbar/(m_g c)$ can fluctuate from Mpc order in the case of cosmological scales to kpc order in the case of galactic scales. Using kpc order for $\lambda$, we observe a similar result in the shadow radius, i.e., a change of $|\delta R_{\rm SH}|\sim 10^{-8}$--$10^{-9}$ compared to the Schwarzschild black hole shadow.

\subsection{Can we constrain $\alpha$?}
\begin{figure*}
      \includegraphics[scale=0.7]{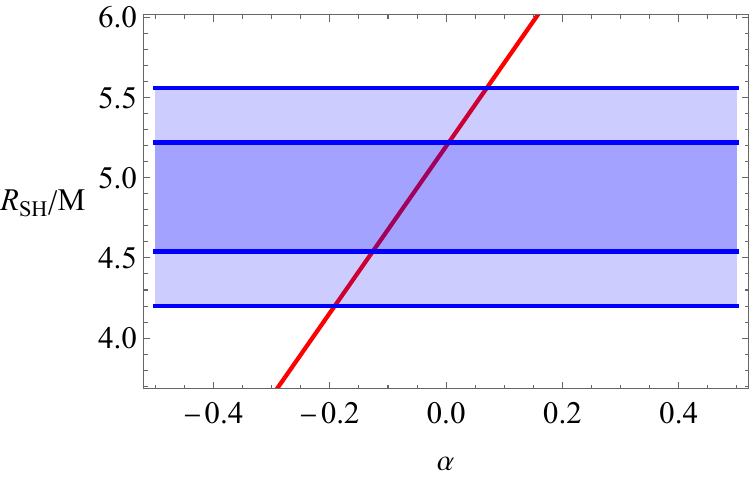}
       \includegraphics[scale=0.7]{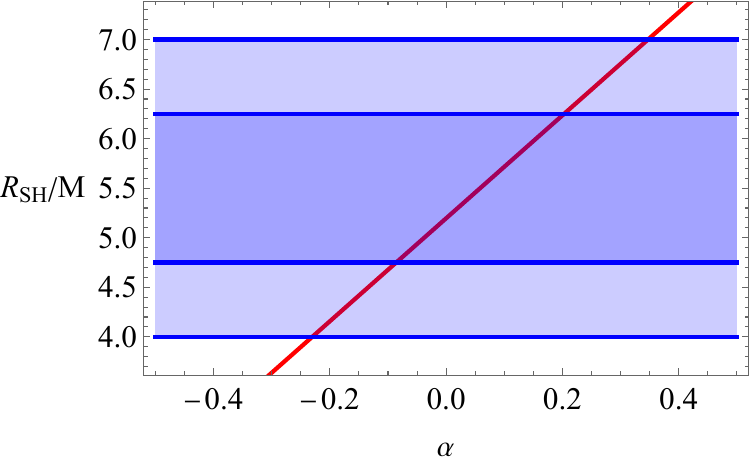}
    \caption{Plots of the shadow radius in units of the black hole bare mass $M$ as a function of $\alpha$ and 1$\sigma$ and 2$\sigma$ regions, respectively, for Sgr A* (left) and M87 (right).}
    \label{FIG12}
\end{figure*}

Note that in our discussion above we have identified the black hole mass with $\mathcal{M}=M_{\rm BH}^{\rm Sgr A}$, since as we have argued $\mathcal{M}=M+\alpha M$. This means that the total mass, as measured by an observer located far away from the black hole, is modified due to the extra term which mimics the effect of dark matter, i.e., $M_{\rm dark\, matter}=\alpha M$. We can interpret, on the other hand, $M$ as the “bare” mass of the black hole, while the extra term $M_{\rm dark\, matter}$ is only an apparent effect which arises from the modification of Newton's law of gravity in large distances but changes $M\to \mathcal{M}$ as measured by an observer located far away. A natural question arises: can we probe the “bare” mass $M$? First, we can easily see that the shadow radius can be approximated as $R_{SH}=3 \sqrt{3} M (1+\alpha)+\mathcal{O}(\Lambda,\alpha,\lambda^2)$, and by keeping only the first term, we can apply the same approach as in \cite{Vagnozzi:2022moj}. Using observations near the black hole, we can effectively probe $M$ and
not $\mathcal{M}$, for example, by studying the motion of S-stars orbiting
Sgr A*. In addition, there is uncertainty about the mass, which does not disappear when we
later fix $M = 1$, it is part of what allows us to constrain
the dark matter effect $\alpha$: so we can fix the bare mass and ask how
much of a change in the total mass/energy of the system
we can tolerate when varying the parameter $\alpha$. Using the  EHT observations for Sgr A*, it was shown  that within  $2\sigma$ constraints \cite{Vagnozzi:2022moj},
\begin{eqnarray}
4.21 \lesssim R_{ sh}/M \lesssim 5.56\,,
\end{eqnarray}
which gives the interval $-0.19 \lesssim \alpha \lesssim 0.07$ (within 2$\sigma$). On the other hand, for the M87 black hole it was shown within  $2\sigma$ constraints \cite{Allahyari:2019jqz},
\begin{eqnarray}
4 \lesssim R_{ sh}/M \lesssim 7,
\end{eqnarray}
which gives the interval $-0.23 \lesssim \alpha \lesssim 0.35$ (within 2$\sigma$). This fact is also shown in Fig.  \ref{FIG12}. We see that $\alpha$ can be negative too, but in the present paper we considered only $\alpha>0$. This shows that nearly $0.35 M$ (for M87) and $0.07 M$ (Sgr A*) of the total mass can be attributed to the apparent dark matter effect. \\

{In the next section we will make the connection between our results demonstrating in an analytical way that the Yukawa corrected black hole also obeys the correspondence between shadow radius and QNMs in the eikonal limit.}

\subsection{The galactic flat rotating curves}
Let us elaborate briefly on how the modification of the gravitational potential can lead to interesting results related to the flat rotating galactic curves. In such a case, the black hole mass does not play an important role and should be replaced by the mass distribution of the surrounding matter. Assuming a constant mass distribution for simplicity, in the limit of large radial distance we can write,
\begin{eqnarray}
f(r)\simeq 1+\frac{\mathcal{M} \alpha r}{(1+\alpha)\lambda^2}
\end{eqnarray}
where we have neglected the second term $2\mathcal{M}/r$, which is small for a very large radial distance, and the last term $\Lambda r^2/3$ since the cosmological constant is small and does not play an important role in galactic scales. The second term in the last equation plays the role of dark matter. For the case of a spherically symmetric spacetime ansatz with pure
dark matter in Schwarzschild coordinates the tangential
velocity using the radial function $f(r)$ can be calculated from the
following equation,
\begin{eqnarray}
v_0^2=\frac{r}{\sqrt{f(r)}}\frac{d}{dr}\left(\sqrt{f(r)}\right)
\end{eqnarray}
which describes a region with a nearly constant tangential
speed of the star. From the last equation we get \cite{Saurabh:2020zqg},
\begin{eqnarray}
    f(r)=\left(\frac{r}{r_0}\right)^{2 v_0^2}\simeq 1+2 v_0^2 \ln(r/r_0),
\end{eqnarray}
where $r_0$ is some integration constant. For the outer part of the galaxy we can take $\ln(r/r_0)\sim 1$ and $r\sim \lambda$, that leads to,
\begin{eqnarray}
    v_0=\sqrt{\frac{G \mathcal{M} \alpha}{2(1+\alpha)\lambda}}=\rm constant, 
\end{eqnarray}
where we have inserted the Newton's constant $G$. We already pointed out that $\lambda$ in cosmological scales was found to be of Mpc order. However, as was pointed out recently in \cite{Jusufi:2024ifp,Jusufi:2024rba}, the graviton mass related to $\lambda$ by $\lambda=\hbar/(m_g c)$ can fluctuate --for galactic scales $\lambda$ indeed can be of kpc order-- and, hence, the assumption for the outer part of the galaxy $r \sim \lambda$ is justified. As an example, one can take the total mass enclosed inside the galaxy to be $\mathcal{M} \sim  10^{12}$ solar masses along with $\lambda \sim 1 $ kpc order and $\alpha \sim 0.5$ which yields $v_0\sim 200$ km/s.  The idea of a mass-varying mechanism for the graviton is yet to be explored in the near future.

\section{Eikonal quasinormal modes and shadow radius correspondence}
\label{sectVIII}

As it is well known, QNMs are characterized by complex
frequencies, $\omega=\omega_{\Re}-i \omega_{\Im}$, with a real part
representing the oscillation of the wave and an imaginary part 
proportional to the damping of a given mode. Moreover, both numbers
depend only on the features of the metric being perturbed and contain
essential information related to the kind of perturbation being used.   
In the last decades several discussions around the nature of
these vibrations emerged. But Goebel pointed out a possible
relation with the photon sphere trajectories for the first
time~\cite{1972ApJ...172L..95G}.  
Later on, Cardoso et al. \cite{Cardoso:2008bp} showed that the real part
of the QNMs in the eikonal regime is related to the angular velocity
of the outermost photon orbit $\Omega_c$, while the imaginary part is
proportional to the Lyapunov exponent $\tilde\lambda$, which determines
the instability timescale of the orbit,
\begin{equation}
\omega_{QNM}=\Omega_c l -i \left(n+\frac{1}{2}\right)|\tilde{\lambda}|\,.
\end{equation}
This result was proven to be valid not only for static spherically symmetric
spacetimes but also for equatorial orbits in the geometry of rotating
black holes solutions. On the other hand, Stefanov et
al.~\cite{Stefanov:2010xz} found a connection between black-hole eikonal
quasinormal modes and gravitational lensing observables (flux ratio
$\tilde r$ and minimal impact angle $\theta_\infty$) in the
strong-deflection regime, 
\begin{equation}
\Omega_c=\frac{c}{\theta_\infty D_{OL}}\,,\qquad \tilde{\lambda}=\frac{c\,\ln \tilde{r}}{2 \pi \theta_\infty D_{OL}}\,,
\end{equation}
where $D_{OL}$ is the distance between the observer and the
lens. Afterwards, a connection with the shadow radius of a black hole
was put forward in~\cite{Jusufi:2019ltj} and analytically proven
in~\cite{Cuadros-Melgar:2020kqn}. This correspondence shows that the real part of the quasinormal frequency is proportional to the inverse
of the shadow radius of the same black hole. 

Nevertheless, we should stress that there are counterexamples to the
correspondence. 
Even for high multipole numbers $l$, the correspondence is
not guaranteed for any field. As stated in
~\cite{Konoplya:2017wot}, the correspondence works only for test fields whose effective potential is positive definite, has a barrier, and goes to a
constant value both at the event horizon and at infinity (or dS horizon),
such that a WKB method can be applied. However, even under these
conditions, some frequencies could be missed when using the standard
WKB method~\cite{Konoplya:2022gjp}.

To see if the correspondence holds for the black hole
corrected by the Yukawa potential, we are going to calculate the
quasinormal frequencies in the eikonal limit. 
Following~\cite{Cuadros-Melgar:2020kqn}, we consider the effective potential
for the scalar perturbation given in
Eq. \eqref{effectivepotenrtial}. Using the 6$^{th}$ order WKB method we
can write the complex frequency $\omega$ as,
\begin{eqnarray}\label{wkbfreq}
\omega^2 &=& V_0 + \frac{V_4}{8V_2}\left(\nu^2+\frac{1}{4}\right) -
\frac{(7+60\nu^2)V_3^2}{288V_2^2}  \nonumber \\
&& + i\nu\sqrt{-2 V_2}
\left[\frac{1}{2V_2^2}\left( \frac{5 V_3^4(77+188\nu^2)}{6912 V_2^3} \right.\right.
  \nonumber \\
&&- \frac{V_3^2 V_4 (51+100\nu^2)}{384 V_2^2} + \frac{V_4^2
  (67+68\nu^2)}{2304 V_2} \nonumber \\
&& \left.\left.+ \frac{V_5 V_3 (19+28\nu^2)}{288 V_2} +
\frac{V_6 (5+4\nu^2)}{288} \right) -1\right]\,,  \nonumber \\
\end{eqnarray}
where $V_i$ represents the $i^{th}$ derivative of the potential with
respect to $r$, evaluated at the position of the maximum value of the
potential $r=r_p$ and $\nu=n+1/2$, being $n$ the overtone
number. If we expand this expression for large $l$,  i.e., we are
taking the eikonal limit, we obtain a simple expression for the real
part of the quasinormal frequency, 
\begin{equation}\label{omre1}
\omega_{\Re} = \left(l+\frac{1}{2}\right) \frac{\sqrt{f(r_p)}}{r_p}\,.
\end{equation}
Now, to find the position of the peak of the potential, we
notice that in the eikonal limit Eq. \eqref{effectivepotenrtial} has
the same form as the lightlike geodesic potential shown in
Eq. \eqref{potgeo}, provided that $l(l+1)$ is identified with
$L^2$. Thus, the position of the maximum of the scalar perturbation
potential corresponds to the photon sphere radius found in
Eq. \eqref{rps}, $r_p=r_c$. Then, since the shadow radius $R_{\rm SH}$
is given by Eq. \eqref{eq:rshnotasymptoticallyflat}, we can rewrite Eq.  \eqref{omre1} as
\begin{equation}\label{omre2}
\omega_{\Re} = \left(l+\frac{1}{2}\right) \frac{\sqrt{f(r_O)}}{R_{\rm SH}}\,.
\end{equation}
Therefore, the correspondence still holds, but we have an additional
factor that depends on the radial coordinate of the distant observer
$r_O$. Furthermore, if we use real astrophysical values for the Sgr
A$^*$, for example, as was done in the previous section, $f(r_O)$
approaches unity, and we obtain the standard result found
in~\cite{Cuadros-Melgar:2020kqn,Jusufi:2019ltj}.

\begin{table*}[h]
\begin{center}
\begin{tabular}{c c c c c c c } 
 \hline\hline
 $\mathcal{M}$ & $\alpha$ & $\Lambda$ [m$^{-2}$] & $\lambda/\mathcal{M}$ & $l$ & $\omega_{\Re}$  & $\delta \omega_{\Re}$ \\ [0.2ex] 
 \hline 
  1.0 & 1.0 & $10^{-52}$ & $10^{10}$ &  1 & 0.288675134  & $6.4952 \times 10^{-21}$ \\ 

  1.0 & 1.0 & $10^{-52}$ & $10^{10}$ &  2 & 0.481125224  & $1.0825 \times 10^{-20}$ \\ 

  1.0 & 1.0 & $10^{-52}$ & $10^{10}$ &  3 & 0.673575314 & $1.5155 \times 10^{-20}$  \\ 
  
 1.0 & 1.0 & $10^{-52}$ & $10^{10}$ &  10 & 2.020725942  & $4.5466 \times 10^{-20}$  \\ 
 
 1.0 & 1.0 & $10^{-52}$ & $10^{10}$ &  20 &  3.945226840  & $1.5155 \times 10^{-20}$  \\ 

 1.0 & 1.0 & $10^{-52}$ & $10^{10}$  & 50 &   9.718729530 & $2.1867 \times 10^{-19}$ \\
 
 1.0 & 1.0 & $10^{-52}$ &  $10^{10}$ & 100 &  19.34123402 & $4.3518 \times 10^{-19}$ \\
 
 1.0 & 1.0 & $10^{-52}$ &  $10^{10}$ & 200 &  38.58624300 & $8.6819 \times 10^{-19}$ \\
 
 1.0 & 1.0 & $10^{-52}$ &  $10^{10}$ & 500 &  96.32126990 & $2.1672 \times 10^{-18}$ \\

 1.0 & 1.0 & $10^{-52}$ &  $10^{10}$ & 1000 &  192.5463148 & $4.3323 \times 10^{-18}$ \\
 [0.2ex] 
 \hline \hline
\end{tabular}
\caption{Numerical values for the eikonal QNMs for different $l$ using Eq.  \eqref{omre1} as well as $\delta \omega_{\Re}$, which gives the difference compared to the Schwarzschild black hole case, $\delta \omega_{\Re}=\omega_{\Re}-\omega_{\Re}^{\rm Schwarzschild.}$. }
\label{TABV}
\end{center}
\end{table*}

In Table  \ref{TABV}, we present our numerical result when computing the real part of eikonal QNMs. The results show that the effect of Yukawa corrections on astrophysical data is extremely small. Specifically, from this table we can see that comparing the QNMs to the Schwarzschild black hole case, the signature of Yukawa corrections that might mimic the dark matter effect is very small, \textit{e.g}, an increase for the QNMs of order $10^{-18}-10^{-21}$. In other words, from a practical point of view, it is almost impossible to detect such corrections using gravitational waves. \\


\section{Conclusion}
\label{sectIX}

In this paper we obtained a black hole solution with non--singular Yukawa--like potential given by \eqref{exact-sol-full}, which generalizes the metric presented in \cite{Gonzalez:2023rsd}. The second term of this solution modifies the geometry due to $\ell_0$ and plays an important role in short--range distances to cure the black hole singularity. The third term is due to the apparent dark matter effect and the fourth term is the contribution due to the cosmological constant. Since $\ell_0$ is of Planck length order, i.e., $\ell_0\sim 10^{-35}\,m$  \cite{Nicolini:2019irw}, in large distances and for astrophysical black holes with large mass $M$ we must have $r\gg\ell_0$, then, we can neglect the effect of $\ell_0$. In short, the spacetime geometry was modified in short distances due to the deformed parameter $\ell_0$ and, most importantly, in large distances due to both parameters $\alpha$ and $\lambda$. The large distance corrections were of particular interest since $\alpha$ modifies Newton's law of gravity and could mimic the dark matter effect. We also investigated the accuracy of the approximate solution \eqref{approximated-solution} using perturbation methods in differential equations.

The Yukawa corrections could be classified in two categories: i) corrections due to $\alpha$ could be significant since the $\alpha$ parameter modifies the total mass of the black hole via $\mathcal{M}=M+\alpha M$, or ii) corrections due to $\lambda$ were very small and could practically be neglected. Keeping this in mind, we subsequently explored the phenomenological aspects of the solution in large distances for astrophysical black holes. 

We performed a thermodynamic analysis by calculating \textit{Hawking} temperature, entropy, and heat capacity and we showed that this black hole solution underwent first--order thermodynamical phase transitions. In addition, we studied the \textit{quasinormal} frequencies for scalar, electromagnetic, and gravitational field perturbations. For doing so, we utilized the WKB method to probe the \textit{quasinormal} modes numerically. Since it was more natural to measure $\mathcal{M}$, we analyzed the QNMs using specific values of the parameters that dictate the structure of the black hole, namely, $\mathcal{M}$, $\alpha$, $\lambda$, and $\Lambda$.

Also we used astrophysical data for the parameters and an optically thin radiating and infalling gas surrounding a black hole to model the black hole shadow image. We considered as an example the Sgr A$^*$ black hole and we found that the shadow radius changes by order of $10^{-9}$, meaning that the shadow radius of a black hole with Yukawa potential practically gave the same results as those ones encountered in Schwarzschild black hole.

On the other hand, in the eikonal regime, using astrophysical data for Yukawa parameters, we showed that the value of the real part of the QNMs frequencies changes by $10^{-18}$. As we pointed out, the major effect via $\alpha$ was encoded in $\mathcal{M}$. However, through astrophysical observations, it is  simpler to probe $\mathcal{M}$ than $M$. In principle, one can probe $M$ with specific observations near the black hole such as the motion of S--stars. In that case we can constrain $\alpha$ and we can calculate how much the total mass changes by working in units of $M$. In general, we have a screening effect due to $\alpha$ and if we work in units of $\mathcal{M}$, the Yukawa--like corrections are difficult to measure by observations with the current technology, for example, using gravitational waves. \\

\section*{CRediT authorship contribution statement}

All the authors equally contributed to Conceptualization, Methodology, Software, Validation, Formal analysis, Investigation, Writing – original draft, Writing – review \& editing, Visualization, Investigation, Formal analysis, Funding acquisition, Supervision, and Project administration. 

\section*{Declaration of competing interest}

The authors declare that they have no known competing financial interests or personal relationships that could have appeared to influence the work reported in this paper.

\section*{Acknowledgements}

A. A. Araújo Filho would like to thank Fundação de Apoio à Pesquisa do Estado da Paraíba (FAPESQ) and Conselho Nacional de Desenvolvimento Cientíıfico e Tecnológico (CNPq)  -- [200486/2022-5] and [150891/2023-7] for the financial support. Genly Leon thanks Vicerrectoría de Investigación y Desarrollo Tecnológico (VRIDT) at Universidad Católica del Norte for financial support through Resolución VRIDT N°026/2023, Resolución VRIDT N°027/2023, Núcleo de Investigación Geometría Diferencial y Aplicaciones (Resolución VRIDT N°096/2022), Proyecto de Investigación Pro Fondecyt Regular 2023 (Resolución VRIDT N°076/2023), Resolución VRIDT N°09/2024 and Agencia Nacional de Investigación y Desarrollo (ANID) through Proyecto Fondecyt Regular 2024,  Folio 1240514, Etapa 2024. Genly Leon would like to express his gratitude towards faculty member Alan Coley and staff members Anna Maria Davis, Nora Amaro, Jeanne Clyburne, and Mark Monk for their warm hospitality during the implementation of the final details of the research in the Department of Mathematics and Statistics at Dalhousie University. 

\section*{Data availability}
No data was used or created for the research described in the article.

%


\begin{thebibliography}{136}%
\makeatletter
\providecommand \@ifxundefined [1]{%
 \@ifx{#1\undefined}
}%
\providecommand \@ifnum [1]{%
 \ifnum #1\expandafter \@firstoftwo
 \else \expandafter \@secondoftwo
 \fi
}%
\providecommand \@ifx [1]{%
 \ifx #1\expandafter \@firstoftwo
 \else \expandafter \@secondoftwo
 \fi
}%
\providecommand \natexlab [1]{#1}%
\providecommand \enquote  [1]{``#1''}%
\providecommand \bibnamefont  [1]{#1}%
\providecommand \bibfnamefont [1]{#1}%
\providecommand \citenamefont [1]{#1}%
\providecommand \href@noop [0]{\@secondoftwo}%
\providecommand \href [0]{\begingroup \@sanitize@url \@href}%
\providecommand \@href[1]{\@@startlink{#1}\@@href}%
\providecommand \@@href[1]{\endgroup#1\@@endlink}%
\providecommand \@sanitize@url [0]{\catcode `\\12\catcode `\$12\catcode `\&12\catcode `\#12\catcode `\^12\catcode `\_12\catcode `\%12\relax}%
\providecommand \@@startlink[1]{}%
\providecommand \@@endlink[0]{}%
\providecommand \url  [0]{\begingroup\@sanitize@url \@url }%
\providecommand \@url [1]{\endgroup\@href {#1}{\urlprefix }}%
\providecommand \urlprefix  [0]{URL }%
\providecommand \Eprint [0]{\href }%
\providecommand \doibase [0]{http://dx.doi.org/}%
\providecommand \selectlanguage [0]{\@gobble}%
\providecommand \bibinfo  [0]{\@secondoftwo}%
\providecommand \bibfield  [0]{\@secondoftwo}%
\providecommand \translation [1]{[#1]}%
\providecommand \BibitemOpen [0]{}%
\providecommand \bibitemStop [0]{}%
\providecommand \bibitemNoStop [0]{.\EOS\space}%
\providecommand \EOS [0]{\spacefactor3000\relax}%
\providecommand \BibitemShut  [1]{\csname bibitem#1\endcsname}%
\let\auto@bib@innerbib\@empty
\bibitem [{\citenamefont {Gonz\'alez}\ \emph {et~al.}(2023)\citenamefont {Gonz\'alez}, \citenamefont {Jusufi}, \citenamefont {Leon},\ and\ \citenamefont {Saridakis}}]{Gonzalez:2023rsd}%
  \BibitemOpen
  \bibfield  {author} {\bibinfo {author} {\bibfnamefont {E.}~\bibnamefont {Gonz\'alez}}, \bibinfo {author} {\bibfnamefont {K.}~\bibnamefont {Jusufi}}, \bibinfo {author} {\bibfnamefont {G.}~\bibnamefont {Leon}}, \ and\ \bibinfo {author} {\bibfnamefont {E.~N.}\ \bibnamefont {Saridakis}},\ }\href {\doibase 10.1016/j.dark.2023.101304} {\bibfield  {journal} {\bibinfo  {journal} {Phys. Dark Univ.}\ }\textbf {\bibinfo {volume} {42}},\ \bibinfo {pages} {101304} (\bibinfo {year} {2023})},\ \Eprint {http://arxiv.org/abs/2305.14305} {arXiv:2305.14305 [astro-ph.CO]} \BibitemShut {NoStop}%
\bibitem [{\citenamefont {Bond}\ and\ \citenamefont {Efstathiou}(1984)}]{Bond:1984fp}%
  \BibitemOpen
  \bibfield  {author} {\bibinfo {author} {\bibfnamefont {J.~R.}\ \bibnamefont {Bond}}\ and\ \bibinfo {author} {\bibfnamefont {G.}~\bibnamefont {Efstathiou}},\ }\href {\doibase 10.1086/184362} {\bibfield  {journal} {\bibinfo  {journal} {Astrophys. J. Lett.}\ }\textbf {\bibinfo {volume} {285}},\ \bibinfo {pages} {L45} (\bibinfo {year} {1984})}\BibitemShut {NoStop}%
\bibitem [{\citenamefont {Trimble}(1987)}]{Trimble:1987ee}%
  \BibitemOpen
  \bibfield  {author} {\bibinfo {author} {\bibfnamefont {V.}~\bibnamefont {Trimble}},\ }\href {\doibase 10.1146/annurev.aa.25.090187.002233} {\bibfield  {journal} {\bibinfo  {journal} {Ann. Rev. Astron. Astrophys.}\ }\textbf {\bibinfo {volume} {25}},\ \bibinfo {pages} {425} (\bibinfo {year} {1987})}\BibitemShut {NoStop}%
\bibitem [{\citenamefont {Carroll}\ \emph {et~al.}(1992)\citenamefont {Carroll}, \citenamefont {Press},\ and\ \citenamefont {Turner}}]{Carroll:1991mt}%
  \BibitemOpen
  \bibfield  {author} {\bibinfo {author} {\bibfnamefont {S.~M.}\ \bibnamefont {Carroll}}, \bibinfo {author} {\bibfnamefont {W.~H.}\ \bibnamefont {Press}}, \ and\ \bibinfo {author} {\bibfnamefont {E.~L.}\ \bibnamefont {Turner}},\ }\href {\doibase 10.1146/annurev.aa.30.090192.002435} {\bibfield  {journal} {\bibinfo  {journal} {Ann. Rev. Astron. Astrophys.}\ }\textbf {\bibinfo {volume} {30}},\ \bibinfo {pages} {499} (\bibinfo {year} {1992})}\BibitemShut {NoStop}%
\bibitem [{\citenamefont {Perlmutter}\ \emph {et~al.}(1998)\citenamefont {Perlmutter}, \citenamefont {Aldering}, \citenamefont {Valle}, \citenamefont {Deustua}, \citenamefont {Ellis}, \citenamefont {Fabbro}, \citenamefont {Fruchter}, \citenamefont {Goldhaber}, \citenamefont {Groom}, \citenamefont {Hook} \emph {et~al.}}]{SupernovaCosmologyProject:1997zqe}%
  \BibitemOpen
  \bibfield  {author} {\bibinfo {author} {\bibfnamefont {S.}~\bibnamefont {Perlmutter}}, \bibinfo {author} {\bibfnamefont {G.}~\bibnamefont {Aldering}}, \bibinfo {author} {\bibfnamefont {M.~D.}\ \bibnamefont {Valle}}, \bibinfo {author} {\bibfnamefont {S.}~\bibnamefont {Deustua}}, \bibinfo {author} {\bibfnamefont {R.}~\bibnamefont {Ellis}}, \bibinfo {author} {\bibfnamefont {S.}~\bibnamefont {Fabbro}}, \bibinfo {author} {\bibfnamefont {A.}~\bibnamefont {Fruchter}}, \bibinfo {author} {\bibfnamefont {G.}~\bibnamefont {Goldhaber}}, \bibinfo {author} {\bibfnamefont {D.}~\bibnamefont {Groom}}, \bibinfo {author} {\bibfnamefont {I.}~\bibnamefont {Hook}},  \emph {et~al.} (\bibinfo {collaboration} {Supernova Cosmology Project}),\ }\href {\doibase 10.1038/34124} {\bibfield  {journal} {\bibinfo  {journal} {Nature}\ }\textbf {\bibinfo {volume} {391}},\ \bibinfo {pages} {51} (\bibinfo {year} {1998})},\ \Eprint {http://arxiv.org/abs/astro-ph/9712212} {arXiv:astro-ph/9712212} \BibitemShut {NoStop}%
\bibitem [{\citenamefont {Riess}\ \emph {et~al.}(1998)\citenamefont {Riess}, \citenamefont {Filippenko},\ and\ \citenamefont {Challis}}]{SupernovaSearchTeam:1998fmf}%
  \BibitemOpen
  \bibfield  {author} {\bibinfo {author} {\bibfnamefont {A.~G.}\ \bibnamefont {Riess}}, \bibinfo {author} {\bibfnamefont {A.~V.}\ \bibnamefont {Filippenko}}, \ and\ \bibinfo {author} {\bibfnamefont {P.}~\bibnamefont {Challis}} (\bibinfo {collaboration} {Supernova Search Team}),\ }\href {\doibase 10.1086/300499} {\bibfield  {journal} {\bibinfo  {journal} {Astron. J.}\ }\textbf {\bibinfo {volume} {116}},\ \bibinfo {pages} {1009} (\bibinfo {year} {1998})},\ \Eprint {http://arxiv.org/abs/astro-ph/9805201} {arXiv:astro-ph/9805201} \BibitemShut {NoStop}%
\bibitem [{\citenamefont {Perlmutter}\ \emph {et~al.}(1999)\citenamefont {Perlmutter}, \citenamefont {Aldering}, \citenamefont {Goldhaber}, \citenamefont {Knop}, \citenamefont {Nugent}, \citenamefont {Castro}, \citenamefont {Deustua}, \citenamefont {Fabbro}, \citenamefont {Goobar}, \citenamefont {Groom} \emph {et~al.}}]{SupernovaCosmologyProject:1998vns}%
  \BibitemOpen
  \bibfield  {author} {\bibinfo {author} {\bibfnamefont {S.}~\bibnamefont {Perlmutter}}, \bibinfo {author} {\bibfnamefont {G.}~\bibnamefont {Aldering}}, \bibinfo {author} {\bibfnamefont {G.}~\bibnamefont {Goldhaber}}, \bibinfo {author} {\bibfnamefont {R.}~\bibnamefont {Knop}}, \bibinfo {author} {\bibfnamefont {P.}~\bibnamefont {Nugent}}, \bibinfo {author} {\bibfnamefont {P.~G.}\ \bibnamefont {Castro}}, \bibinfo {author} {\bibfnamefont {S.}~\bibnamefont {Deustua}}, \bibinfo {author} {\bibfnamefont {S.}~\bibnamefont {Fabbro}}, \bibinfo {author} {\bibfnamefont {A.}~\bibnamefont {Goobar}}, \bibinfo {author} {\bibfnamefont {D.~E.}\ \bibnamefont {Groom}},  \emph {et~al.} (\bibinfo {collaboration} {Supernova Cosmology Project}),\ }\href {\doibase 10.1086/307221} {\bibfield  {journal} {\bibinfo  {journal} {Astrophys. J.}\ }\textbf {\bibinfo {volume} {517}},\ \bibinfo {pages} {565} (\bibinfo {year} {1999})},\ \Eprint {http://arxiv.org/abs/astro-ph/9812133} {arXiv:astro-ph/9812133} \BibitemShut {NoStop}%
\bibitem [{\citenamefont {Aghanim}\ \emph {et~al.}(2020)\citenamefont {Aghanim}, \citenamefont {Akrami}, \citenamefont {Ashdown}, \citenamefont {Aumont}, \citenamefont {Baccigalupi}, \citenamefont {Ballardini}, \citenamefont {Banday}, \citenamefont {Barreiro}, \citenamefont {Bartolo}, \citenamefont {Basak} \emph {et~al.}}]{Planck:2018vyg}%
  \BibitemOpen
  \bibfield  {author} {\bibinfo {author} {\bibfnamefont {N.}~\bibnamefont {Aghanim}}, \bibinfo {author} {\bibfnamefont {Y.}~\bibnamefont {Akrami}}, \bibinfo {author} {\bibfnamefont {M.}~\bibnamefont {Ashdown}}, \bibinfo {author} {\bibfnamefont {J.}~\bibnamefont {Aumont}}, \bibinfo {author} {\bibfnamefont {C.}~\bibnamefont {Baccigalupi}}, \bibinfo {author} {\bibfnamefont {M.}~\bibnamefont {Ballardini}}, \bibinfo {author} {\bibfnamefont {A.}~\bibnamefont {Banday}}, \bibinfo {author} {\bibfnamefont {R.}~\bibnamefont {Barreiro}}, \bibinfo {author} {\bibfnamefont {N.}~\bibnamefont {Bartolo}}, \bibinfo {author} {\bibfnamefont {S.}~\bibnamefont {Basak}},  \emph {et~al.} (\bibinfo {collaboration} {Planck}),\ }\href {\doibase 10.1051/0004-6361/201833910} {\bibfield  {journal} {\bibinfo  {journal} {Astron. Astrophys.}\ }\textbf {\bibinfo {volume} {641}},\ \bibinfo {pages} {A6} (\bibinfo {year} {2020})},\ \bibinfo {note} {[Erratum: Astron.Astrophys. 652, C4 (2021)]},\ \Eprint {http://arxiv.org/abs/1807.06209}
  {arXiv:1807.06209 [astro-ph.CO]} \BibitemShut {NoStop}%
\bibitem [{\citenamefont {Guth}(1981)}]{Guth:1980zm}%
  \BibitemOpen
  \bibfield  {author} {\bibinfo {author} {\bibfnamefont {A.~H.}\ \bibnamefont {Guth}},\ }\href {\doibase 10.1103/PhysRevD.23.347} {\bibfield  {journal} {\bibinfo  {journal} {Phys. Rev. D}\ }\textbf {\bibinfo {volume} {23}},\ \bibinfo {pages} {347} (\bibinfo {year} {1981})}\BibitemShut {NoStop}%
\bibitem [{\citenamefont {Ratra}\ and\ \citenamefont {Peebles}(1988)}]{Ratra:1987rm}%
  \BibitemOpen
  \bibfield  {author} {\bibinfo {author} {\bibfnamefont {B.}~\bibnamefont {Ratra}}\ and\ \bibinfo {author} {\bibfnamefont {P.~J.~E.}\ \bibnamefont {Peebles}},\ }\href {\doibase 10.1103/PhysRevD.37.3406} {\bibfield  {journal} {\bibinfo  {journal} {Phys. Rev. D}\ }\textbf {\bibinfo {volume} {37}},\ \bibinfo {pages} {3406} (\bibinfo {year} {1988})}\BibitemShut {NoStop}%
\bibitem [{\citenamefont {Parsons}\ and\ \citenamefont {Barrow}(1995)}]{Parsons:1995kt}%
  \BibitemOpen
  \bibfield  {author} {\bibinfo {author} {\bibfnamefont {P.}~\bibnamefont {Parsons}}\ and\ \bibinfo {author} {\bibfnamefont {J.~D.}\ \bibnamefont {Barrow}},\ }\href {\doibase 10.1088/0264-9381/12/7/013} {\bibfield  {journal} {\bibinfo  {journal} {Class. Quant. Grav.}\ }\textbf {\bibinfo {volume} {12}},\ \bibinfo {pages} {1715} (\bibinfo {year} {1995})}\BibitemShut {NoStop}%
\bibitem [{\citenamefont {Rubano}\ and\ \citenamefont {Barrow}(2001)}]{Rubano:2001xi}%
  \BibitemOpen
  \bibfield  {author} {\bibinfo {author} {\bibfnamefont {C.}~\bibnamefont {Rubano}}\ and\ \bibinfo {author} {\bibfnamefont {J.~D.}\ \bibnamefont {Barrow}},\ }\href {\doibase 10.1103/PhysRevD.64.127301} {\bibfield  {journal} {\bibinfo  {journal} {Phys. Rev. D}\ }\textbf {\bibinfo {volume} {64}},\ \bibinfo {pages} {127301} (\bibinfo {year} {2001})},\ \Eprint {http://arxiv.org/abs/gr-qc/0105037} {arXiv:gr-qc/0105037} \BibitemShut {NoStop}%
\bibitem [{\citenamefont {Saridakis}(2009)}]{Saridakis:2008fy}%
  \BibitemOpen
  \bibfield  {author} {\bibinfo {author} {\bibfnamefont {E.~N.}\ \bibnamefont {Saridakis}},\ }\href {\doibase 10.1016/j.physletb.2009.04.065} {\bibfield  {journal} {\bibinfo  {journal} {Phys. Lett. B}\ }\textbf {\bibinfo {volume} {676}},\ \bibinfo {pages} {7} (\bibinfo {year} {2009})},\ \Eprint {http://arxiv.org/abs/0811.1333} {arXiv:0811.1333 [hep-th]} \BibitemShut {NoStop}%
\bibitem [{\citenamefont {Cai}\ \emph {et~al.}(2010)\citenamefont {Cai}, \citenamefont {Saridakis}, \citenamefont {Setare},\ and\ \citenamefont {Xia}}]{Cai:2009zp}%
  \BibitemOpen
  \bibfield  {author} {\bibinfo {author} {\bibfnamefont {Y.-F.}\ \bibnamefont {Cai}}, \bibinfo {author} {\bibfnamefont {E.~N.}\ \bibnamefont {Saridakis}}, \bibinfo {author} {\bibfnamefont {M.~R.}\ \bibnamefont {Setare}}, \ and\ \bibinfo {author} {\bibfnamefont {J.-Q.}\ \bibnamefont {Xia}},\ }\href {\doibase 10.1016/j.physrep.2010.04.001} {\bibfield  {journal} {\bibinfo  {journal} {Phys. Rept.}\ }\textbf {\bibinfo {volume} {493}},\ \bibinfo {pages} {1} (\bibinfo {year} {2010})},\ \Eprint {http://arxiv.org/abs/0909.2776} {arXiv:0909.2776 [hep-th]} \BibitemShut {NoStop}%
\bibitem [{\citenamefont {Wali~Hossain}\ \emph {et~al.}(2015)\citenamefont {Wali~Hossain}, \citenamefont {Myrzakulov}, \citenamefont {Sami},\ and\ \citenamefont {Saridakis}}]{WaliHossain:2014usl}%
  \BibitemOpen
  \bibfield  {author} {\bibinfo {author} {\bibfnamefont {M.}~\bibnamefont {Wali~Hossain}}, \bibinfo {author} {\bibfnamefont {R.}~\bibnamefont {Myrzakulov}}, \bibinfo {author} {\bibfnamefont {M.}~\bibnamefont {Sami}}, \ and\ \bibinfo {author} {\bibfnamefont {E.~N.}\ \bibnamefont {Saridakis}},\ }\href {\doibase 10.1142/S0218271815300141} {\bibfield  {journal} {\bibinfo  {journal} {Int. J. Mod. Phys. D}\ }\textbf {\bibinfo {volume} {24}},\ \bibinfo {pages} {1530014} (\bibinfo {year} {2015})},\ \Eprint {http://arxiv.org/abs/1410.6100} {arXiv:1410.6100 [gr-qc]} \BibitemShut {NoStop}%
\bibitem [{\citenamefont {Barrow}\ and\ \citenamefont {Paliathanasis}(2016)}]{Barrow:2016qkh}%
  \BibitemOpen
  \bibfield  {author} {\bibinfo {author} {\bibfnamefont {J.~D.}\ \bibnamefont {Barrow}}\ and\ \bibinfo {author} {\bibfnamefont {A.}~\bibnamefont {Paliathanasis}},\ }\href {\doibase 10.1103/PhysRevD.94.083518} {\bibfield  {journal} {\bibinfo  {journal} {Phys. Rev. D}\ }\textbf {\bibinfo {volume} {94}},\ \bibinfo {pages} {083518} (\bibinfo {year} {2016})},\ \Eprint {http://arxiv.org/abs/1609.01126} {arXiv:1609.01126 [gr-qc]} \BibitemShut {NoStop}%
\bibitem [{\citenamefont {Elizalde}\ \emph {et~al.}(2004)\citenamefont {Elizalde}, \citenamefont {Nojiri},\ and\ \citenamefont {Odintsov}}]{Elizalde:2004mq}%
  \BibitemOpen
  \bibfield  {author} {\bibinfo {author} {\bibfnamefont {E.}~\bibnamefont {Elizalde}}, \bibinfo {author} {\bibfnamefont {S.}~\bibnamefont {Nojiri}}, \ and\ \bibinfo {author} {\bibfnamefont {S.~D.}\ \bibnamefont {Odintsov}},\ }\href {\doibase 10.1103/PhysRevD.70.043539} {\bibfield  {journal} {\bibinfo  {journal} {Phys. Rev. D}\ }\textbf {\bibinfo {volume} {70}},\ \bibinfo {pages} {043539} (\bibinfo {year} {2004})},\ \Eprint {http://arxiv.org/abs/hep-th/0405034} {arXiv:hep-th/0405034} \BibitemShut {NoStop}%
\bibitem [{\citenamefont {Elizalde}\ \emph {et~al.}(2008)\citenamefont {Elizalde}, \citenamefont {Nojiri}, \citenamefont {Odintsov}, \citenamefont {Saez-Gomez},\ and\ \citenamefont {Faraoni}}]{Elizalde:2008yf}%
  \BibitemOpen
  \bibfield  {author} {\bibinfo {author} {\bibfnamefont {E.}~\bibnamefont {Elizalde}}, \bibinfo {author} {\bibfnamefont {S.}~\bibnamefont {Nojiri}}, \bibinfo {author} {\bibfnamefont {S.~D.}\ \bibnamefont {Odintsov}}, \bibinfo {author} {\bibfnamefont {D.}~\bibnamefont {Saez-Gomez}}, \ and\ \bibinfo {author} {\bibfnamefont {V.}~\bibnamefont {Faraoni}},\ }\href {\doibase 10.1103/PhysRevD.77.106005} {\bibfield  {journal} {\bibinfo  {journal} {Phys. Rev. D}\ }\textbf {\bibinfo {volume} {77}},\ \bibinfo {pages} {106005} (\bibinfo {year} {2008})},\ \Eprint {http://arxiv.org/abs/0803.1311} {arXiv:0803.1311 [hep-th]} \BibitemShut {NoStop}%
\bibitem [{\citenamefont {Skugoreva}\ \emph {et~al.}(2015)\citenamefont {Skugoreva}, \citenamefont {Saridakis},\ and\ \citenamefont {Toporensky}}]{Skugoreva:2014ena}%
  \BibitemOpen
  \bibfield  {author} {\bibinfo {author} {\bibfnamefont {M.~A.}\ \bibnamefont {Skugoreva}}, \bibinfo {author} {\bibfnamefont {E.~N.}\ \bibnamefont {Saridakis}}, \ and\ \bibinfo {author} {\bibfnamefont {A.~V.}\ \bibnamefont {Toporensky}},\ }\href {\doibase 10.1103/PhysRevD.91.044023} {\bibfield  {journal} {\bibinfo  {journal} {Phys. Rev. D}\ }\textbf {\bibinfo {volume} {91}},\ \bibinfo {pages} {044023} (\bibinfo {year} {2015})},\ \Eprint {http://arxiv.org/abs/1412.1502} {arXiv:1412.1502 [gr-qc]} \BibitemShut {NoStop}%
\bibitem [{\citenamefont {Saridakis}\ and\ \citenamefont {Tsoukalas}(2016)}]{Saridakis:2016mjd}%
  \BibitemOpen
  \bibfield  {author} {\bibinfo {author} {\bibfnamefont {E.~N.}\ \bibnamefont {Saridakis}}\ and\ \bibinfo {author} {\bibfnamefont {M.}~\bibnamefont {Tsoukalas}},\ }\href {\doibase 10.1088/1475-7516/2016/04/017} {\bibfield  {journal} {\bibinfo  {journal} {JCAP}\ }\textbf {\bibinfo {volume} {04}},\ \bibinfo {pages} {017} (\bibinfo {year} {2016})},\ \Eprint {http://arxiv.org/abs/1602.06890} {arXiv:1602.06890 [gr-qc]} \BibitemShut {NoStop}%
\bibitem [{\citenamefont {Paliathanasis}(2019)}]{Paliathanasis:2019luv}%
  \BibitemOpen
  \bibfield  {author} {\bibinfo {author} {\bibfnamefont {A.}~\bibnamefont {Paliathanasis}},\ }\href {\doibase 10.1007/s10714-019-2585-3} {\bibfield  {journal} {\bibinfo  {journal} {Gen. Rel. Grav.}\ }\textbf {\bibinfo {volume} {51}},\ \bibinfo {pages} {101} (\bibinfo {year} {2019})},\ \Eprint {http://arxiv.org/abs/1907.12261} {arXiv:1907.12261 [gr-qc]} \BibitemShut {NoStop}%
\bibitem [{\citenamefont {Banerjee}\ \emph {et~al.}(2023)\citenamefont {Banerjee}, \citenamefont {Petronikolou},\ and\ \citenamefont {Saridakis}}]{Banerjee:2022ynv}%
  \BibitemOpen
  \bibfield  {author} {\bibinfo {author} {\bibfnamefont {S.}~\bibnamefont {Banerjee}}, \bibinfo {author} {\bibfnamefont {M.}~\bibnamefont {Petronikolou}}, \ and\ \bibinfo {author} {\bibfnamefont {E.~N.}\ \bibnamefont {Saridakis}},\ }\href {\doibase 10.1103/PhysRevD.108.024012} {\bibfield  {journal} {\bibinfo  {journal} {Phys. Rev. D}\ }\textbf {\bibinfo {volume} {108}},\ \bibinfo {pages} {024012} (\bibinfo {year} {2023})},\ \Eprint {http://arxiv.org/abs/2209.02426} {arXiv:2209.02426 [gr-qc]} \BibitemShut {NoStop}%
\bibitem [{\citenamefont {Santos}\ \emph {et~al.}(2023)\citenamefont {Santos}, \citenamefont {Pourhassan},\ and\ \citenamefont {Saridakis}}]{Santos:2023eqp}%
  \BibitemOpen
  \bibfield  {author} {\bibinfo {author} {\bibfnamefont {F.~F.}\ \bibnamefont {Santos}}, \bibinfo {author} {\bibfnamefont {B.}~\bibnamefont {Pourhassan}}, \ and\ \bibinfo {author} {\bibfnamefont {E.}~\bibnamefont {Saridakis}},\ }\href@noop {} {\enquote {\bibinfo {title} {{de Sitter versus anti-de Sitter in Horndeski-like gravity}},}\ } (\bibinfo {year} {2023}),\ \Eprint {http://arxiv.org/abs/2305.05794} {arXiv:2305.05794 [hep-th]} \BibitemShut {NoStop}%
\bibitem [{\citenamefont {Leon}\ \emph {et~al.}(2022)\citenamefont {Leon}, \citenamefont {Paliathanasis}, \citenamefont {Saridakis},\ and\ \citenamefont {Basilakos}}]{Leon:2022oyy}%
  \BibitemOpen
  \bibfield  {author} {\bibinfo {author} {\bibfnamefont {G.}~\bibnamefont {Leon}}, \bibinfo {author} {\bibfnamefont {A.}~\bibnamefont {Paliathanasis}}, \bibinfo {author} {\bibfnamefont {E.~N.}\ \bibnamefont {Saridakis}}, \ and\ \bibinfo {author} {\bibfnamefont {S.}~\bibnamefont {Basilakos}},\ }\href {\doibase 10.1103/PhysRevD.106.024055} {\bibfield  {journal} {\bibinfo  {journal} {Phys. Rev. D}\ }\textbf {\bibinfo {volume} {106}},\ \bibinfo {pages} {024055} (\bibinfo {year} {2022})},\ \Eprint {http://arxiv.org/abs/2203.14866} {arXiv:2203.14866 [gr-qc]} \BibitemShut {NoStop}%
\bibitem [{\citenamefont {Akrami}\ \emph {et~al.}(2021)\citenamefont {Akrami} \emph {et~al.}}]{CANTATA:2021ktz}%
  \BibitemOpen
  \bibfield  {author} {\bibinfo {author} {\bibfnamefont {Y.}~\bibnamefont {Akrami}} \emph {et~al.} (\bibinfo {collaboration} {CANTATA}),\ }\href {\doibase 10.1007/978-3-030-83715-0} {\emph {\bibinfo {title} {{Modified Gravity and Cosmology}: {An Update by the CANTATA Network}}}},\ edited by\ \bibinfo {editor} {\bibfnamefont {E.~N.}\ \bibnamefont {Saridakis}}, \bibinfo {editor} {\bibfnamefont {R.}~\bibnamefont {Lazkoz}}, \bibinfo {editor} {\bibfnamefont {V.}~\bibnamefont {Salzano}}, \bibinfo {editor} {\bibfnamefont {P.}~\bibnamefont {Vargas~Moniz}}, \bibinfo {editor} {\bibfnamefont {S.}~\bibnamefont {Capozziello}}, \bibinfo {editor} {\bibfnamefont {J.}~\bibnamefont {Beltr\'an~Jim\'enez}}, \bibinfo {editor} {\bibfnamefont {M.}~\bibnamefont {De~Laurentis}}, \ and\ \bibinfo {editor} {\bibfnamefont {G.~J.}\ \bibnamefont {Olmo}}\ (\bibinfo  {publisher} {Springer},\ \bibinfo {year} {2021})\ \Eprint {http://arxiv.org/abs/2105.12582} {arXiv:2105.12582 [gr-qc]} \BibitemShut {NoStop}%
\bibitem [{\citenamefont {Leon}\ and\ \citenamefont {Saridakis}(2009)}]{Leon:2009rc}%
  \BibitemOpen
  \bibfield  {author} {\bibinfo {author} {\bibfnamefont {G.}~\bibnamefont {Leon}}\ and\ \bibinfo {author} {\bibfnamefont {E.~N.}\ \bibnamefont {Saridakis}},\ }\href {\doibase 10.1088/1475-7516/2009/11/006} {\bibfield  {journal} {\bibinfo  {journal} {JCAP}\ }\textbf {\bibinfo {volume} {11}},\ \bibinfo {pages} {006} (\bibinfo {year} {2009})},\ \Eprint {http://arxiv.org/abs/0909.3571} {arXiv:0909.3571 [hep-th]} \BibitemShut {NoStop}%
\bibitem [{\citenamefont {De~Felice}\ and\ \citenamefont {Tsujikawa}(2010)}]{DeFelice:2010aj}%
  \BibitemOpen
  \bibfield  {author} {\bibinfo {author} {\bibfnamefont {A.}~\bibnamefont {De~Felice}}\ and\ \bibinfo {author} {\bibfnamefont {S.}~\bibnamefont {Tsujikawa}},\ }\href {\doibase 10.12942/lrr-2010-3} {\bibfield  {journal} {\bibinfo  {journal} {Living Rev. Rel.}\ }\textbf {\bibinfo {volume} {13}},\ \bibinfo {pages} {3} (\bibinfo {year} {2010})},\ \Eprint {http://arxiv.org/abs/1002.4928} {arXiv:1002.4928 [gr-qc]} \BibitemShut {NoStop}%
\bibitem [{\citenamefont {Clifton}\ \emph {et~al.}(2012)\citenamefont {Clifton}, \citenamefont {Ferreira}, \citenamefont {Padilla},\ and\ \citenamefont {Skordis}}]{Clifton:2011jh}%
  \BibitemOpen
  \bibfield  {author} {\bibinfo {author} {\bibfnamefont {T.}~\bibnamefont {Clifton}}, \bibinfo {author} {\bibfnamefont {P.~G.}\ \bibnamefont {Ferreira}}, \bibinfo {author} {\bibfnamefont {A.}~\bibnamefont {Padilla}}, \ and\ \bibinfo {author} {\bibfnamefont {C.}~\bibnamefont {Skordis}},\ }\href {\doibase 10.1016/j.physrep.2012.01.001} {\bibfield  {journal} {\bibinfo  {journal} {Phys. Rept.}\ }\textbf {\bibinfo {volume} {513}},\ \bibinfo {pages} {1} (\bibinfo {year} {2012})},\ \Eprint {http://arxiv.org/abs/1106.2476} {arXiv:1106.2476 [astro-ph.CO]} \BibitemShut {NoStop}%
\bibitem [{\citenamefont {Capozziello}\ and\ \citenamefont {De~Laurentis}(2011)}]{Capozziello:2011et}%
  \BibitemOpen
  \bibfield  {author} {\bibinfo {author} {\bibfnamefont {S.}~\bibnamefont {Capozziello}}\ and\ \bibinfo {author} {\bibfnamefont {M.}~\bibnamefont {De~Laurentis}},\ }\href {\doibase 10.1016/j.physrep.2011.09.003} {\bibfield  {journal} {\bibinfo  {journal} {Phys. Rept.}\ }\textbf {\bibinfo {volume} {509}},\ \bibinfo {pages} {167} (\bibinfo {year} {2011})},\ \Eprint {http://arxiv.org/abs/1108.6266} {arXiv:1108.6266 [gr-qc]} \BibitemShut {NoStop}%
\bibitem [{\citenamefont {De~Felice}\ and\ \citenamefont {Tsujikawa}(2012)}]{DeFelice:2011bh}%
  \BibitemOpen
  \bibfield  {author} {\bibinfo {author} {\bibfnamefont {A.}~\bibnamefont {De~Felice}}\ and\ \bibinfo {author} {\bibfnamefont {S.}~\bibnamefont {Tsujikawa}},\ }\href {\doibase 10.1088/1475-7516/2012/02/007} {\bibfield  {journal} {\bibinfo  {journal} {JCAP}\ }\textbf {\bibinfo {volume} {02}},\ \bibinfo {pages} {007} (\bibinfo {year} {2012})},\ \Eprint {http://arxiv.org/abs/1110.3878} {arXiv:1110.3878 [gr-qc]} \BibitemShut {NoStop}%
\bibitem [{\citenamefont {Xu}\ \emph {et~al.}(2012)\citenamefont {Xu}, \citenamefont {Saridakis},\ and\ \citenamefont {Leon}}]{Xu:2012jf}%
  \BibitemOpen
  \bibfield  {author} {\bibinfo {author} {\bibfnamefont {C.}~\bibnamefont {Xu}}, \bibinfo {author} {\bibfnamefont {E.~N.}\ \bibnamefont {Saridakis}}, \ and\ \bibinfo {author} {\bibfnamefont {G.}~\bibnamefont {Leon}},\ }\href {\doibase 10.1088/1475-7516/2012/07/005} {\bibfield  {journal} {\bibinfo  {journal} {JCAP}\ }\textbf {\bibinfo {volume} {07}},\ \bibinfo {pages} {005} (\bibinfo {year} {2012})},\ \Eprint {http://arxiv.org/abs/1202.3781} {arXiv:1202.3781 [gr-qc]} \BibitemShut {NoStop}%
\bibitem [{\citenamefont {Bamba}\ \emph {et~al.}(2012)\citenamefont {Bamba}, \citenamefont {Capozziello}, \citenamefont {Nojiri},\ and\ \citenamefont {Odintsov}}]{Bamba:2012cp}%
  \BibitemOpen
  \bibfield  {author} {\bibinfo {author} {\bibfnamefont {K.}~\bibnamefont {Bamba}}, \bibinfo {author} {\bibfnamefont {S.}~\bibnamefont {Capozziello}}, \bibinfo {author} {\bibfnamefont {S.}~\bibnamefont {Nojiri}}, \ and\ \bibinfo {author} {\bibfnamefont {S.~D.}\ \bibnamefont {Odintsov}},\ }\href {\doibase 10.1007/s10509-012-1181-8} {\bibfield  {journal} {\bibinfo  {journal} {Astrophys. Space Sci.}\ }\textbf {\bibinfo {volume} {342}},\ \bibinfo {pages} {155} (\bibinfo {year} {2012})},\ \Eprint {http://arxiv.org/abs/1205.3421} {arXiv:1205.3421 [gr-qc]} \BibitemShut {NoStop}%
\bibitem [{\citenamefont {Leon}\ and\ \citenamefont {Saridakis}(2013)}]{Leon:2012mt}%
  \BibitemOpen
  \bibfield  {author} {\bibinfo {author} {\bibfnamefont {G.}~\bibnamefont {Leon}}\ and\ \bibinfo {author} {\bibfnamefont {E.~N.}\ \bibnamefont {Saridakis}},\ }\href {\doibase 10.1088/1475-7516/2013/03/025} {\bibfield  {journal} {\bibinfo  {journal} {JCAP}\ }\textbf {\bibinfo {volume} {03}},\ \bibinfo {pages} {025} (\bibinfo {year} {2013})},\ \Eprint {http://arxiv.org/abs/1211.3088} {arXiv:1211.3088 [astro-ph.CO]} \BibitemShut {NoStop}%
\bibitem [{\citenamefont {Kofinas}\ \emph {et~al.}(2014)\citenamefont {Kofinas}, \citenamefont {Leon},\ and\ \citenamefont {Saridakis}}]{Kofinas:2014aka}%
  \BibitemOpen
  \bibfield  {author} {\bibinfo {author} {\bibfnamefont {G.}~\bibnamefont {Kofinas}}, \bibinfo {author} {\bibfnamefont {G.}~\bibnamefont {Leon}}, \ and\ \bibinfo {author} {\bibfnamefont {E.~N.}\ \bibnamefont {Saridakis}},\ }\href {\doibase 10.1088/0264-9381/31/17/175011} {\bibfield  {journal} {\bibinfo  {journal} {Class. Quant. Grav.}\ }\textbf {\bibinfo {volume} {31}},\ \bibinfo {pages} {175011} (\bibinfo {year} {2014})},\ \Eprint {http://arxiv.org/abs/1404.7100} {arXiv:1404.7100 [gr-qc]} \BibitemShut {NoStop}%
\bibitem [{\citenamefont {Bahamonde}\ \emph {et~al.}(2015)\citenamefont {Bahamonde}, \citenamefont {B\"ohmer},\ and\ \citenamefont {Wright}}]{Bahamonde:2015zma}%
  \BibitemOpen
  \bibfield  {author} {\bibinfo {author} {\bibfnamefont {S.}~\bibnamefont {Bahamonde}}, \bibinfo {author} {\bibfnamefont {C.~G.}\ \bibnamefont {B\"ohmer}}, \ and\ \bibinfo {author} {\bibfnamefont {M.}~\bibnamefont {Wright}},\ }\href {\doibase 10.1103/PhysRevD.92.104042} {\bibfield  {journal} {\bibinfo  {journal} {Phys. Rev. D}\ }\textbf {\bibinfo {volume} {92}},\ \bibinfo {pages} {104042} (\bibinfo {year} {2015})},\ \Eprint {http://arxiv.org/abs/1508.05120} {arXiv:1508.05120 [gr-qc]} \BibitemShut {NoStop}%
\bibitem [{\citenamefont {Momeni}\ and\ \citenamefont {Myrzakulov}(2015)}]{Momeni:2015uwx}%
  \BibitemOpen
  \bibfield  {author} {\bibinfo {author} {\bibfnamefont {D.}~\bibnamefont {Momeni}}\ and\ \bibinfo {author} {\bibfnamefont {R.}~\bibnamefont {Myrzakulov}},\ }\href {\doibase 10.1007/s10509-015-2546-6} {\bibfield  {journal} {\bibinfo  {journal} {Astrophys. Space Sci.}\ }\textbf {\bibinfo {volume} {360}},\ \bibinfo {pages} {28} (\bibinfo {year} {2015})},\ \Eprint {http://arxiv.org/abs/1511.01205} {arXiv:1511.01205 [physics.gen-ph]} \BibitemShut {NoStop}%
\bibitem [{\citenamefont {Cai}\ \emph {et~al.}(2016)\citenamefont {Cai}, \citenamefont {Capozziello}, \citenamefont {De~Laurentis},\ and\ \citenamefont {Saridakis}}]{Cai:2015emx}%
  \BibitemOpen
  \bibfield  {author} {\bibinfo {author} {\bibfnamefont {Y.-F.}\ \bibnamefont {Cai}}, \bibinfo {author} {\bibfnamefont {S.}~\bibnamefont {Capozziello}}, \bibinfo {author} {\bibfnamefont {M.}~\bibnamefont {De~Laurentis}}, \ and\ \bibinfo {author} {\bibfnamefont {E.~N.}\ \bibnamefont {Saridakis}},\ }\href {\doibase 10.1088/0034-4885/79/10/106901} {\bibfield  {journal} {\bibinfo  {journal} {Rept. Prog. Phys.}\ }\textbf {\bibinfo {volume} {79}},\ \bibinfo {pages} {106901} (\bibinfo {year} {2016})},\ \Eprint {http://arxiv.org/abs/1511.07586} {arXiv:1511.07586 [gr-qc]} \BibitemShut {NoStop}%
\bibitem [{\citenamefont {Krssak}\ \emph {et~al.}(2019)\citenamefont {Krssak}, \citenamefont {van~den Hoogen}, \citenamefont {Pereira}, \citenamefont {B\"ohmer},\ and\ \citenamefont {Coley}}]{Krssak:2018ywd}%
  \BibitemOpen
  \bibfield  {author} {\bibinfo {author} {\bibfnamefont {M.}~\bibnamefont {Krssak}}, \bibinfo {author} {\bibfnamefont {R.~J.}\ \bibnamefont {van~den Hoogen}}, \bibinfo {author} {\bibfnamefont {J.~G.}\ \bibnamefont {Pereira}}, \bibinfo {author} {\bibfnamefont {C.~G.}\ \bibnamefont {B\"ohmer}}, \ and\ \bibinfo {author} {\bibfnamefont {A.~A.}\ \bibnamefont {Coley}},\ }\href {\doibase 10.1088/1361-6382/ab2e1f} {\bibfield  {journal} {\bibinfo  {journal} {Class. Quant. Grav.}\ }\textbf {\bibinfo {volume} {36}},\ \bibinfo {pages} {183001} (\bibinfo {year} {2019})},\ \Eprint {http://arxiv.org/abs/1810.12932} {arXiv:1810.12932 [gr-qc]} \BibitemShut {NoStop}%
\bibitem [{\citenamefont {Dehghani}\ \emph {et~al.}(2023)\citenamefont {Dehghani}, \citenamefont {Pourhassan}, \citenamefont {Zarepour},\ and\ \citenamefont {Saridakis}}]{Dehghani:2023yph}%
  \BibitemOpen
  \bibfield  {author} {\bibinfo {author} {\bibfnamefont {A.}~\bibnamefont {Dehghani}}, \bibinfo {author} {\bibfnamefont {B.}~\bibnamefont {Pourhassan}}, \bibinfo {author} {\bibfnamefont {S.}~\bibnamefont {Zarepour}}, \ and\ \bibinfo {author} {\bibfnamefont {E.~N.}\ \bibnamefont {Saridakis}},\ }\href@noop {} {\enquote {\bibinfo {title} {{Thermodynamic schemes of charged BTZ-like black holes in arbitrary dimensions}},}\ } (\bibinfo {year} {2023}),\ \Eprint {http://arxiv.org/abs/2305.08219} {arXiv:2305.08219 [hep-th]} \BibitemShut {NoStop}%
\bibitem [{\citenamefont {Salucci}(2018)}]{Salucci:2018eie}%
  \BibitemOpen
  \bibfield  {author} {\bibinfo {author} {\bibfnamefont {P.}~\bibnamefont {Salucci}},\ }\href {\doibase 10.1007/s10701-018-0209-5} {\bibfield  {journal} {\bibinfo  {journal} {Found. Phys.}\ }\textbf {\bibinfo {volume} {48}},\ \bibinfo {pages} {1517} (\bibinfo {year} {2018})},\ \Eprint {http://arxiv.org/abs/1807.08541} {arXiv:1807.08541 [astro-ph.GA]} \BibitemShut {NoStop}%
\bibitem [{\citenamefont {Milgrom}(1983)}]{Milgrom:1983ca}%
  \BibitemOpen
  \bibfield  {author} {\bibinfo {author} {\bibfnamefont {M.}~\bibnamefont {Milgrom}},\ }\href {\doibase 10.1086/161130} {\bibfield  {journal} {\bibinfo  {journal} {Astrophys. J.}\ }\textbf {\bibinfo {volume} {270}},\ \bibinfo {pages} {365} (\bibinfo {year} {1983})}\BibitemShut {NoStop}%
\bibitem [{\citenamefont {Ferreira}\ and\ \citenamefont {Starkmann}(2009)}]{Ferreira:2009eg}%
  \BibitemOpen
  \bibfield  {author} {\bibinfo {author} {\bibfnamefont {P.~G.}\ \bibnamefont {Ferreira}}\ and\ \bibinfo {author} {\bibfnamefont {G.}~\bibnamefont {Starkmann}},\ }\href {\doibase 10.1126/science.1172245} {\bibfield  {journal} {\bibinfo  {journal} {Science}\ }\textbf {\bibinfo {volume} {326}},\ \bibinfo {pages} {812} (\bibinfo {year} {2009})},\ \Eprint {http://arxiv.org/abs/0911.1212} {arXiv:0911.1212 [astro-ph.CO]} \BibitemShut {NoStop}%
\bibitem [{\citenamefont {Milgrom}\ and\ \citenamefont {Sanders}(2003)}]{Milgrom:2003}%
  \BibitemOpen
  \bibfield  {author} {\bibinfo {author} {\bibfnamefont {M.}~\bibnamefont {Milgrom}}\ and\ \bibinfo {author} {\bibfnamefont {R.~H.}\ \bibnamefont {Sanders}},\ }\href {\doibase 10.1086/381138} {\bibfield  {journal} {\bibinfo  {journal} {The Astrophysical Journal}\ }\textbf {\bibinfo {volume} {599}},\ \bibinfo {pages} {L25} (\bibinfo {year} {2003})}\BibitemShut {NoStop}%
\bibitem [{\citenamefont {Tiret}\ \emph {et~al.}(2007)\citenamefont {Tiret}, \citenamefont {Combes}, \citenamefont {Angus}, \citenamefont {Famaey},\ and\ \citenamefont {Zhao}}]{Tiret:2007kq}%
  \BibitemOpen
  \bibfield  {author} {\bibinfo {author} {\bibfnamefont {O.}~\bibnamefont {Tiret}}, \bibinfo {author} {\bibfnamefont {F.}~\bibnamefont {Combes}}, \bibinfo {author} {\bibfnamefont {G.~W.}\ \bibnamefont {Angus}}, \bibinfo {author} {\bibfnamefont {B.}~\bibnamefont {Famaey}}, \ and\ \bibinfo {author} {\bibfnamefont {H.}~\bibnamefont {Zhao}},\ }\href {\doibase 10.1051/0004-6361:20078569} {\bibfield  {journal} {\bibinfo  {journal} {Astron. Astrophys.}\ }\textbf {\bibinfo {volume} {476}},\ \bibinfo {pages} {L1} (\bibinfo {year} {2007})},\ \Eprint {http://arxiv.org/abs/0710.4070} {arXiv:0710.4070 [astro-ph]} \BibitemShut {NoStop}%
\bibitem [{\citenamefont {Kroupa}\ \emph {et~al.}(2010)\citenamefont {Kroupa}, \citenamefont {Famaey}, \citenamefont {de~Boer}, \citenamefont {Dabringhausen}, \citenamefont {Pawlowski}, \citenamefont {Boily}, \citenamefont {Jerjen}, \citenamefont {Forbes}, \citenamefont {Hensler},\ and\ \citenamefont {Metz}}]{Kroupa:2010hf}%
  \BibitemOpen
  \bibfield  {author} {\bibinfo {author} {\bibfnamefont {P.}~\bibnamefont {Kroupa}}, \bibinfo {author} {\bibfnamefont {B.}~\bibnamefont {Famaey}}, \bibinfo {author} {\bibfnamefont {K.~S.}\ \bibnamefont {de~Boer}}, \bibinfo {author} {\bibfnamefont {J.}~\bibnamefont {Dabringhausen}}, \bibinfo {author} {\bibfnamefont {M.~S.}\ \bibnamefont {Pawlowski}}, \bibinfo {author} {\bibfnamefont {C.~M.}\ \bibnamefont {Boily}}, \bibinfo {author} {\bibfnamefont {H.}~\bibnamefont {Jerjen}}, \bibinfo {author} {\bibfnamefont {D.}~\bibnamefont {Forbes}}, \bibinfo {author} {\bibfnamefont {G.}~\bibnamefont {Hensler}}, \ and\ \bibinfo {author} {\bibfnamefont {M.}~\bibnamefont {Metz}},\ }\href {\doibase 10.1051/0004-6361/201014892} {\bibfield  {journal} {\bibinfo  {journal} {Astron. Astrophys.}\ }\textbf {\bibinfo {volume} {523}},\ \bibinfo {pages} {A32} (\bibinfo {year} {2010})},\ \Eprint {http://arxiv.org/abs/1006.1647} {arXiv:1006.1647 [astro-ph.CO]} \BibitemShut {NoStop}%
\bibitem [{\citenamefont {Cardone}\ \emph {et~al.}(2011)\citenamefont {Cardone}, \citenamefont {Angus}, \citenamefont {Diaferio}, \citenamefont {Tortora},\ and\ \citenamefont {Molinaro}}]{Cardone:2010ru}%
  \BibitemOpen
  \bibfield  {author} {\bibinfo {author} {\bibfnamefont {V.~F.}\ \bibnamefont {Cardone}}, \bibinfo {author} {\bibfnamefont {G.}~\bibnamefont {Angus}}, \bibinfo {author} {\bibfnamefont {A.}~\bibnamefont {Diaferio}}, \bibinfo {author} {\bibfnamefont {C.}~\bibnamefont {Tortora}}, \ and\ \bibinfo {author} {\bibfnamefont {R.}~\bibnamefont {Molinaro}},\ }\href {\doibase 10.1111/j.1365-2966.2010.18081.x} {\bibfield  {journal} {\bibinfo  {journal} {Mon. Not. Roy. Astron. Soc.}\ }\textbf {\bibinfo {volume} {412}},\ \bibinfo {pages} {2617} (\bibinfo {year} {2011})},\ \Eprint {http://arxiv.org/abs/1011.5741} {arXiv:1011.5741 [astro-ph.CO]} \BibitemShut {NoStop}%
\bibitem [{\citenamefont {Richtler}\ \emph {et~al.}(2011)\citenamefont {Richtler}, \citenamefont {Salinas}, \citenamefont {Misgeld}, \citenamefont {Hilker}, \citenamefont {Hau}, \citenamefont {Romanowsky}, \citenamefont {Schuberth},\ and\ \citenamefont {Spolaor}}]{Richtler:2011zk}%
  \BibitemOpen
  \bibfield  {author} {\bibinfo {author} {\bibfnamefont {T.}~\bibnamefont {Richtler}}, \bibinfo {author} {\bibfnamefont {R.}~\bibnamefont {Salinas}}, \bibinfo {author} {\bibfnamefont {I.}~\bibnamefont {Misgeld}}, \bibinfo {author} {\bibfnamefont {M.}~\bibnamefont {Hilker}}, \bibinfo {author} {\bibfnamefont {G.~K.~T.}\ \bibnamefont {Hau}}, \bibinfo {author} {\bibfnamefont {A.~J.}\ \bibnamefont {Romanowsky}}, \bibinfo {author} {\bibfnamefont {Y.}~\bibnamefont {Schuberth}}, \ and\ \bibinfo {author} {\bibfnamefont {M.}~\bibnamefont {Spolaor}},\ }\href {\doibase 10.1051/0004-6361/201015948} {\bibfield  {journal} {\bibinfo  {journal} {Astron. Astrophys.}\ }\textbf {\bibinfo {volume} {531}},\ \bibinfo {pages} {A119} (\bibinfo {year} {2011})},\ \Eprint {http://arxiv.org/abs/1103.2053} {arXiv:1103.2053 [astro-ph.CO]} \BibitemShut {NoStop}%
\bibitem [{\citenamefont {Berezhiani}\ and\ \citenamefont {Khoury}(2015)}]{Berezhiani:2015bqa}%
  \BibitemOpen
  \bibfield  {author} {\bibinfo {author} {\bibfnamefont {L.}~\bibnamefont {Berezhiani}}\ and\ \bibinfo {author} {\bibfnamefont {J.}~\bibnamefont {Khoury}},\ }\href {\doibase 10.1103/PhysRevD.92.103510} {\bibfield  {journal} {\bibinfo  {journal} {Phys. Rev. D}\ }\textbf {\bibinfo {volume} {92}},\ \bibinfo {pages} {103510} (\bibinfo {year} {2015})},\ \Eprint {http://arxiv.org/abs/1507.01019} {arXiv:1507.01019 [astro-ph.CO]} \BibitemShut {NoStop}%
\bibitem [{\citenamefont {Boehmer}\ and\ \citenamefont {Harko}(2007)}]{Boehmer:2007um}%
  \BibitemOpen
  \bibfield  {author} {\bibinfo {author} {\bibfnamefont {C.~G.}\ \bibnamefont {Boehmer}}\ and\ \bibinfo {author} {\bibfnamefont {T.}~\bibnamefont {Harko}},\ }\href {\doibase 10.1088/1475-7516/2007/06/025} {\bibfield  {journal} {\bibinfo  {journal} {JCAP}\ }\textbf {\bibinfo {volume} {06}},\ \bibinfo {pages} {025} (\bibinfo {year} {2007})},\ \Eprint {http://arxiv.org/abs/0705.4158} {arXiv:0705.4158 [astro-ph]} \BibitemShut {NoStop}%
\bibitem [{\citenamefont {Takahashi}(2004)}]{Takahashi:2004xh}%
  \BibitemOpen
  \bibfield  {author} {\bibinfo {author} {\bibfnamefont {R.}~\bibnamefont {Takahashi}},\ }\href {\doibase 10.1086/422403} {\bibfield  {journal} {\bibinfo  {journal} {J. Korean Phys. Soc.}\ }\textbf {\bibinfo {volume} {45}},\ \bibinfo {pages} {S1808} (\bibinfo {year} {2004})},\ \Eprint {http://arxiv.org/abs/astro-ph/0405099} {arXiv:astro-ph/0405099} \BibitemShut {NoStop}%
\bibitem [{\citenamefont {Hioki}\ and\ \citenamefont {Maeda}(2009)}]{Hioki:2009na}%
  \BibitemOpen
  \bibfield  {author} {\bibinfo {author} {\bibfnamefont {K.}~\bibnamefont {Hioki}}\ and\ \bibinfo {author} {\bibfnamefont {K.-i.}\ \bibnamefont {Maeda}},\ }\href {\doibase 10.1103/PhysRevD.80.024042} {\bibfield  {journal} {\bibinfo  {journal} {Phys. Rev. D}\ }\textbf {\bibinfo {volume} {80}},\ \bibinfo {pages} {024042} (\bibinfo {year} {2009})},\ \Eprint {http://arxiv.org/abs/0904.3575} {arXiv:0904.3575 [astro-ph.HE]} \BibitemShut {NoStop}%
\bibitem [{\citenamefont {Brito}\ \emph {et~al.}(2015)\citenamefont {Brito}, \citenamefont {Cardoso},\ and\ \citenamefont {Pani}}]{Brito:2015oca}%
  \BibitemOpen
  \bibfield  {author} {\bibinfo {author} {\bibfnamefont {R.}~\bibnamefont {Brito}}, \bibinfo {author} {\bibfnamefont {V.}~\bibnamefont {Cardoso}}, \ and\ \bibinfo {author} {\bibfnamefont {P.}~\bibnamefont {Pani}},\ }\href {\doibase 10.1007/978-3-319-19000-6} {\bibfield  {journal} {\bibinfo  {journal} {Lect. Notes Phys.}\ }\textbf {\bibinfo {volume} {906}},\ \bibinfo {pages} {pp.1} (\bibinfo {year} {2015})},\ \Eprint {http://arxiv.org/abs/1501.06570} {arXiv:1501.06570 [gr-qc]} \BibitemShut {NoStop}%
\bibitem [{\citenamefont {Cunha}\ \emph {et~al.}(2015)\citenamefont {Cunha}, \citenamefont {Herdeiro}, \citenamefont {Radu},\ and\ \citenamefont {Runarsson}}]{Cunha:2015yba}%
  \BibitemOpen
  \bibfield  {author} {\bibinfo {author} {\bibfnamefont {P.~V.~P.}\ \bibnamefont {Cunha}}, \bibinfo {author} {\bibfnamefont {C.~A.~R.}\ \bibnamefont {Herdeiro}}, \bibinfo {author} {\bibfnamefont {E.}~\bibnamefont {Radu}}, \ and\ \bibinfo {author} {\bibfnamefont {H.~F.}\ \bibnamefont {Runarsson}},\ }\href {\doibase 10.1103/PhysRevLett.115.211102} {\bibfield  {journal} {\bibinfo  {journal} {Phys. Rev. Lett.}\ }\textbf {\bibinfo {volume} {115}},\ \bibinfo {pages} {211102} (\bibinfo {year} {2015})},\ \Eprint {http://arxiv.org/abs/1509.00021} {arXiv:1509.00021 [gr-qc]} \BibitemShut {NoStop}%
\bibitem [{\citenamefont {Ohgami}\ and\ \citenamefont {Sakai}(2015)}]{Ohgami:2015nra}%
  \BibitemOpen
  \bibfield  {author} {\bibinfo {author} {\bibfnamefont {T.}~\bibnamefont {Ohgami}}\ and\ \bibinfo {author} {\bibfnamefont {N.}~\bibnamefont {Sakai}},\ }\href {\doibase 10.1103/PhysRevD.91.124020} {\bibfield  {journal} {\bibinfo  {journal} {Phys. Rev. D}\ }\textbf {\bibinfo {volume} {91}},\ \bibinfo {pages} {124020} (\bibinfo {year} {2015})},\ \Eprint {http://arxiv.org/abs/1704.07065} {arXiv:1704.07065 [gr-qc]} \BibitemShut {NoStop}%
\bibitem [{\citenamefont {Moffat}(2015)}]{Moffat:2015kva}%
  \BibitemOpen
  \bibfield  {author} {\bibinfo {author} {\bibfnamefont {J.~W.}\ \bibnamefont {Moffat}},\ }\href {\doibase 10.1140/epjc/s10052-015-3352-6} {\bibfield  {journal} {\bibinfo  {journal} {Eur. Phys. J. C}\ }\textbf {\bibinfo {volume} {75}},\ \bibinfo {pages} {130} (\bibinfo {year} {2015})},\ \Eprint {http://arxiv.org/abs/1502.01677} {arXiv:1502.01677 [gr-qc]} \BibitemShut {NoStop}%
\bibitem [{\citenamefont {Abdujabbarov}\ \emph {et~al.}(2016)\citenamefont {Abdujabbarov}, \citenamefont {Amir}, \citenamefont {Ahmedov},\ and\ \citenamefont {Ghosh}}]{Abdujabbarov:2016hnw}%
  \BibitemOpen
  \bibfield  {author} {\bibinfo {author} {\bibfnamefont {A.}~\bibnamefont {Abdujabbarov}}, \bibinfo {author} {\bibfnamefont {M.}~\bibnamefont {Amir}}, \bibinfo {author} {\bibfnamefont {B.}~\bibnamefont {Ahmedov}}, \ and\ \bibinfo {author} {\bibfnamefont {S.~G.}\ \bibnamefont {Ghosh}},\ }\href {\doibase 10.1103/PhysRevD.93.104004} {\bibfield  {journal} {\bibinfo  {journal} {Phys. Rev. D}\ }\textbf {\bibinfo {volume} {93}},\ \bibinfo {pages} {104004} (\bibinfo {year} {2016})},\ \Eprint {http://arxiv.org/abs/1604.03809} {arXiv:1604.03809 [gr-qc]} \BibitemShut {NoStop}%
\bibitem [{\citenamefont {Cunha}\ and\ \citenamefont {Herdeiro}(2018)}]{Cunha:2018acu}%
  \BibitemOpen
  \bibfield  {author} {\bibinfo {author} {\bibfnamefont {P.~V.~P.}\ \bibnamefont {Cunha}}\ and\ \bibinfo {author} {\bibfnamefont {C.~A.~R.}\ \bibnamefont {Herdeiro}},\ }\href {\doibase 10.1007/s10714-018-2361-9} {\bibfield  {journal} {\bibinfo  {journal} {Gen. Rel. Grav.}\ }\textbf {\bibinfo {volume} {50}},\ \bibinfo {pages} {42} (\bibinfo {year} {2018})},\ \Eprint {http://arxiv.org/abs/1801.00860} {arXiv:1801.00860 [gr-qc]} \BibitemShut {NoStop}%
\bibitem [{\citenamefont {Mizuno}\ \emph {et~al.}(2018)\citenamefont {Mizuno}, \citenamefont {Younsi}, \citenamefont {Fromm}, \citenamefont {Porth}, \citenamefont {De~Laurentis}, \citenamefont {Olivares}, \citenamefont {Falcke}, \citenamefont {Kramer},\ and\ \citenamefont {Rezzolla}}]{Mizuno:2018lxz}%
  \BibitemOpen
  \bibfield  {author} {\bibinfo {author} {\bibfnamefont {Y.}~\bibnamefont {Mizuno}}, \bibinfo {author} {\bibfnamefont {Z.}~\bibnamefont {Younsi}}, \bibinfo {author} {\bibfnamefont {C.~M.}\ \bibnamefont {Fromm}}, \bibinfo {author} {\bibfnamefont {O.}~\bibnamefont {Porth}}, \bibinfo {author} {\bibfnamefont {M.}~\bibnamefont {De~Laurentis}}, \bibinfo {author} {\bibfnamefont {H.}~\bibnamefont {Olivares}}, \bibinfo {author} {\bibfnamefont {H.}~\bibnamefont {Falcke}}, \bibinfo {author} {\bibfnamefont {M.}~\bibnamefont {Kramer}}, \ and\ \bibinfo {author} {\bibfnamefont {L.}~\bibnamefont {Rezzolla}},\ }\href {\doibase 10.1038/s41550-018-0449-5} {\bibfield  {journal} {\bibinfo  {journal} {Nature Astron.}\ }\textbf {\bibinfo {volume} {2}},\ \bibinfo {pages} {585} (\bibinfo {year} {2018})},\ \Eprint {http://arxiv.org/abs/1804.05812} {arXiv:1804.05812 [astro-ph.GA]} \BibitemShut {NoStop}%
\bibitem [{\citenamefont {Tsukamoto}(2018)}]{Tsukamoto:2017fxq}%
  \BibitemOpen
  \bibfield  {author} {\bibinfo {author} {\bibfnamefont {N.}~\bibnamefont {Tsukamoto}},\ }\href {\doibase 10.1103/PhysRevD.97.064021} {\bibfield  {journal} {\bibinfo  {journal} {Phys. Rev. D}\ }\textbf {\bibinfo {volume} {97}},\ \bibinfo {pages} {064021} (\bibinfo {year} {2018})},\ \Eprint {http://arxiv.org/abs/1708.07427} {arXiv:1708.07427 [gr-qc]} \BibitemShut {NoStop}%
\bibitem [{\citenamefont {Psaltis}(2019)}]{Psaltis:2018xkc}%
  \BibitemOpen
  \bibfield  {author} {\bibinfo {author} {\bibfnamefont {D.}~\bibnamefont {Psaltis}},\ }\href {\doibase 10.1007/s10714-019-2611-5} {\bibfield  {journal} {\bibinfo  {journal} {Gen. Rel. Grav.}\ }\textbf {\bibinfo {volume} {51}},\ \bibinfo {pages} {137} (\bibinfo {year} {2019})},\ \Eprint {http://arxiv.org/abs/1806.09740} {arXiv:1806.09740 [astro-ph.HE]} \BibitemShut {NoStop}%
\bibitem [{\citenamefont {Amir}\ \emph {et~al.}(2019)\citenamefont {Amir}, \citenamefont {Jusufi}, \citenamefont {Banerjee},\ and\ \citenamefont {Hansraj}}]{Amir:2018pcu}%
  \BibitemOpen
  \bibfield  {author} {\bibinfo {author} {\bibfnamefont {M.}~\bibnamefont {Amir}}, \bibinfo {author} {\bibfnamefont {K.}~\bibnamefont {Jusufi}}, \bibinfo {author} {\bibfnamefont {A.}~\bibnamefont {Banerjee}}, \ and\ \bibinfo {author} {\bibfnamefont {S.}~\bibnamefont {Hansraj}},\ }\href {\doibase 10.1088/1361-6382/ab42be} {\bibfield  {journal} {\bibinfo  {journal} {Class. Quant. Grav.}\ }\textbf {\bibinfo {volume} {36}},\ \bibinfo {pages} {215007} (\bibinfo {year} {2019})},\ \Eprint {http://arxiv.org/abs/1806.07782} {arXiv:1806.07782 [gr-qc]} \BibitemShut {NoStop}%
\bibitem [{\citenamefont {Gralla}\ \emph {et~al.}(2019)\citenamefont {Gralla}, \citenamefont {Holz},\ and\ \citenamefont {Wald}}]{Gralla:2019xty}%
  \BibitemOpen
  \bibfield  {author} {\bibinfo {author} {\bibfnamefont {S.~E.}\ \bibnamefont {Gralla}}, \bibinfo {author} {\bibfnamefont {D.~E.}\ \bibnamefont {Holz}}, \ and\ \bibinfo {author} {\bibfnamefont {R.~M.}\ \bibnamefont {Wald}},\ }\href {\doibase 10.1103/PhysRevD.100.024018} {\bibfield  {journal} {\bibinfo  {journal} {Phys. Rev. D}\ }\textbf {\bibinfo {volume} {100}},\ \bibinfo {pages} {024018} (\bibinfo {year} {2019})},\ \Eprint {http://arxiv.org/abs/1906.00873} {arXiv:1906.00873 [astro-ph.HE]} \BibitemShut {NoStop}%
\bibitem [{\citenamefont {Bambi}\ \emph {et~al.}(2019)\citenamefont {Bambi}, \citenamefont {Freese}, \citenamefont {Vagnozzi},\ and\ \citenamefont {Visinelli}}]{Bambi:2019tjh}%
  \BibitemOpen
  \bibfield  {author} {\bibinfo {author} {\bibfnamefont {C.}~\bibnamefont {Bambi}}, \bibinfo {author} {\bibfnamefont {K.}~\bibnamefont {Freese}}, \bibinfo {author} {\bibfnamefont {S.}~\bibnamefont {Vagnozzi}}, \ and\ \bibinfo {author} {\bibfnamefont {L.}~\bibnamefont {Visinelli}},\ }\href {\doibase 10.1103/PhysRevD.100.044057} {\bibfield  {journal} {\bibinfo  {journal} {Phys. Rev. D}\ }\textbf {\bibinfo {volume} {100}},\ \bibinfo {pages} {044057} (\bibinfo {year} {2019})},\ \Eprint {http://arxiv.org/abs/1904.12983} {arXiv:1904.12983 [gr-qc]} \BibitemShut {NoStop}%
\bibitem [{\citenamefont {Cunha}\ \emph {et~al.}(2019)\citenamefont {Cunha}, \citenamefont {Herdeiro},\ and\ \citenamefont {Radu}}]{Cunha:2019ikd}%
  \BibitemOpen
  \bibfield  {author} {\bibinfo {author} {\bibfnamefont {P.~V.~P.}\ \bibnamefont {Cunha}}, \bibinfo {author} {\bibfnamefont {C.~A.~R.}\ \bibnamefont {Herdeiro}}, \ and\ \bibinfo {author} {\bibfnamefont {E.}~\bibnamefont {Radu}},\ }\href {\doibase 10.3390/universe5120220} {\bibfield  {journal} {\bibinfo  {journal} {Universe}\ }\textbf {\bibinfo {volume} {5}},\ \bibinfo {pages} {220} (\bibinfo {year} {2019})},\ \Eprint {http://arxiv.org/abs/1909.08039} {arXiv:1909.08039 [gr-qc]} \BibitemShut {NoStop}%
\bibitem [{\citenamefont {Khodadi}\ \emph {et~al.}(2020)\citenamefont {Khodadi}, \citenamefont {Allahyari}, \citenamefont {Vagnozzi},\ and\ \citenamefont {Mota}}]{Khodadi:2020jij}%
  \BibitemOpen
  \bibfield  {author} {\bibinfo {author} {\bibfnamefont {M.}~\bibnamefont {Khodadi}}, \bibinfo {author} {\bibfnamefont {A.}~\bibnamefont {Allahyari}}, \bibinfo {author} {\bibfnamefont {S.}~\bibnamefont {Vagnozzi}}, \ and\ \bibinfo {author} {\bibfnamefont {D.~F.}\ \bibnamefont {Mota}},\ }\href {\doibase 10.1088/1475-7516/2020/09/026} {\bibfield  {journal} {\bibinfo  {journal} {JCAP}\ }\textbf {\bibinfo {volume} {09}},\ \bibinfo {pages} {026} (\bibinfo {year} {2020})},\ \Eprint {http://arxiv.org/abs/2005.05992} {arXiv:2005.05992 [gr-qc]} \BibitemShut {NoStop}%
\bibitem [{\citenamefont {Perlick}\ and\ \citenamefont {Tsupko}(2022)}]{Perlick:2021aok}%
  \BibitemOpen
  \bibfield  {author} {\bibinfo {author} {\bibfnamefont {V.}~\bibnamefont {Perlick}}\ and\ \bibinfo {author} {\bibfnamefont {O.~Y.}\ \bibnamefont {Tsupko}},\ }\href {\doibase 10.1016/j.physrep.2021.10.004} {\bibfield  {journal} {\bibinfo  {journal} {Phys. Rept.}\ }\textbf {\bibinfo {volume} {947}},\ \bibinfo {pages} {1} (\bibinfo {year} {2022})},\ \Eprint {http://arxiv.org/abs/2105.07101} {arXiv:2105.07101 [gr-qc]} \BibitemShut {NoStop}%
\bibitem [{\citenamefont {Vagnozzi}\ \emph {et~al.}(2023)\citenamefont {Vagnozzi} \emph {et~al.}}]{Vagnozzi:2022moj}%
  \BibitemOpen
  \bibfield  {author} {\bibinfo {author} {\bibfnamefont {S.}~\bibnamefont {Vagnozzi}} \emph {et~al.},\ }\href {\doibase 10.1088/1361-6382/acd97b} {\bibfield  {journal} {\bibinfo  {journal} {Class. Quant. Grav.}\ }\textbf {\bibinfo {volume} {40}},\ \bibinfo {pages} {165007} (\bibinfo {year} {2023})},\ \Eprint {http://arxiv.org/abs/2205.07787} {arXiv:2205.07787 [gr-qc]} \BibitemShut {NoStop}%
\bibitem [{\citenamefont {Saurabh}\ and\ \citenamefont {Jusufi}(2021)}]{Saurabh:2020zqg}%
  \BibitemOpen
  \bibfield  {author} {\bibinfo {author} {\bibfnamefont {K.}~\bibnamefont {Saurabh}}\ and\ \bibinfo {author} {\bibfnamefont {K.}~\bibnamefont {Jusufi}},\ }\href {\doibase 10.1140/epjc/s10052-021-09280-9} {\bibfield  {journal} {\bibinfo  {journal} {Eur. Phys. J. C}\ }\textbf {\bibinfo {volume} {81}},\ \bibinfo {pages} {490} (\bibinfo {year} {2021})},\ \Eprint {http://arxiv.org/abs/2009.10599} {arXiv:2009.10599 [gr-qc]} \BibitemShut {NoStop}%
\bibitem [{\citenamefont {Jusufi}\ \emph {et~al.}(2020)\citenamefont {Jusufi}, \citenamefont {Jamil},\ and\ \citenamefont {Zhu}}]{Jusufi:2020cpn}%
  \BibitemOpen
  \bibfield  {author} {\bibinfo {author} {\bibfnamefont {K.}~\bibnamefont {Jusufi}}, \bibinfo {author} {\bibfnamefont {M.}~\bibnamefont {Jamil}}, \ and\ \bibinfo {author} {\bibfnamefont {T.}~\bibnamefont {Zhu}},\ }\href {\doibase 10.1140/epjc/s10052-020-7899-5} {\bibfield  {journal} {\bibinfo  {journal} {Eur. Phys. J. C}\ }\textbf {\bibinfo {volume} {80}},\ \bibinfo {pages} {354} (\bibinfo {year} {2020})},\ \Eprint {http://arxiv.org/abs/2005.05299} {arXiv:2005.05299 [gr-qc]} \BibitemShut {NoStop}%
\bibitem [{\citenamefont {Tsupko}\ \emph {et~al.}(2020)\citenamefont {Tsupko}, \citenamefont {Fan},\ and\ \citenamefont {Bisnovatyi-Kogan}}]{Tsupko:2019pzg}%
  \BibitemOpen
  \bibfield  {author} {\bibinfo {author} {\bibfnamefont {O.~Y.}\ \bibnamefont {Tsupko}}, \bibinfo {author} {\bibfnamefont {Z.}~\bibnamefont {Fan}}, \ and\ \bibinfo {author} {\bibfnamefont {G.~S.}\ \bibnamefont {Bisnovatyi-Kogan}},\ }\href {\doibase 10.1088/1361-6382/ab6f7d} {\bibfield  {journal} {\bibinfo  {journal} {Class. Quant. Grav.}\ }\textbf {\bibinfo {volume} {37}},\ \bibinfo {pages} {065016} (\bibinfo {year} {2020})},\ \Eprint {http://arxiv.org/abs/1905.10509} {arXiv:1905.10509 [gr-qc]} \BibitemShut {NoStop}%
\bibitem [{\citenamefont {Akiyama}\ \emph {et~al.}(2019)\citenamefont {Akiyama} \emph {et~al.}}]{EventHorizonTelescope:2019dse}%
  \BibitemOpen
  \bibfield  {author} {\bibinfo {author} {\bibfnamefont {K.}~\bibnamefont {Akiyama}} \emph {et~al.} (\bibinfo {collaboration} {Event Horizon Telescope}),\ }\href {\doibase 10.3847/2041-8213/ab0ec7} {\bibfield  {journal} {\bibinfo  {journal} {Astrophys. J. Lett.}\ }\textbf {\bibinfo {volume} {875}},\ \bibinfo {pages} {L1} (\bibinfo {year} {2019})},\ \Eprint {http://arxiv.org/abs/1906.11238} {arXiv:1906.11238 [astro-ph.GA]} \BibitemShut {NoStop}%
\bibitem [{\citenamefont {Psaltis}\ \emph {et~al.}(2020)\citenamefont {Psaltis} \emph {et~al.}}]{EventHorizonTelescope:2020qrl}%
  \BibitemOpen
  \bibfield  {author} {\bibinfo {author} {\bibfnamefont {D.}~\bibnamefont {Psaltis}} \emph {et~al.} (\bibinfo {collaboration} {Event Horizon Telescope}),\ }\href {\doibase 10.1103/PhysRevLett.125.141104} {\bibfield  {journal} {\bibinfo  {journal} {Phys. Rev. Lett.}\ }\textbf {\bibinfo {volume} {125}},\ \bibinfo {pages} {141104} (\bibinfo {year} {2020})},\ \Eprint {http://arxiv.org/abs/2010.01055} {arXiv:2010.01055 [gr-qc]} \BibitemShut {NoStop}%
\bibitem [{\citenamefont {Akiyama}\ \emph {et~al.}(2021)\citenamefont {Akiyama} \emph {et~al.}}]{EventHorizonTelescope:2021srq}%
  \BibitemOpen
  \bibfield  {author} {\bibinfo {author} {\bibfnamefont {K.}~\bibnamefont {Akiyama}} \emph {et~al.} (\bibinfo {collaboration} {Event Horizon Telescope}),\ }\href {\doibase 10.3847/2041-8213/abe4de} {\bibfield  {journal} {\bibinfo  {journal} {Astrophys. J. Lett.}\ }\textbf {\bibinfo {volume} {910}},\ \bibinfo {pages} {L13} (\bibinfo {year} {2021})},\ \Eprint {http://arxiv.org/abs/2105.01173} {arXiv:2105.01173 [astro-ph.HE]} \BibitemShut {NoStop}%
\bibitem [{\citenamefont {Akiyama}\ \emph {et~al.}(2022)\citenamefont {Akiyama} \emph {et~al.}}]{EventHorizonTelescope:2022wkp}%
  \BibitemOpen
  \bibfield  {author} {\bibinfo {author} {\bibfnamefont {K.}~\bibnamefont {Akiyama}} \emph {et~al.} (\bibinfo {collaboration} {Event Horizon Telescope}),\ }\href {\doibase 10.3847/2041-8213/ac6674} {\bibfield  {journal} {\bibinfo  {journal} {Astrophys. J. Lett.}\ }\textbf {\bibinfo {volume} {930}},\ \bibinfo {pages} {L12} (\bibinfo {year} {2022})}\BibitemShut {NoStop}%
\bibitem [{\citenamefont {Jusufi}\ \emph {et~al.}(2023)\citenamefont {Jusufi}, \citenamefont {Leon},\ and\ \citenamefont {Millano}}]{Jusufi:2023xoa}%
  \BibitemOpen
  \bibfield  {author} {\bibinfo {author} {\bibfnamefont {K.}~\bibnamefont {Jusufi}}, \bibinfo {author} {\bibfnamefont {G.}~\bibnamefont {Leon}}, \ and\ \bibinfo {author} {\bibfnamefont {A.~D.}\ \bibnamefont {Millano}},\ }\href {\doibase 10.1016/j.dark.2023.101318} {\bibfield  {journal} {\bibinfo  {journal} {Phys. Dark Univ.}\ }\textbf {\bibinfo {volume} {42}},\ \bibinfo {pages} {101318} (\bibinfo {year} {2023})},\ \Eprint {http://arxiv.org/abs/2304.11492} {arXiv:2304.11492 [gr-qc]} \BibitemShut {NoStop}%
\bibitem [{\citenamefont {Loeb}\ and\ \citenamefont {Weiner}(2011)}]{Loeb:2010gj}%
  \BibitemOpen
  \bibfield  {author} {\bibinfo {author} {\bibfnamefont {A.}~\bibnamefont {Loeb}}\ and\ \bibinfo {author} {\bibfnamefont {N.}~\bibnamefont {Weiner}},\ }\href {\doibase 10.1103/PhysRevLett.106.171302} {\bibfield  {journal} {\bibinfo  {journal} {Phys. Rev. Lett.}\ }\textbf {\bibinfo {volume} {106}},\ \bibinfo {pages} {171302} (\bibinfo {year} {2011})},\ \Eprint {http://arxiv.org/abs/1011.6374} {arXiv:1011.6374 [astro-ph.CO]} \BibitemShut {NoStop}%
\bibitem [{\citenamefont {Anderson}\ \emph {et~al.}(2017)\citenamefont {Anderson}, \citenamefont {Bergin}, \citenamefont {Blake}, \citenamefont {Ciesla}, \citenamefont {Visser},\ and\ \citenamefont {Lee}}]{anderson2017destruction}%
  \BibitemOpen
  \bibfield  {author} {\bibinfo {author} {\bibfnamefont {D.~E.}\ \bibnamefont {Anderson}}, \bibinfo {author} {\bibfnamefont {E.~A.}\ \bibnamefont {Bergin}}, \bibinfo {author} {\bibfnamefont {G.~A.}\ \bibnamefont {Blake}}, \bibinfo {author} {\bibfnamefont {F.~J.}\ \bibnamefont {Ciesla}}, \bibinfo {author} {\bibfnamefont {R.}~\bibnamefont {Visser}}, \ and\ \bibinfo {author} {\bibfnamefont {J.-E.}\ \bibnamefont {Lee}},\ }\href@noop {} {\bibfield  {journal} {\bibinfo  {journal} {The Astrophysical Journal}\ }\textbf {\bibinfo {volume} {845}},\ \bibinfo {pages} {13} (\bibinfo {year} {2017})}\BibitemShut {NoStop}%
\bibitem [{\citenamefont {Kanzi}\ \emph {et~al.}(2020)\citenamefont {Kanzi}, \citenamefont {Mazharimousavi},\ and\ \citenamefont {Sakall\i{}}}]{Kanzi:2020cyv}%
  \BibitemOpen
  \bibfield  {author} {\bibinfo {author} {\bibfnamefont {S.}~\bibnamefont {Kanzi}}, \bibinfo {author} {\bibfnamefont {S.~H.}\ \bibnamefont {Mazharimousavi}}, \ and\ \bibinfo {author} {\bibfnamefont {I.}~\bibnamefont {Sakall\i{}}},\ }\href {\doibase 10.1016/j.aop.2020.168301} {\bibfield  {journal} {\bibinfo  {journal} {Annals Phys.}\ }\textbf {\bibinfo {volume} {422}},\ \bibinfo {pages} {168301} (\bibinfo {year} {2020})},\ \Eprint {http://arxiv.org/abs/2007.05814} {arXiv:2007.05814 [hep-th]} \BibitemShut {NoStop}%
\bibitem [{\citenamefont {Cede\~no}\ \emph {et~al.}(2017)\citenamefont {Cede\~no}, \citenamefont {Gonz\'alez-Morales},\ and\ \citenamefont {Ure\~na L\'opez}}]{Cedeno:2017sou}%
  \BibitemOpen
  \bibfield  {author} {\bibinfo {author} {\bibfnamefont {F.~X.~L.}\ \bibnamefont {Cede\~no}}, \bibinfo {author} {\bibfnamefont {A.~X.}\ \bibnamefont {Gonz\'alez-Morales}}, \ and\ \bibinfo {author} {\bibfnamefont {L.~A.}\ \bibnamefont {Ure\~na L\'opez}},\ }\href {\doibase 10.1103/PhysRevD.96.061301} {\bibfield  {journal} {\bibinfo  {journal} {Phys. Rev. D}\ }\textbf {\bibinfo {volume} {96}},\ \bibinfo {pages} {061301} (\bibinfo {year} {2017})},\ \Eprint {http://arxiv.org/abs/1703.10180} {arXiv:1703.10180 [gr-qc]} \BibitemShut {NoStop}%
\bibitem [{\citenamefont {Berezhiani}\ \emph {et~al.}(2009)\citenamefont {Berezhiani}, \citenamefont {Nesti}, \citenamefont {Pilo},\ and\ \citenamefont {Rossi}}]{Berezhiani:2009kv}%
  \BibitemOpen
  \bibfield  {author} {\bibinfo {author} {\bibfnamefont {Z.}~\bibnamefont {Berezhiani}}, \bibinfo {author} {\bibfnamefont {F.}~\bibnamefont {Nesti}}, \bibinfo {author} {\bibfnamefont {L.}~\bibnamefont {Pilo}}, \ and\ \bibinfo {author} {\bibfnamefont {N.}~\bibnamefont {Rossi}},\ }\href {\doibase 10.1088/1126-6708/2009/07/083} {\bibfield  {journal} {\bibinfo  {journal} {JHEP}\ }\textbf {\bibinfo {volume} {07}},\ \bibinfo {pages} {083} (\bibinfo {year} {2009})},\ \Eprint {http://arxiv.org/abs/0902.0144} {arXiv:0902.0144 [hep-th]} \BibitemShut {NoStop}%
\bibitem [{\citenamefont {Sakall\i{}}\ \emph {et~al.}(2018)\citenamefont {Sakall\i{}}, \citenamefont {Jusufi},\ and\ \citenamefont {\"Ovg\"un}}]{Sakalli:2018nug}%
  \BibitemOpen
  \bibfield  {author} {\bibinfo {author} {\bibfnamefont {I.}~\bibnamefont {Sakall\i{}}}, \bibinfo {author} {\bibfnamefont {K.}~\bibnamefont {Jusufi}}, \ and\ \bibinfo {author} {\bibfnamefont {A.}~\bibnamefont {\"Ovg\"un}},\ }\href {\doibase 10.1007/s10714-018-2455-4} {\bibfield  {journal} {\bibinfo  {journal} {Gen. Rel. Grav.}\ }\textbf {\bibinfo {volume} {50}},\ \bibinfo {pages} {125} (\bibinfo {year} {2018})},\ \Eprint {http://arxiv.org/abs/1803.10583} {arXiv:1803.10583 [gr-qc]} \BibitemShut {NoStop}%
\bibitem [{\citenamefont {Nicolini}\ \emph {et~al.}(2019)\citenamefont {Nicolini}, \citenamefont {Spallucci},\ and\ \citenamefont {Wondrak}}]{Nicolini:2019irw}%
  \BibitemOpen
  \bibfield  {author} {\bibinfo {author} {\bibfnamefont {P.}~\bibnamefont {Nicolini}}, \bibinfo {author} {\bibfnamefont {E.}~\bibnamefont {Spallucci}}, \ and\ \bibinfo {author} {\bibfnamefont {M.~F.}\ \bibnamefont {Wondrak}},\ }\href {\doibase 10.1016/j.physletb.2019.134888} {\bibfield  {journal} {\bibinfo  {journal} {Phys. Lett. B}\ }\textbf {\bibinfo {volume} {797}},\ \bibinfo {pages} {134888} (\bibinfo {year} {2019})},\ \Eprint {http://arxiv.org/abs/1902.11242} {arXiv:1902.11242 [gr-qc]} \BibitemShut {NoStop}%
\bibitem [{\citenamefont {Laurentis}\ and\ \citenamefont {Salucci}(2022)}]{DeLaurentis_2022}%
  \BibitemOpen
  \bibfield  {author} {\bibinfo {author} {\bibfnamefont {M.~D.}\ \bibnamefont {Laurentis}}\ and\ \bibinfo {author} {\bibfnamefont {P.}~\bibnamefont {Salucci}},\ }\href {\doibase 10.3847/1538-4357/ac54b9} {\bibfield  {journal} {\bibinfo  {journal} {The Astrophysical Journal}\ }\textbf {\bibinfo {volume} {929}},\ \bibinfo {pages} {17} (\bibinfo {year} {2022})}\BibitemShut {NoStop}%
\bibitem [{\citenamefont {de~Rham}\ \emph {et~al.}(2011)\citenamefont {de~Rham}, \citenamefont {Gabadadze},\ and\ \citenamefont {Tolley}}]{deRham:2010kj}%
  \BibitemOpen
  \bibfield  {author} {\bibinfo {author} {\bibfnamefont {C.}~\bibnamefont {de~Rham}}, \bibinfo {author} {\bibfnamefont {G.}~\bibnamefont {Gabadadze}}, \ and\ \bibinfo {author} {\bibfnamefont {A.~J.}\ \bibnamefont {Tolley}},\ }\href {\doibase 10.1103/PhysRevLett.106.231101} {\bibfield  {journal} {\bibinfo  {journal} {Phys. Rev. Lett.}\ }\textbf {\bibinfo {volume} {106}},\ \bibinfo {pages} {231101} (\bibinfo {year} {2011})},\ \Eprint {http://arxiv.org/abs/1011.1232} {arXiv:1011.1232 [hep-th]} \BibitemShut {NoStop}%
\bibitem [{\citenamefont {Ghosh}\ \emph {et~al.}(2016)\citenamefont {Ghosh}, \citenamefont {Tannukij},\ and\ \citenamefont {Wongjun}}]{Ghosh:2015cva}%
  \BibitemOpen
  \bibfield  {author} {\bibinfo {author} {\bibfnamefont {S.~G.}\ \bibnamefont {Ghosh}}, \bibinfo {author} {\bibfnamefont {L.}~\bibnamefont {Tannukij}}, \ and\ \bibinfo {author} {\bibfnamefont {P.}~\bibnamefont {Wongjun}},\ }\href {\doibase 10.1140/epjc/s10052-016-3943-x} {\bibfield  {journal} {\bibinfo  {journal} {Eur. Phys. J. C}\ }\textbf {\bibinfo {volume} {76}},\ \bibinfo {pages} {119} (\bibinfo {year} {2016})},\ \Eprint {http://arxiv.org/abs/1506.07119} {arXiv:1506.07119 [gr-qc]} \BibitemShut {NoStop}%
\bibitem [{\citenamefont {Simpson}\ and\ \citenamefont {Visser}(2019)}]{Simpson:2018tsi}%
  \BibitemOpen
  \bibfield  {author} {\bibinfo {author} {\bibfnamefont {A.}~\bibnamefont {Simpson}}\ and\ \bibinfo {author} {\bibfnamefont {M.}~\bibnamefont {Visser}},\ }\href {\doibase 10.1088/1475-7516/2019/02/042} {\bibfield  {journal} {\bibinfo  {journal} {JCAP}\ }\textbf {\bibinfo {volume} {02}},\ \bibinfo {pages} {042} (\bibinfo {year} {2019})},\ \Eprint {http://arxiv.org/abs/1812.07114} {arXiv:1812.07114 [gr-qc]} \BibitemShut {NoStop}%
\bibitem [{\citenamefont {Visser}(1995)}]{alma991027056009703276}%
  \BibitemOpen
  \bibfield  {author} {\bibinfo {author} {\bibfnamefont {M.}~\bibnamefont {Visser}},\ }\href {https://books.google.cl/books?id=Zo_vAAAAMAAJ} {\emph {\bibinfo {title} {Lorentzian Wormholes: From Einstein to Hawking}}},\ Computational and Mathematical Physics\ (\bibinfo  {publisher} {American Inst. of Physics},\ \bibinfo {year} {1995})\BibitemShut {NoStop}%
\bibitem [{\citenamefont {Visser}(2020)}]{Visser:2019brz}%
  \BibitemOpen
  \bibfield  {author} {\bibinfo {author} {\bibfnamefont {M.}~\bibnamefont {Visser}},\ }\href {\doibase 10.1088/1361-6382/ab60b8} {\bibfield  {journal} {\bibinfo  {journal} {Class. Quant. Grav.}\ }\textbf {\bibinfo {volume} {37}},\ \bibinfo {pages} {045001} (\bibinfo {year} {2020})},\ \Eprint {http://arxiv.org/abs/1908.11058} {arXiv:1908.11058 [gr-qc]} \BibitemShut {NoStop}%
\bibitem [{\citenamefont {Boonserm}\ \emph {et~al.}(2020)\citenamefont {Boonserm}, \citenamefont {Ngampitipan}, \citenamefont {Simpson},\ and\ \citenamefont {Visser}}]{Boonserm:2019phw}%
  \BibitemOpen
  \bibfield  {author} {\bibinfo {author} {\bibfnamefont {P.}~\bibnamefont {Boonserm}}, \bibinfo {author} {\bibfnamefont {T.}~\bibnamefont {Ngampitipan}}, \bibinfo {author} {\bibfnamefont {A.}~\bibnamefont {Simpson}}, \ and\ \bibinfo {author} {\bibfnamefont {M.}~\bibnamefont {Visser}},\ }\href {\doibase 10.1103/PhysRevD.101.024022} {\bibfield  {journal} {\bibinfo  {journal} {Phys. Rev. D}\ }\textbf {\bibinfo {volume} {101}},\ \bibinfo {pages} {024022} (\bibinfo {year} {2020})},\ \Eprint {http://arxiv.org/abs/1910.08008} {arXiv:1910.08008 [gr-qc]} \BibitemShut {NoStop}%
\bibitem [{\citenamefont {Fenichel}(1979)}]{fenichel}%
  \BibitemOpen
  \bibfield  {author} {\bibinfo {author} {\bibfnamefont {N.}~\bibnamefont {Fenichel}},\ }\href@noop {} {\bibfield  {journal} {\bibinfo  {journal} {Journal of Differential Equations}\ }\textbf {\bibinfo {volume} {31}},\ \bibinfo {pages} {53} (\bibinfo {year} {1979})}\BibitemShut {NoStop}%
\bibitem [{\citenamefont {Fusco}\ and\ \citenamefont {Hale}(1989)}]{Fusco1989SlowmotionMD}%
  \BibitemOpen
  \bibfield  {author} {\bibinfo {author} {\bibfnamefont {G.}~\bibnamefont {Fusco}}\ and\ \bibinfo {author} {\bibfnamefont {J.~K.}\ \bibnamefont {Hale}},\ }\href@noop {} {\bibfield  {journal} {\bibinfo  {journal} {Journal of Dynamics and Differential Equations}\ }\textbf {\bibinfo {volume} {1}},\ \bibinfo {pages} {75} (\bibinfo {year} {1989})}\BibitemShut {NoStop}%
\bibitem [{\citenamefont {Berglund}\ and\ \citenamefont {Gentz}(2006)}]{Berglund}%
  \BibitemOpen
  \bibfield  {author} {\bibinfo {author} {\bibfnamefont {N.}~\bibnamefont {Berglund}}\ and\ \bibinfo {author} {\bibfnamefont {B.}~\bibnamefont {Gentz}},\ }\href@noop {} {\emph {\bibinfo {title} {Noise-Induced Phenomena in Slow-Fast Dynamical Systems}}}\ (\bibinfo  {publisher} {Springer-Verlag},\ \bibinfo {year} {2006})\BibitemShut {NoStop}%
\bibitem [{\citenamefont {Verhulst}(2006)}]{verhulst2006method}%
  \BibitemOpen
  \bibfield  {author} {\bibinfo {author} {\bibfnamefont {F.}~\bibnamefont {Verhulst}},\ }\href@noop {} {\emph {\bibinfo {title} {Methods and Applications of Singular Perturbations: Boundary Layers and Multiple Timescale Dynamics}}},\ Vol.~\bibinfo {volume} {50}\ (\bibinfo  {publisher} {Springer Science \& Business Media},\ \bibinfo {year} {2006})\BibitemShut {NoStop}%
\bibitem [{\citenamefont {Holmes}(2012)}]{holmes2012introduction}%
  \BibitemOpen
  \bibfield  {author} {\bibinfo {author} {\bibfnamefont {M.}~\bibnamefont {Holmes}},\ }\href@noop {} {\emph {\bibinfo {title} {Introduction to Perturbation Methods}}},\ Texts in Applied Mathematics\ (\bibinfo  {publisher} {Springer New York},\ \bibinfo {year} {2012})\BibitemShut {NoStop}%
\bibitem [{\citenamefont {Kevorkian}\ and\ \citenamefont {Cole}(2013)}]{kevorkian2013perturbation}%
  \BibitemOpen
  \bibfield  {author} {\bibinfo {author} {\bibfnamefont {J.}~\bibnamefont {Kevorkian}}\ and\ \bibinfo {author} {\bibfnamefont {J.}~\bibnamefont {Cole}},\ }\href@noop {} {\emph {\bibinfo {title} {Perturbation Methods in Applied Mathematics}}},\ Applied Mathematical Sciences\ (\bibinfo  {publisher} {Springer New York},\ \bibinfo {year} {2013})\BibitemShut {NoStop}%
\bibitem [{\citenamefont {Awrejcewicz}(2014)}]{AwrejcewiczJan2014ANDS}%
  \BibitemOpen
  \bibfield  {author} {\bibinfo {author} {\bibfnamefont {J.}~\bibnamefont {Awrejcewicz}},\ }\href@noop {} {\emph {\bibinfo {title} {Applied Non-Linear Dynamical Systems}}},\ \bibinfo {series} {Springer Proceedings in Mathematics \& Statistics}, Vol.~\bibinfo {volume} {93}\ (\bibinfo  {publisher} {Springer International Publishing AG},\ \bibinfo {address} {Cham},\ \bibinfo {year} {2014})\BibitemShut {NoStop}%
\bibitem [{\citenamefont {Hawking}(1975{\natexlab{a}})}]{hawking1975particle}%
  \BibitemOpen
  \bibfield  {author} {\bibinfo {author} {\bibfnamefont {S.~W.}\ \bibnamefont {Hawking}},\ }in\ \href@noop {} {\emph {\bibinfo {booktitle} {Euclidean quantum gravity}}}\ (\bibinfo  {publisher} {World Scientific},\ \bibinfo {year} {1975})\ pp.\ \bibinfo {pages} {167--188}\BibitemShut {NoStop}%
\bibitem [{\citenamefont {Hawking}(1975{\natexlab{b}})}]{Hawking:1975vcx}%
  \BibitemOpen
  \bibfield  {author} {\bibinfo {author} {\bibfnamefont {S.~W.}\ \bibnamefont {Hawking}},\ }\href {\doibase 10.1007/BF02345020} {\bibfield  {journal} {\bibinfo  {journal} {Commun. Math. Phys.}\ }\textbf {\bibinfo {volume} {43}},\ \bibinfo {pages} {199} (\bibinfo {year} {1975}{\natexlab{b}})},\ \bibinfo {note} {[Erratum: Commun.Math.Phys. 46, 206 (1976)]}\BibitemShut {NoStop}%
\bibitem [{\citenamefont {Filho}\ \emph {et~al.}(2023{\natexlab{a}})\citenamefont {Filho}, \citenamefont {Zare}, \citenamefont {Porf\'\i{}rio}, \citenamefont {K\v{r}\'\i{}\v{z}},\ and\ \citenamefont {Hassanabadi}}]{Filho:2022zdh}%
  \BibitemOpen
  \bibfield  {author} {\bibinfo {author} {\bibfnamefont {A.~A.~A.}\ \bibnamefont {Filho}}, \bibinfo {author} {\bibfnamefont {S.}~\bibnamefont {Zare}}, \bibinfo {author} {\bibfnamefont {P.~J.}\ \bibnamefont {Porf\'\i{}rio}}, \bibinfo {author} {\bibfnamefont {J.}~\bibnamefont {K\v{r}\'\i{}\v{z}}}, \ and\ \bibinfo {author} {\bibfnamefont {H.}~\bibnamefont {Hassanabadi}},\ }\href {\doibase 10.1016/j.physletb.2023.137744} {\bibfield  {journal} {\bibinfo  {journal} {Phys. Lett. B}\ }\textbf {\bibinfo {volume} {838}},\ \bibinfo {pages} {137744} (\bibinfo {year} {2023}{\natexlab{a}})},\ \Eprint {http://arxiv.org/abs/2212.03134} {arXiv:2212.03134 [hep-th]} \BibitemShut {NoStop}%
\bibitem [{\citenamefont {Sedaghatnia}\ \emph {et~al.}(2023)\citenamefont {Sedaghatnia}, \citenamefont {Hassanabadi}, \citenamefont {Filho}, \citenamefont {Porf\'\i{}rio},\ and\ \citenamefont {Chung}}]{Sedaghatnia:2023fod}%
  \BibitemOpen
  \bibfield  {author} {\bibinfo {author} {\bibfnamefont {P.}~\bibnamefont {Sedaghatnia}}, \bibinfo {author} {\bibfnamefont {H.}~\bibnamefont {Hassanabadi}}, \bibinfo {author} {\bibfnamefont {A.~A.~A.}\ \bibnamefont {Filho}}, \bibinfo {author} {\bibfnamefont {P.~J.~P.}\ \bibnamefont {Porf\'\i{}rio}}, \ and\ \bibinfo {author} {\bibfnamefont {W.~S.}\ \bibnamefont {Chung}},\ }\href@noop {} {\enquote {\bibinfo {title} {{Thermodynamical properties of a deformed Schwarzschild black hole via Dunkl generalization}},}\ } (\bibinfo {year} {2023}),\ \Eprint {http://arxiv.org/abs/2302.11460} {arXiv:2302.11460 [gr-qc]} \BibitemShut {NoStop}%
\bibitem [{\citenamefont {Filho}\ \emph {et~al.}(2023{\natexlab{b}})\citenamefont {Filho}, \citenamefont {Furtado}, \citenamefont {Hassanabadi},\ and\ \citenamefont {Reis}}]{Filho:2023ydb}%
  \BibitemOpen
  \bibfield  {author} {\bibinfo {author} {\bibfnamefont {A.~A.~A.}\ \bibnamefont {Filho}}, \bibinfo {author} {\bibfnamefont {J.}~\bibnamefont {Furtado}}, \bibinfo {author} {\bibfnamefont {H.}~\bibnamefont {Hassanabadi}}, \ and\ \bibinfo {author} {\bibfnamefont {J.~A. A.~S.}\ \bibnamefont {Reis}},\ }\href {\doibase 10.1016/j.dark.2023.101310} {\enquote {\bibinfo {title} {{Thermal analysis of photon-like particles in rainbow gravity}},}\ } (\bibinfo {year} {2023}{\natexlab{b}}),\ \Eprint {http://arxiv.org/abs/2305.08587} {arXiv:2305.08587 [gr-qc]} \BibitemShut {NoStop}%
\bibitem [{\citenamefont {Filho}\ \emph {et~al.}(2023{\natexlab{c}})\citenamefont {Filho}, \citenamefont {Furtado}, \citenamefont {Reis},\ and\ \citenamefont {Silva}}]{Filho:2023yly}%
  \BibitemOpen
  \bibfield  {author} {\bibinfo {author} {\bibfnamefont {A.~A.~A.}\ \bibnamefont {Filho}}, \bibinfo {author} {\bibfnamefont {J.}~\bibnamefont {Furtado}}, \bibinfo {author} {\bibfnamefont {J.~A. A.~S.}\ \bibnamefont {Reis}}, \ and\ \bibinfo {author} {\bibfnamefont {J.~E.~G.}\ \bibnamefont {Silva}},\ }\href {\doibase 10.1088/1361-6382/ad0421} {\enquote {\bibinfo {title} {{Thermodynamical properties of an ideal gas in a traversable wormhole}},}\ } (\bibinfo {year} {2023}{\natexlab{c}}),\ \Eprint {http://arxiv.org/abs/2302.05492} {arXiv:2302.05492 [gr-qc]} \BibitemShut {NoStop}%
\bibitem [{\citenamefont {Ma}\ and\ \citenamefont {Zhao}(2014)}]{Ma:2014qma}%
  \BibitemOpen
  \bibfield  {author} {\bibinfo {author} {\bibfnamefont {M.-S.}\ \bibnamefont {Ma}}\ and\ \bibinfo {author} {\bibfnamefont {R.}~\bibnamefont {Zhao}},\ }\href {\doibase 10.1088/0264-9381/31/24/245014} {\bibfield  {journal} {\bibinfo  {journal} {Class. Quant. Grav.}\ }\textbf {\bibinfo {volume} {31}},\ \bibinfo {pages} {245014} (\bibinfo {year} {2014})},\ \Eprint {http://arxiv.org/abs/1411.0833} {arXiv:1411.0833 [gr-qc]} \BibitemShut {NoStop}%
\bibitem [{\citenamefont {Maluf}\ and\ \citenamefont {Neves}(2018)}]{Maluf:2018lyu}%
  \BibitemOpen
  \bibfield  {author} {\bibinfo {author} {\bibfnamefont {R.~V.}\ \bibnamefont {Maluf}}\ and\ \bibinfo {author} {\bibfnamefont {J.~C.~S.}\ \bibnamefont {Neves}},\ }\href {\doibase 10.1103/PhysRevD.97.104015} {\bibfield  {journal} {\bibinfo  {journal} {Phys. Rev. D}\ }\textbf {\bibinfo {volume} {97}},\ \bibinfo {pages} {104015} (\bibinfo {year} {2018})},\ \Eprint {http://arxiv.org/abs/1801.02661} {arXiv:1801.02661 [gr-qc]} \BibitemShut {NoStop}%
\bibitem [{\citenamefont {Iyer}\ and\ \citenamefont {Will}(1987)}]{Iyer:1986np}%
  \BibitemOpen
  \bibfield  {author} {\bibinfo {author} {\bibfnamefont {S.}~\bibnamefont {Iyer}}\ and\ \bibinfo {author} {\bibfnamefont {C.~M.}\ \bibnamefont {Will}},\ }\href {\doibase 10.1103/PhysRevD.35.3621} {\bibfield  {journal} {\bibinfo  {journal} {Phys. Rev. D}\ }\textbf {\bibinfo {volume} {35}},\ \bibinfo {pages} {3621} (\bibinfo {year} {1987})}\BibitemShut {NoStop}%
\bibitem [{\citenamefont {Iyer}(1987)}]{Iyer:1986nq}%
  \BibitemOpen
  \bibfield  {author} {\bibinfo {author} {\bibfnamefont {S.}~\bibnamefont {Iyer}},\ }\href {\doibase 10.1103/PhysRevD.35.3632} {\bibfield  {journal} {\bibinfo  {journal} {Phys. Rev. D}\ }\textbf {\bibinfo {volume} {35}},\ \bibinfo {pages} {3632} (\bibinfo {year} {1987})}\BibitemShut {NoStop}%
\bibitem [{\citenamefont {Konoplya}(2003)}]{Konoplya:2003ii}%
  \BibitemOpen
  \bibfield  {author} {\bibinfo {author} {\bibfnamefont {R.~A.}\ \bibnamefont {Konoplya}},\ }\href {\doibase 10.1103/PhysRevD.68.024018} {\bibfield  {journal} {\bibinfo  {journal} {Phys. Rev. D}\ }\textbf {\bibinfo {volume} {68}},\ \bibinfo {pages} {024018} (\bibinfo {year} {2003})},\ \Eprint {http://arxiv.org/abs/gr-qc/0303052} {arXiv:gr-qc/0303052} \BibitemShut {NoStop}%
\bibitem [{\citenamefont {Matyjasek}\ and\ \citenamefont {Opala}(2017)}]{Matyjasek:2017psv}%
  \BibitemOpen
  \bibfield  {author} {\bibinfo {author} {\bibfnamefont {J.}~\bibnamefont {Matyjasek}}\ and\ \bibinfo {author} {\bibfnamefont {M.}~\bibnamefont {Opala}},\ }\href {\doibase 10.1103/PhysRevD.96.024011} {\bibfield  {journal} {\bibinfo  {journal} {Phys. Rev. D}\ }\textbf {\bibinfo {volume} {96}},\ \bibinfo {pages} {024011} (\bibinfo {year} {2017})},\ \Eprint {http://arxiv.org/abs/1704.00361} {arXiv:1704.00361 [gr-qc]} \BibitemShut {NoStop}%
\bibitem [{\citenamefont {{Schutz}}\ and\ \citenamefont {{Will}}(1985)}]{1985ApJ...291L..33S}%
  \BibitemOpen
  \bibfield  {author} {\bibinfo {author} {\bibfnamefont {B.~F.}\ \bibnamefont {{Schutz}}}\ and\ \bibinfo {author} {\bibfnamefont {C.~M.}\ \bibnamefont {{Will}}},\ }\href {\doibase 10.1086/184453} {\bibfield  {journal} {\bibinfo  {journal} {APJL}\ }\textbf {\bibinfo {volume} {291}},\ \bibinfo {pages} {L33} (\bibinfo {year} {1985})}\BibitemShut {NoStop}%
\bibitem [{\citenamefont {Konoplya}(2004)}]{Konoplya:2004ip}%
  \BibitemOpen
  \bibfield  {author} {\bibinfo {author} {\bibfnamefont {R.~A.}\ \bibnamefont {Konoplya}},\ }\href@noop {} {\bibfield  {journal} {\bibinfo  {journal} {J. Phys. Stud.}\ }\textbf {\bibinfo {volume} {8}},\ \bibinfo {pages} {93} (\bibinfo {year} {2004})}\BibitemShut {NoStop}%
\bibitem [{\citenamefont {Santos}\ \emph {et~al.}(2016)\citenamefont {Santos}, \citenamefont {Maluf},\ and\ \citenamefont {Almeida}}]{Santos:2015gja}%
  \BibitemOpen
  \bibfield  {author} {\bibinfo {author} {\bibfnamefont {V.}~\bibnamefont {Santos}}, \bibinfo {author} {\bibfnamefont {R.~V.}\ \bibnamefont {Maluf}}, \ and\ \bibinfo {author} {\bibfnamefont {C.~A.~S.}\ \bibnamefont {Almeida}},\ }\href {\doibase 10.1103/PhysRevD.93.084047} {\bibfield  {journal} {\bibinfo  {journal} {Phys. Rev. D}\ }\textbf {\bibinfo {volume} {93}},\ \bibinfo {pages} {084047} (\bibinfo {year} {2016})},\ \Eprint {http://arxiv.org/abs/1509.04306} {arXiv:1509.04306 [gr-qc]} \BibitemShut {NoStop}%
\bibitem [{\citenamefont {Konoplya}\ and\ \citenamefont {Zhidenko}(2011)}]{Konoplya:2011qq}%
  \BibitemOpen
  \bibfield  {author} {\bibinfo {author} {\bibfnamefont {R.~A.}\ \bibnamefont {Konoplya}}\ and\ \bibinfo {author} {\bibfnamefont {A.}~\bibnamefont {Zhidenko}},\ }\href {\doibase 10.1103/RevModPhys.83.793} {\bibfield  {journal} {\bibinfo  {journal} {Rev. Mod. Phys.}\ }\textbf {\bibinfo {volume} {83}},\ \bibinfo {pages} {793} (\bibinfo {year} {2011})},\ \Eprint {http://arxiv.org/abs/1102.4014} {arXiv:1102.4014 [gr-qc]} \BibitemShut {NoStop}%
\bibitem [{\citenamefont {Berti}\ \emph {et~al.}(2009)\citenamefont {Berti}, \citenamefont {Cardoso},\ and\ \citenamefont {Starinets}}]{Berti:2009kk}%
  \BibitemOpen
  \bibfield  {author} {\bibinfo {author} {\bibfnamefont {E.}~\bibnamefont {Berti}}, \bibinfo {author} {\bibfnamefont {V.}~\bibnamefont {Cardoso}}, \ and\ \bibinfo {author} {\bibfnamefont {A.~O.}\ \bibnamefont {Starinets}},\ }\href {\doibase 10.1088/0264-9381/26/16/163001} {\bibfield  {journal} {\bibinfo  {journal} {Class. Quant. Grav.}\ }\textbf {\bibinfo {volume} {26}},\ \bibinfo {pages} {163001} (\bibinfo {year} {2009})},\ \Eprint {http://arxiv.org/abs/0905.2975} {arXiv:0905.2975 [gr-qc]} \BibitemShut {NoStop}%
\bibitem [{\citenamefont {Heidari}\ \emph {et~al.}(2023{\natexlab{a}})\citenamefont {Heidari}, \citenamefont {Hassanabadi}, \citenamefont {Filho}, \citenamefont {Kur\'\i{}uz}, \citenamefont {Zare},\ and\ \citenamefont {Porf\'\i{}rio}}]{Heidari:2023bww}%
  \BibitemOpen
  \bibfield  {author} {\bibinfo {author} {\bibfnamefont {N.}~\bibnamefont {Heidari}}, \bibinfo {author} {\bibfnamefont {H.}~\bibnamefont {Hassanabadi}}, \bibinfo {author} {\bibfnamefont {A.~A.~A.}\ \bibnamefont {Filho}}, \bibinfo {author} {\bibfnamefont {J.}~\bibnamefont {Kur\'\i{}uz}}, \bibinfo {author} {\bibfnamefont {S.}~\bibnamefont {Zare}}, \ and\ \bibinfo {author} {\bibfnamefont {P.~J.}\ \bibnamefont {Porf\'\i{}rio}},\ }\href@noop {} {\enquote {\bibinfo {title} {{Gravitational signatures of a non--commutative stable black hole}},}\ } (\bibinfo {year} {2023}{\natexlab{a}}),\ \Eprint {http://arxiv.org/abs/2305.06838} {arXiv:2305.06838 [gr-qc]} \BibitemShut {NoStop}%
\bibitem [{\citenamefont {{Chen}}\ \emph {et~al.}(2023)\citenamefont {{Chen}}, \citenamefont {{Sathiyaraj}}, \citenamefont {{Hassanabadi}}, \citenamefont {{Yang}}, \citenamefont {{Long}},\ and\ \citenamefont {{Tu}}}]{2023InJPh.tmp..228C}%
  \BibitemOpen
  \bibfield  {author} {\bibinfo {author} {\bibfnamefont {H.}~\bibnamefont {{Chen}}}, \bibinfo {author} {\bibfnamefont {T.}~\bibnamefont {{Sathiyaraj}}}, \bibinfo {author} {\bibfnamefont {H.}~\bibnamefont {{Hassanabadi}}}, \bibinfo {author} {\bibfnamefont {Y.}~\bibnamefont {{Yang}}}, \bibinfo {author} {\bibfnamefont {Z.~W.}\ \bibnamefont {{Long}}}, \ and\ \bibinfo {author} {\bibfnamefont {F.~Q.}\ \bibnamefont {{Tu}}},\ }\href {\doibase 10.1007/s12648-023-02734-8} {\bibfield  {journal} {\bibinfo  {journal} {Indian Journal of Physics}\ } (\bibinfo {year} {2023}),\ 10.1007/s12648-023-02734-8}\BibitemShut {NoStop}%
\bibitem [{\citenamefont {Toshmatov}\ \emph {et~al.}(2017)\citenamefont {Toshmatov}, \citenamefont {Bambi}, \citenamefont {Ahmedov}, \citenamefont {Stuchl\'\i{}k},\ and\ \citenamefont {Schee}}]{Toshmatov:2017bpx}%
  \BibitemOpen
  \bibfield  {author} {\bibinfo {author} {\bibfnamefont {B.}~\bibnamefont {Toshmatov}}, \bibinfo {author} {\bibfnamefont {C.}~\bibnamefont {Bambi}}, \bibinfo {author} {\bibfnamefont {B.}~\bibnamefont {Ahmedov}}, \bibinfo {author} {\bibfnamefont {Z.}~\bibnamefont {Stuchl\'\i{}k}}, \ and\ \bibinfo {author} {\bibfnamefont {J.}~\bibnamefont {Schee}},\ }\href {\doibase 10.1103/PhysRevD.96.06402} {\bibfield  {journal} {\bibinfo  {journal} {Phys. Rev. D}\ }\textbf {\bibinfo {volume} {96}},\ \bibinfo {pages} {064028} (\bibinfo {year} {2017})},\ \Eprint {http://arxiv.org/abs/1705.03654} {arXiv:1705.03654 [gr-qc]} \BibitemShut {NoStop}%
\bibitem [{\citenamefont {Kim}(2004)}]{Kim:2004ju}%
  \BibitemOpen
  \bibfield  {author} {\bibinfo {author} {\bibfnamefont {S.-W.}\ \bibnamefont {Kim}},\ }\href@noop {} {\enquote {\bibinfo {title} {{Gravitational perturbation of traversable wormhole}},}\ } (\bibinfo {year} {2004}),\ \Eprint {http://arxiv.org/abs/gr-qc/0401007} {arXiv:gr-qc/0401007} \BibitemShut {NoStop}%
\bibitem [{\citenamefont {Filho}(2023)}]{Filho:2023qxu}%
  \BibitemOpen
  \bibfield  {author} {\bibinfo {author} {\bibfnamefont {A.~A.~A.}\ \bibnamefont {Filho}},\ }\href@noop {} {\enquote {\bibinfo {title} {{Implications of a Simpson-Visser solution in Verlinde's framework}},}\ } (\bibinfo {year} {2023}),\ \Eprint {http://arxiv.org/abs/2308.04939} {arXiv:2308.04939 [gr-qc]} \BibitemShut {NoStop}%
\bibitem [{\citenamefont {Tsukamoto}(2021)}]{Tsukamoto:2021caq}%
  \BibitemOpen
  \bibfield  {author} {\bibinfo {author} {\bibfnamefont {N.}~\bibnamefont {Tsukamoto}},\ }\href {\doibase 10.1103/PhysRevD.104.064022} {\bibfield  {journal} {\bibinfo  {journal} {Phys. Rev. D}\ }\textbf {\bibinfo {volume} {104}},\ \bibinfo {pages} {064022} (\bibinfo {year} {2021})},\ \Eprint {http://arxiv.org/abs/2105.14336} {arXiv:2105.14336 [gr-qc]} \BibitemShut {NoStop}%
\bibitem [{\citenamefont {Tsukamoto}(2022)}]{Tsukamoto:2022vkt}%
  \BibitemOpen
  \bibfield  {author} {\bibinfo {author} {\bibfnamefont {N.}~\bibnamefont {Tsukamoto}},\ }\href {\doibase 10.1103/PhysRevD.105.084036} {\bibfield  {journal} {\bibinfo  {journal} {Phys. Rev. D}\ }\textbf {\bibinfo {volume} {105}},\ \bibinfo {pages} {084036} (\bibinfo {year} {2022})},\ \Eprint {http://arxiv.org/abs/2202.09641} {arXiv:2202.09641 [gr-qc]} \BibitemShut {NoStop}%
\bibitem [{\citenamefont {Guerrero}\ \emph {et~al.}(2022)\citenamefont {Guerrero}, \citenamefont {Olmo}, \citenamefont {Rubiera-Garcia},\ and\ \citenamefont {S\'aez-Chill\'on~G\'omez}}]{Guerrero:2022msp}%
  \BibitemOpen
  \bibfield  {author} {\bibinfo {author} {\bibfnamefont {M.}~\bibnamefont {Guerrero}}, \bibinfo {author} {\bibfnamefont {G.~J.}\ \bibnamefont {Olmo}}, \bibinfo {author} {\bibfnamefont {D.}~\bibnamefont {Rubiera-Garcia}}, \ and\ \bibinfo {author} {\bibfnamefont {D.}~\bibnamefont {S\'aez-Chill\'on~G\'omez}},\ }\href {\doibase 10.1103/PhysRevD.106.044070} {\bibfield  {journal} {\bibinfo  {journal} {Phys. Rev. D}\ }\textbf {\bibinfo {volume} {106}},\ \bibinfo {pages} {044070} (\bibinfo {year} {2022})},\ \Eprint {http://arxiv.org/abs/2205.12147} {arXiv:2205.12147 [gr-qc]} \BibitemShut {NoStop}%
\bibitem [{\citenamefont {Singh}\ and\ \citenamefont {Ghosh}(2018)}]{Singh:2017vfr}%
  \BibitemOpen
  \bibfield  {author} {\bibinfo {author} {\bibfnamefont {B.~P.}\ \bibnamefont {Singh}}\ and\ \bibinfo {author} {\bibfnamefont {S.~G.}\ \bibnamefont {Ghosh}},\ }\href {\doibase 10.1016/j.aop.2018.05.010} {\bibfield  {journal} {\bibinfo  {journal} {Annals Phys.}\ }\textbf {\bibinfo {volume} {395}},\ \bibinfo {pages} {127} (\bibinfo {year} {2018})},\ \Eprint {http://arxiv.org/abs/1707.07125} {arXiv:1707.07125 [gr-qc]} \BibitemShut {NoStop}%
\bibitem [{\citenamefont {Filho}\ \emph {et~al.}(2023{\natexlab{d}})\citenamefont {Filho}, \citenamefont {Reis},\ and\ \citenamefont {Hassanabadi}}]{Filho:2023ycx}%
  \BibitemOpen
  \bibfield  {author} {\bibinfo {author} {\bibfnamefont {A.~A.~A.}\ \bibnamefont {Filho}}, \bibinfo {author} {\bibfnamefont {J.~A. A.~S.}\ \bibnamefont {Reis}}, \ and\ \bibinfo {author} {\bibfnamefont {H.}~\bibnamefont {Hassanabadi}},\ }\href@noop {} {\enquote {\bibinfo {title} {{Exploring antisymmetric tensor effects on black hole shadows and quasinormal frequencies}},}\ } (\bibinfo {year} {2023}{\natexlab{d}}),\ \Eprint {http://arxiv.org/abs/2309.15778} {arXiv:2309.15778 [gr-qc]} \BibitemShut {NoStop}%
\bibitem [{\citenamefont {Heidari}\ \emph {et~al.}(2023{\natexlab{b}})\citenamefont {Heidari}, \citenamefont {Hassanabadi}, \citenamefont {Filho},\ and\ \citenamefont {Kur\'\i{}uz}}]{Heidari:2023egu}%
  \BibitemOpen
  \bibfield  {author} {\bibinfo {author} {\bibfnamefont {N.}~\bibnamefont {Heidari}}, \bibinfo {author} {\bibfnamefont {H.}~\bibnamefont {Hassanabadi}}, \bibinfo {author} {\bibfnamefont {A.~A.~A.}\ \bibnamefont {Filho}}, \ and\ \bibinfo {author} {\bibfnamefont {J.}~\bibnamefont {Kur\'\i{}uz}},\ }\href@noop {} {\enquote {\bibinfo {title} {{Exploring Non--commutativity as a Perturbation in the Schwarzschild Black Hole: Quasinormal Modes, Scattering, and Shadows}},}\ } (\bibinfo {year} {2023}{\natexlab{b}}),\ \Eprint {http://arxiv.org/abs/2308.03284} {arXiv:2308.03284 [gr-qc]} \BibitemShut {NoStop}%
\bibitem [{\citenamefont {Filho}\ \emph {et~al.}(2023{\natexlab{e}})\citenamefont {Filho}, \citenamefont {Hassanabadi}, \citenamefont {Heidari}, \citenamefont {Kr\'\i{}z}, \citenamefont {Porf\'\i{}rio},\ and\ \citenamefont {Zare}}]{Filho:2023etf}%
  \BibitemOpen
  \bibfield  {author} {\bibinfo {author} {\bibfnamefont {A.~A.~A.}\ \bibnamefont {Filho}}, \bibinfo {author} {\bibfnamefont {H.}~\bibnamefont {Hassanabadi}}, \bibinfo {author} {\bibfnamefont {N.}~\bibnamefont {Heidari}}, \bibinfo {author} {\bibfnamefont {J.}~\bibnamefont {Kr\'\i{}z}}, \bibinfo {author} {\bibfnamefont {P.~J.}\ \bibnamefont {Porf\'\i{}rio}}, \ and\ \bibinfo {author} {\bibfnamefont {S.}~\bibnamefont {Zare}},\ }\href@noop {} {\enquote {\bibinfo {title} {{Gravitational traces of bumblebee gravity in metric-affine formalism}},}\ } (\bibinfo {year} {2023}{\natexlab{e}}),\ \Eprint {http://arxiv.org/abs/2305.18871} {arXiv:2305.18871 [gr-qc]} \BibitemShut {NoStop}%
\bibitem [{\citenamefont {Bambi}(2013)}]{Bambi:2013nla}%
  \BibitemOpen
  \bibfield  {author} {\bibinfo {author} {\bibfnamefont {C.}~\bibnamefont {Bambi}},\ }\href {\doibase 10.1103/PhysRevD.87.107501} {\bibfield  {journal} {\bibinfo  {journal} {Phys. Rev. D}\ }\textbf {\bibinfo {volume} {87}},\ \bibinfo {pages} {107501} (\bibinfo {year} {2013})},\ \Eprint {http://arxiv.org/abs/1304.5691} {arXiv:1304.5691 [gr-qc]} \BibitemShut {NoStop}%
\bibitem [{\citenamefont {Jusufi}\ \emph {et~al.}(2024)\citenamefont {Jusufi}, \citenamefont {Gonz\'alez},\ and\ \citenamefont {Leon}}]{Jusufi:2024ifp}%
  \BibitemOpen
  \bibfield  {author} {\bibinfo {author} {\bibfnamefont {K.}~\bibnamefont {Jusufi}}, \bibinfo {author} {\bibfnamefont {E.}~\bibnamefont {Gonz\'alez}}, \ and\ \bibinfo {author} {\bibfnamefont {G.}~\bibnamefont {Leon}},\ }\href@noop {} {\enquote {\bibinfo {title} {{Addressing the Hubble tension in Yukawa cosmology?}}}\ } (\bibinfo {year} {2024}),\ \Eprint {http://arxiv.org/abs/2402.02512} {arXiv:2402.02512 [astro-ph.CO]} \BibitemShut {NoStop}%
\bibitem [{\citenamefont {Jusufi}\ and\ \citenamefont {Sheykhi}(2024)}]{Jusufi:2024rba}%
  \BibitemOpen
  \bibfield  {author} {\bibinfo {author} {\bibfnamefont {K.}~\bibnamefont {Jusufi}}\ and\ \bibinfo {author} {\bibfnamefont {A.}~\bibnamefont {Sheykhi}},\ }\href@noop {} {\enquote {\bibinfo {title} {{Apparent Dark Matter Inspired by Einstein Equation of State}},}\ } (\bibinfo {year} {2024}),\ \Eprint {http://arxiv.org/abs/2402.00785} {arXiv:2402.00785 [gr-qc]} \BibitemShut {NoStop}%
\bibitem [{\citenamefont {Allahyari}\ \emph {et~al.}(2020)\citenamefont {Allahyari}, \citenamefont {Khodadi}, \citenamefont {Vagnozzi},\ and\ \citenamefont {Mota}}]{Allahyari:2019jqz}%
  \BibitemOpen
  \bibfield  {author} {\bibinfo {author} {\bibfnamefont {A.}~\bibnamefont {Allahyari}}, \bibinfo {author} {\bibfnamefont {M.}~\bibnamefont {Khodadi}}, \bibinfo {author} {\bibfnamefont {S.}~\bibnamefont {Vagnozzi}}, \ and\ \bibinfo {author} {\bibfnamefont {D.~F.}\ \bibnamefont {Mota}},\ }\href {\doibase 10.1088/1475-7516/2020/02/003} {\bibfield  {journal} {\bibinfo  {journal} {JCAP}\ }\textbf {\bibinfo {volume} {02}},\ \bibinfo {pages} {003} (\bibinfo {year} {2020})},\ \Eprint {http://arxiv.org/abs/1912.08231} {arXiv:1912.08231 [gr-qc]} \BibitemShut {NoStop}%
\bibitem [{\citenamefont {{Goebel}}(1972)}]{1972ApJ...172L..95G}%
  \BibitemOpen
  \bibfield  {author} {\bibinfo {author} {\bibfnamefont {C.~J.}\ \bibnamefont {{Goebel}}},\ }\href {\doibase 10.1086/180898} {\bibfield  {journal} {\bibinfo  {journal} {APJL}\ }\textbf {\bibinfo {volume} {172}},\ \bibinfo {pages} {L95} (\bibinfo {year} {1972})}\BibitemShut {NoStop}%
\bibitem [{\citenamefont {Cardoso}\ \emph {et~al.}(2009)\citenamefont {Cardoso}, \citenamefont {Miranda}, \citenamefont {Berti}, \citenamefont {Witek},\ and\ \citenamefont {Zanchin}}]{Cardoso:2008bp}%
  \BibitemOpen
  \bibfield  {author} {\bibinfo {author} {\bibfnamefont {V.}~\bibnamefont {Cardoso}}, \bibinfo {author} {\bibfnamefont {A.~S.}\ \bibnamefont {Miranda}}, \bibinfo {author} {\bibfnamefont {E.}~\bibnamefont {Berti}}, \bibinfo {author} {\bibfnamefont {H.}~\bibnamefont {Witek}}, \ and\ \bibinfo {author} {\bibfnamefont {V.~T.}\ \bibnamefont {Zanchin}},\ }\href {\doibase 10.1103/PhysRevD.79.064016} {\bibfield  {journal} {\bibinfo  {journal} {Phys. Rev. D}\ }\textbf {\bibinfo {volume} {79}},\ \bibinfo {pages} {064016} (\bibinfo {year} {2009})},\ \Eprint {http://arxiv.org/abs/0812.1806} {arXiv:0812.1806 [hep-th]} \BibitemShut {NoStop}%
\bibitem [{\citenamefont {Stefanov}\ \emph {et~al.}(2010)\citenamefont {Stefanov}, \citenamefont {Yazadjiev},\ and\ \citenamefont {Gyulchev}}]{Stefanov:2010xz}%
  \BibitemOpen
  \bibfield  {author} {\bibinfo {author} {\bibfnamefont {I.~Z.}\ \bibnamefont {Stefanov}}, \bibinfo {author} {\bibfnamefont {S.~S.}\ \bibnamefont {Yazadjiev}}, \ and\ \bibinfo {author} {\bibfnamefont {G.~G.}\ \bibnamefont {Gyulchev}},\ }\href {\doibase 10.1103/PhysRevLett.104.251103} {\bibfield  {journal} {\bibinfo  {journal} {Phys. Rev. Lett.}\ }\textbf {\bibinfo {volume} {104}},\ \bibinfo {pages} {251103} (\bibinfo {year} {2010})},\ \Eprint {http://arxiv.org/abs/1003.1609} {arXiv:1003.1609 [gr-qc]} \BibitemShut {NoStop}%
\bibitem [{\citenamefont {Jusufi}(2020)}]{Jusufi:2019ltj}%
  \BibitemOpen
  \bibfield  {author} {\bibinfo {author} {\bibfnamefont {K.}~\bibnamefont {Jusufi}},\ }\href {\doibase 10.1103/PhysRevD.101.084055} {\bibfield  {journal} {\bibinfo  {journal} {Phys. Rev. D}\ }\textbf {\bibinfo {volume} {101}},\ \bibinfo {pages} {084055} (\bibinfo {year} {2020})},\ \Eprint {http://arxiv.org/abs/1912.13320} {arXiv:1912.13320 [gr-qc]} \BibitemShut {NoStop}%
\bibitem [{\citenamefont {Cuadros-Melgar}\ \emph {et~al.}(2020)\citenamefont {Cuadros-Melgar}, \citenamefont {Fontana},\ and\ \citenamefont {de~Oliveira}}]{Cuadros-Melgar:2020kqn}%
  \BibitemOpen
  \bibfield  {author} {\bibinfo {author} {\bibfnamefont {B.}~\bibnamefont {Cuadros-Melgar}}, \bibinfo {author} {\bibfnamefont {R.~D.~B.}\ \bibnamefont {Fontana}}, \ and\ \bibinfo {author} {\bibfnamefont {J.}~\bibnamefont {de~Oliveira}},\ }\href {\doibase 10.1016/j.physletb.2020.135966} {\bibfield  {journal} {\bibinfo  {journal} {Phys. Lett. B}\ }\textbf {\bibinfo {volume} {811}},\ \bibinfo {pages} {135966} (\bibinfo {year} {2020})},\ \Eprint {http://arxiv.org/abs/2005.09761} {arXiv:2005.09761 [gr-qc]} \BibitemShut {NoStop}%
\bibitem [{\citenamefont {Konoplya}\ and\ \citenamefont {Stuchl\'\i{}k}(2017)}]{Konoplya:2017wot}%
  \BibitemOpen
  \bibfield  {author} {\bibinfo {author} {\bibfnamefont {R.~A.}\ \bibnamefont {Konoplya}}\ and\ \bibinfo {author} {\bibfnamefont {Z.}~\bibnamefont {Stuchl\'\i{}k}},\ }\href {\doibase 10.1016/j.physletb.2017.06.015} {\bibfield  {journal} {\bibinfo  {journal} {Phys. Lett. B}\ }\textbf {\bibinfo {volume} {771}},\ \bibinfo {pages} {597} (\bibinfo {year} {2017})},\ \Eprint {http://arxiv.org/abs/1705.05928} {arXiv:1705.05928 [gr-qc]} \BibitemShut {NoStop}%
\bibitem [{\citenamefont {Konoplya}(2023)}]{Konoplya:2022gjp}%
  \BibitemOpen
  \bibfield  {author} {\bibinfo {author} {\bibfnamefont {R.~A.}\ \bibnamefont {Konoplya}},\ }\href {\doibase 10.1016/j.physletb.2023.137674} {\bibfield  {journal} {\bibinfo  {journal} {Phys. Lett. B}\ }\textbf {\bibinfo {volume} {838}},\ \bibinfo {pages} {137674} (\bibinfo {year} {2023})},\ \Eprint {http://arxiv.org/abs/2210.08373} {arXiv:2210.08373 [gr-qc]} \BibitemShut {NoStop}%
\end{thebibliography}%






\end{document}